\documentclass[10pt]{article}
\ifx\macrosloaded\relax \else\let\macrosloaded\relax\fi
\newtheorem{theorem}{Theorem}[section]

\newtheorem{Prop}{Proposition}[section]

\newtheorem{Def}{Definition}[section]

\catcode`\@=11
\newcommand{\eqinsec}{\relax\@addtoreset{equation}{section}}
\renewcommand{\theequation}{\ifx\showlabels\iftrue\the\id\else\thesection.\arabic{equation}\fi}
\catcode`\@=13
\newcounter{supeq}
\newenvironment{subeq}
\stepcounter{equation}
\def\theequation{\ifx\showlabels\iftrue\the\id\else\thesection.\arabic{equation}\fi}
\newtoks\id
\newcommand{\eqlabel}[1]{\label{#1}\global\id={(#1)}} 

\newcommand{\tr}{\mbox{tr}}
\newcommand{\be}{\begin{equation}}
\newcommand{\eeq}{\end{equation}}
\newcommand{\bea}{\begin{eqnarray}}
\newcommand{\eea}{\end{eqnarray}}
\newcommand{\beaa}{\begin{eqnarray*}}
\newcommand{\eeaa}{\end{eqnarray*}}
\newcommand{\bseq}{\begin{subeq}}
\newcommand{\eseq}{\end{subeq}}
\newcommand{\ba}{\begin{array}}
\newcommand{\ea}{\end{array}}
\newcommand{\eql}{\eqlabel}

\def \rectangle#1#2{\hbox{\vrule\vbox to #2
{\hrule\hbox to #1{\hfil}\vfil\hrule}\vrule}}\def\square{\,\,\rectangle{7pt}{7 pt}\,\,}
\newcommand{\edd}{\end{document}}

\renewcommand{\c}{\cdot}
\newcommand{\NI}{\noindent}

\newcommand{\Lb}{\underline{L}}

\newcommand{\Si}{\Sigma}
\newcommand{\ga}{\gamma}
\newcommand{\Ga}{\Gamma}

\newcommand{\GGa}{{\bf \Gamma}}

\newcommand{\rr}{\mbox{${\bf R}$}}

\newcommand{\ggg}{\mbox{${\bf g}$}}

\newcommand{\dd}{\mbox{${\bf D}$}}

\newcommand{\omm}{\mbox{${\bf\om}$}}

\newcommand{\lap}{\mbox{$\bigtriangleup$}}
\newcommand{\lapp}{\mbox{$\bigtriangleup  \mkern-13mu / \,$}}
\newcommand{\nab}{\mbox{$\nabla$}}
\newcommand{\nabb}{\mbox{$\nabla \mkern-13mu /$\,}}

\newcommand{\Us}{\mbox{$U\mkern-13mu /$\,}}

\newcommand{\ddb}{\mbox{$\dd \mkern-13mu /$\,}}

\newcommand{\hot}{\widehat{\otimes}}

\newtheorem{Le}{Lemma}[section]

\newcommand{\Lie}{\mbox{$\cal L$}}

\newcommand{\lie}{\hat{\Lie}}

\newcommand{\nn}{\nonumber}

\newcommand{\chib}{\underline{\chi}}

\newcommand{\de}{\delta}

\newcommand{\ep}{\epsilon}
\newcommand{\xib}{\underline{\xi}}
\newcommand{\chih}{\hat{\chi}}

\newcommand{\chibh}{\underline{\hat{\chi}}}

\newcommand{\Fb}{\und{F}}

\newcommand{\Gb}{\und{G}}

\newcommand{\und}[1]{\underline{#1}}

\newcommand{\s}{\mbox{$\mkern 7mu$}}

\newcommand{\Nb}{\und{N}}

\newcommand{\vb}{\und{v}}

\newcommand{\Cb}{\und{C}}

\newcommand{\ub}{{\und{u}}}
\renewcommand{\c}{\cdot}

\newcommand{\M}{{\cal M}}

\renewcommand{\aa}{\underline{\alpha}}
\newcommand{\bb}{\underline{\beta}}
\renewcommand{\a}{\alpha}
\renewcommand{\b}{\beta}
\newcommand{\dual}{\mbox{}^{\star}\!}

\newcommand{\si}{\sigma}

\newcommand{\ro}{\rho}

\newcommand{\ze}{\zeta}
\newcommand{\divv}{\mbox{div}\mkern-19mu /\,\,\,\,}

\newcommand{\curll}{\mbox{curl}\mkern-19mu /\,\,\,\,}

\newcommand{\om}{\omega}
\newcommand{\oom}{\Omega}
\newcommand{\oomb}{\underline{\Omega}}
\newcommand{\omb}{\underline{\omega}}

\newcommand{\etab}{\underline{\eta}}
\newcommand{\la}{\lambda}

\newcommand{\dddd}{{\bf D} \mkern-13mu /\,}

\def\frac#1#2{{{#1}\over{#2}}}

\newcommand{\QQ}{{\cal Q}}
\newcommand{\QQb}{\underline{\cal Q}}

\newcommand{\acc}{\bar{K}}

\eqinsec
\let\showlabels\iffalse
\begin{document}

\title{Global characteristic problem for Einstein vacuum equations with small initial data:
 (I)\ \ \ \ \ \ \ \  The initial data constraints.}
\author{Giulio Caciotta\ \ \ \  Francesco Nicol\`{o}
\footnote{Dipartimento di Matematica, Universit\`{a} degli Studi di Roma
``Tor Vergata", Via della Ricerca Scientifica, 00133-Roma, Italy}
\\Universit\`{a} degli studi di Roma ``Tor Vergata"\date{\today}}\maketitle

\begin {abstract}
{We show how to prescribe the initial data of a characteristic problem satisfying the costraints, the smallness, the regularity and the
asymptotic decay suitable to prove a global existence result. In this paper, the first of two, we show in detail the construction of the initial data and
give a sketch of the existence result. This proof, which mimicks the analogous one for the non characteristic problem in \cite{Kl-Ni:book}, will be the
content of a subsequent paper.}
\end{abstract}
\newpage
\tableofcontents

\newpage
\section{Introduction}\label{S.1}
\subsection{Statement of the problem}\label{SS1.1}
Recently Sergiu Klainerman and one of the present authors (F.N.) gave a new proof of the stability result for the Minkowski spacetime,
\cite{Kl-Ni:book}, previously obtained by D.Christodoulou and S.Klainerman, \cite{C-K:book}. The proof refers to the spacetime outside the domain of
influence of a compact region with initial data on a spacelike hypersurface $\Si_0$. The global existence is proved via a bootstrap mechanism
in a way which avoids the choice of a specific set of coordinates as, for instance, the harmonic ones. In the present paper we
show that similar techniques can be used to solve a characteristic problem for the vacuum Einstein equations.

As in the non characteristic case, the problem is naturally split in two parts. In the first one the goal is showing how the initial data have to
be assigned, which of them can be given freely and which are determined by the costraint equations which, in the characteristic case, are
``ordinary differential" transport equations. In the second part  we prove, given the initial data, a (global) existence theorem. In this paper
we focus our attention on the first part and give only a broad sketch of the existence proof. Therefore while we discuss in a great detail the
way of obtaining the initial data satisfying the ``characteristic problem" we postpone to the second paper a detailed discussion on the minimal
regularity assumptions  required to prove the maximal development existence.

The characteristic problem has been treated by various authors, see for instance F.Cagnac, \cite{Cagnac}, H.Muller Zum Hagen, \cite{Muller}, 
H.Muller Zum Hagen and H.J.Seifert,
\cite{MullerSeifert} and the series of papers by Dossa, see \cite{Dossa} and references therein, but, in our opinion, the more significant paper
is the one by A.Rendall, \cite{rendall:charact}, where a thorough examination is
done to show how to obtain initial data satisfying the costraint equations and the harmonic conditions. In the present work, as our existence
theorem will be an adaptation of the result in \cite{Kl-Ni:book} which is, basically, coordinate independent, we discuss the costraint
equations for the initial data without having to worry about the harmonic conditions. This, we believe, is an advantage and a simplification
with respect to the work of A.Rendall. Moreover as our existence result is a global result, in a sense we are going to specify later on, it
uses, as in \cite{Kl-Ni:book}, a bootstrap mechanism based on some apriori estimates for a family of generalized ``energy integral norms".
Therefore we have to impose that the analogous norms, written in terms of the initial data as $L^2$ integrals on the null initial hypersurface,
be finite (and small). This requires that our initial data, in addition to satisfy the costraints, must have appropriate regularity and
asymptotic decay.

 To apply the bootstrap program and prove the global existence result, we need a separate local existence proof; we can use, to provide it,
the H.Muller Zum Hagen, H.Muller Zum Hagen and H.J.Seifert result, \cite{MullerSeifert}. Their results provide, nevertheless, a local existence proof
for the vacuum Einstein equations written in harmonic coordinates. Therefore, to use our initial data in proving these local existence
results we have, at least locally, to reexpress them in the harmonic gauge. This is done in the second part of the present paper.

 Finally we have to specify the geometry of the null hypersurface on which we prescribe the initial data. In this paper we consider as null
initial hypersurface the union of a portion of an ``outgoing null cone", $C_0$, and a portion of an ``incoming null cone",
$\Cb_0$, 
\footnote{The definition of ``outgoing null cone" and ``incoming null cone" are made precise in subsection 1.3 where we specify the properties of
$C_0$ and $\Cb_0$.} such that their intersection, $S_0$, is a two dimensional surface $S^2$-diffeomorphic,
$S_0\!=\!C_0\cap\Cb_0$. This particular initial hypersurface allows us to prove the existence of a maximal Cauchy development of
$C_0\cup\Cb_0$, using the same techniques developed in
\cite{Kl-Ni:book}. If, nevertheless, we restrict our goal only to proving the existence of initial data satisfying the costraints it turns out
that the procedure developed here can be also used to obtain initial data on a single outgoing cone truncated near the vertex.\footnote{The
truncation is to avoid the singularity problem at the cone vertex.}
Finally let us point out that in the paper by H.Muller Zum Hagen and H.J.Seifert, \cite{MullerSeifert} and also in the one by A.Rendall,
\cite{rendall:charact} the choice of the initial hypersurface is more general. There in fact the analogous of $C_0$ and $\Cb_0$ are two null
hypersurfaces, denoted $N_1,N_2$, intersecting on a codimension 2 surface $S$ which is not necessarily diffeomorphic to $S^2$. In particular
$N_1$ and $N_2$ can be diffeomorphic to null hyperplanes.\footnote{In \cite{MullerSeifert} the case where only one of the two hypersurfaces is
null is also considered.} We believe that our approach can also deal, with minor modifications, the case where  $N_1$ and $N_2$ are
diffeomorphic to null hyperplanes.

\subsection{Structure equations}\label{SS1.2}
Let $(\M,\ggg)$ a manifold with a Lorentzian metric. If we introduce in it a null orthonormal frame, the so called ``rep\`ere mobile" of
Cartan, see \cite{Sp:Spivak}, Vol.2, it is immediate to show that some differential relations between the covariant derivatives of the vector
fields forming the null frame are automatically satisfied. These relations, called ``structure equations", have the following expressions.
Let us write the null frame and its dual basis as
\beaa
\{e_{\alpha}\}=\{e_{1},e_{2},e_{3},e_{4}\}\ \ ,\ \ \{\theta^{\alpha}\}=\{\theta^{1},\theta^{2},\theta^{3},\theta^{4}\}
\eeaa
and define
\bea
\dd_{e_{\a}}e_{\b}\equiv \GGa^{\ga}_{\a\b}e_{\gamma}\ \ ,\ \ \rr(e_{\a},e_{\b})e_{\ga}\equiv \rr^{\de}_{\ \ga\a\b}\ e_{\de}\ ,
\eea
where $\rr$ is the Riemann tensor of the manifold $\M$.
The connection coefficients are the components $\GGa^{\ga}_{\a\b}$. 
The connection 1-form and curvature 2-form are 
\bea
\omm^{\a}_{\b}&\equiv&\GGa^{\a}_{\ga\b}\theta^{\ga}\nn\\
{\bf{\oom}}^{\a}_{\b}&\equiv&\frac{1}{2}\rr^{\a}_{\ \b\ga\de}\theta^{\ga}\wedge\theta^{\de}
\eea
They satisfy
the first and the second structure equations\index{structure equations},
see \cite{Sp:Spivak},\footnote{With the obvious modifications due to the Lorentzian metric.} 
\bea
&&d\theta^{\a}=-\omm^{\a}_{\ga}\wedge\theta^{\ga}\ \ \ \ \nn\\
&&d\om^{\de}_{\ga}=-\om^{\de}_{\si}\wedge\om^{\si}_{\ga}+{\bf\Omega}^{\de}_{\ga}\ .\ \ \ 
\eql{structureq}
\eea

 Although the structure equations are valid in a more general framework, let us assume that it is possible to foliate the manifold
$(\M,\ggg)$ with a double null foliation, that is by two families of null hypersurfaces (the analogous of the null cones in the Minkowski
spacetime) we denote $\{C(\la)\}$ and $\{\Cb(\nu)\}$. Moreover we assume that each two dimensional surface
$S(\la,\nu)\equiv C(\la)\cap\Cb(\nu)$ is diffeomorphic to $S^2$ and that the family of these surfaces defines a foliation of $(\M,\ggg)$.
The null moving frame we consider is a moving frame adapted to this foliation in the sense that $\{e_a\}\ a\in\{1,2\}$ are vector fields tangent
at each point $p\in\M$ to the surface $S(\la,\nu)$ containing this point and $e_4,e_3$ are null vector fields orthogonal to the $e_a$'s, outgoing and
incoming, respectively, which basically means ``tangent" to the null hypersurfaces $C(\la)$ and $\Cb(\nu)$. They can be chosen in such a way that,
denoted $L$ and $\Lb$ the vector fields associated to the null geodesics generating the $C(\la)$ and $\Cb(\nu)$ hypersurfaces, there is a
function $\oom$ such that the following relations hold
\bea
e_4=2\oom L\ \ ,\ \ e_3=2\oom\Lb\ .
\eea    
Under these assumptions the connection coefficients have the following expressions,
\bea
\chi_{ab}&=&\ggg(\dd_{{e_a}}{e_4},e_{b})\ \ \ ,\ \chib_{ab}\ =\ \ggg(\dd_{{e_a}}{e_3},e_{b})\nn\\
\xib_{a}&=&\frac{1}{2}\ggg(\dd_{{e_3}}{e_3},e_{a})=\frac{1}{2}{e_3}(\log\oom)\ggg({e_3},e_{a})=0\nn\\
\xi_{a}&=&\frac{1}{2}\ggg(\dd_{{e_4}}{e_4},e_{a})=\frac{1}{2}{e_4}(\log\oom)\ggg({e_4},e_{a})=0\nn\\
\omb&=&\frac{1}{4}\ggg(\dd_{{e_3}}{e_3},{e_4})=-\frac{1}{2}\dd_3(\log\oom)\nn\\
\om&=&\frac{1}{4}\ggg(\dd_{{e_4}}{e_4},{e_3})=-\frac{1}{2}\dd_4(\log\oom)\eql{1.5f}\\
\etab_{a}&=&-\ze_{a}+\nabb_a\log\oom\ ,\ \eta_{a}=\ze_a+\nabb_a\log\oom\nn\\
\ze_{a}&=&\frac{1}{2}\ggg(\dd_{e_{a}}{e_4},{e_3})\ .\nn
\eea
The first structure equations, see \ref{structureq}, written
explicitely in terms of the connection coefficients, are
\footnote{Hereafter $\dd_{e_a}=\dd_{a}$ and $\dd_{e_{(3,4)}}=\dd_{(3,4)}$.}
\bea
\dd_{a}e_b&=&\nabb_{a}e_b+\frac{1}{2}\chi_{ab}e_3+\frac{1}{2}\chib_{ab}e_4\nn\\
\dd_{a}e_3&=&\chib_{ab}e_b+\ze_{a}e_3\ \ \ ,\ \ \ \dd_{a}e_4=\chi_{ab}e_b-\ze_{a}e_4\nn\\
\dd_{3}e_a&=&\dddd_3 e_a+\eta_{a}e_3\ \ \ ,\ \ \ \dd_{4}e_a=\dddd_4 e_a+\etab_{a}e_4\eql{struct1}\\
\dd_{3}e_3&=&(\dd_3\log\oom)e_3\ \ \ ,\ \ \ \dd_{3}e_4=-(\dd_3\log\oom)e_4+2\eta_{b}e_{b}\nn\\
\dd_{4}e_4&=&(\dd_4\log\oom)e_4\ \ \ ,\ \ \ \dd_{4}e_3=-(\dd_4\log\oom)e_3+2\etab_{b}e_{b}\ .\nn
\eea
Keeping in mind that the structure equations are automatically satisfied in any Lorentzian manifold, $(\M,\ggg)$, 
they can be interpreted simply as a way of rewriting the first order covariant derivatives and the Riemann tensor in terms of the connection
coefficients and their first derivatives. Equations \ref{struct1} show it in a clear way. In the same way the second set
of equations in \ref{structureq} expresses the Riemann tensor in terms of the connection coefficients and their derivatives.

 These relations acquire, nevertheless, a different meaning if we consider them in a vacuum Einstein spacetime, namely in a manifold
$(\M,\ggg)$ where the Ricci part of the Riemann tensor is identically zero. In this case a subset of the second set of equations in
\ref{structureq} can be interpreted as a way of writing the vacuum Einstein equations in terms of the connection coefficients and their
first derivatives.\footnote{As we shall show in the sequel of this section these equations are of two types: evolution equations along the null
directions $e_3$, $e_4$ and elliptic equations of the Hodge type along the two dimensional surfaces $S(\la,\nu)$.}

 As our goal is posing and solving a class of initial data characteristic problems for the vacuum Einstein equations, we will consider initial
data given on a null hypersurface (a ``null outgoing cone" or the union of a ``null outgoing cone"  and an ``incoming" one). Instead of writing
the initial data in terms of the metric tensor components and their partial derivatives we will use as initial data the metric tensor components
and the connection coefficients restricted to the initial hypersurface.\footnote{The possibility of doing this will be discussed in the next
sections.} Therefore the subset of the structure equations defined above  will play the role of costraint equations that our initial data have
to satisfy.

 To better clarify this point let us recall that chosen a specific set of coordinates the initial data of the characteristic problem
can be assigned prescribing the various components of the Lorentzian metric and their first partial derivatives. The characteristic nature of the
problem implies that these quantities cannot be given freely, but have to satisfy some costraints. To associate these costraints to the structure
equations for the connection coefficients we observe that, once we have given on the null initial hypersurface a moving frame and a foliation,
all the first derivatives of the metric components or their Christoffel symbols can be expressed in terms of the connection coefficients and of
the derivatives of the moving frame vector fields (this will be discussed in any detail later on). Therefore the costraint equations for the first
derivatives of the metric components are the immediate consequence of this subset of the structure equations we have previously introduced.

 This approach has the advantage of presenting the costraint equations in a more covariant way, without requiring a specific set of
coordinates. Moreover the choice of a ``gauge" for our problem is associated to specifying a foliation on the initial
hypersurface and, subsequently, in the whole spacetime instead of the more usual choice of ``harmonic coordinates". 
\smallskip

 In the remaining part of this section we give the explicit form of the structure equations which will be interpreted as costraint equations and
in the following section we give the definition of the initial Cauchy characteristic problem. The remaining sections are devote to showing how one
can explicitely obtain a family of initial data satisfying the characteristic costraint equations and some general ``regularity smallness
conditions" which will allow to prove a global existence result.  
\subsubsection{Structure equations for vacuum Einstein manifolds}
We recall here, without derivation, see, for more details, \cite{Kl-Ni:book}, Chapter 3, the form of the Einstein equations written
as a subset of the structure equations.
\medskip

Denoting ${\bf R}_{\a,\b}={\bf Ricci}(e_{\a},e_{\b})$, we have:

 i) ${\bf R}_{44}=0$.
\bea
{\bf R}_{44}=\dd_4{\tr\chi}+\frac{1}{2}(\tr\chi)^2+2\om \tr\chi+|\chih|^2=0\eql{1.73c}
\eea

 ii) ${\bf R}_{33}=0$.
\bea
{\bf R}_{33}=\dd_3{\tr\chib}+\frac{1}{2}(\tr\chib)^2+2\omb \tr\chib+|\chibh|^2=0\eql{1.73de}
\eea

 iii) ${\bf R}_{4a}=0$.
\bea
{\bf R}_{4a}=\dddd_4\zeta+\frac{3}{2}\tr\chi\zeta+\zeta\chih-\divv\chih+\frac{1}{2}\nabb\tr\chi+\ddb_4\nabb\log\oom=0\eql{1.74c}
\eea

 iv) ${\bf R}_{3a}=0$.
\bea
{\bf R}_{3a}=\dddd_3\zeta+\frac{3}{2}\tr\chib\zeta+\zeta\chibh+\divv\chibh-\frac{1}{2}\nabb\tr\chib-\ddb_3\nabb\log\oom=0\eql{1.74cz}
\eea
 v) ${\bf R}_{ab}=0$.\ \ This equation is
\bea
{\bf R}_{ab}=-\frac{1}{2}({\bf R}_{a3b4}+{\bf R}_{b3a4})+{\bf R}_{bcac}=0\ ,\eql{1.77c}
\eea
and, from explicit computation we obtain:
\bea
&&-\frac{1}{2}({\bf R}_{a3b4}+{\bf R}_{b3a4})
=\dddd_4\chib_{ab}+\frac{1}{2}\left((\chi\c\chib)_{ab}+(\chi\c\chib)_{ba}\right)+(\dd_4\log\oom)\chib_{ab}\nn\\
&&\ \ \ \ \ \ \ \ \ \ \ \ \ \ \ \ \ \ \ \ \ \ \ \ -\left((\nabb\etab)_{ab}+(\nabb\etab)_{ba}\right)
-\left((\etab\ \!\etab)_{ab}+(\etab\ \!\etab)_{ba}\right)\\
&&\ \ \ \ {\bf R}_{bcac}=
{^{(2)}{\!\bf R}}_{bcac}+\frac{1}{2}\left(\tr\chi\chib_{ab}+\tr\chib\chi_{ab}-(\chi\c\chib)_{ab}-(\chi\c\chib)_{ba}\right)\nn\\
&&\ \ \ \ \ \ \ \ \ \ \ \ \ ={^{(2)}{\!\bf R}}_{ba}
+\frac{1}{2}\left(\tr\chi\chib_{ab}+\tr\chib\chi_{ab}-(\chi\c\chib)_{ab}-(\chi\c\chib)_{ba}\right)\ .\nn
\eea
Therefore equation \ref{1.77c} becomes
\bea
&&\dddd_4\chib_{ab}+\frac{1}{2}((\chi\c\chib)_{ab}+(\chi\c\chib)_{ba})+(\dd_4\log\oom)\chib_{ab}
-\left((\nabb\etab)_{ab}+(\nabb\etab)_{ba}\right)\eql{1.79c}\\
&&-\left((\etab\ \!\etab)_{ab}+(\etab\ \!\etab)_{ba}\right)=-^{(2)}{\!\bf R}_{ba}-\frac{1}{2}\left(\tr\chi\chib_{ab}
+\tr\chib\chi_{ab}-(\chi\c\chib)_{ab}-(\chi\c\chib)_{ba}\right)\ .\nn
\eea
Decomposing this equation in its trace and traceless part, with respect to the $a,b$ indices\ , we obtain, for the trace part:
\bea
\dddd_4\tr\chib\!+\!\chi\c\chib\!+\!(\dd_4\log\oom)\tr\chib\!-\!2(\divv\etab)\!-\!2|\etab|^2
=-^{(2)}{\!\bf R}\!-\!\frac{1}{2}\tr\chi\tr\chib\!+\!\chih\c\chibh\ ,\ \ \ 
\eea
which we rewrite, recalling that $^{(2)}{\!\bf R}=2{\bf K}$, twice the scalar curvature of the leaves $S(\la,\nu)$ of the null hypersurface
$C(\la)$,
\bea
\dddd_4\tr\chib+\tr\chi\tr\chib -2\om\tr\chib-2(\divv\etab)-2|\etab|^2+2{\bf K}=0\ .\ \ \ \ \ \eql{1.81c}
\eea
The equation associated to the traceless part is:
\bea
\dddd_4\chibh_{ab}+\frac{1}{2}\tr\chi\chibh_{ab}+\frac{1}{2}\tr\chib\chih_{ab}-2\om\chibh_{ab}
-2(\nabb\hot\etab)_{ab}-2(\etab\hot\etab)_{ab}=0.\eql{1.82c}
\eea
Equations \ref{1.81c}, \ref{1.82c} correspond to ${\bf R}_{ab}=0$. They can also be written in a similar way as evolution
equations along the incoming cones, namely:
\bea
&&\dddd_3\tr\chi\!+\!\tr\chib\tr\chi\!-\!2\omb\tr\chi\!-\!2(\divv\eta)\!-\!2|\eta|^2\!+\!2{\bf K}\!=0\ ,\ \ \ \eql{1.83cz}\\
&&\dddd_3\chih_{ab}\!+\!\frac{1}{2}\tr\chib\chih_{ab}\!+\!\frac{1}{2}\tr\chi\chibh_{ab}\!-\!2\omb\chih_{ab}
\!-\!2(\nabb\hot\eta)_{ab}\!-\!2(\eta\hot\eta)_{ab}=0\ .\ \ \ \ \ \ \ \ \ \eql{1.84cz}
\eea
Interpreting these equations as constraint equations it is clear that their expressions \ref{1.81c},
\ref{1.82c} are suitable for the initial data on the outgoing part of the null initial hypersurface, while those provided by
\ref{1.83cz}, \ref{1.84cz} will be used to assigne $\tr\chi$ and $\chih$ on the incoming part of the null initial hypersurface.
\smallskip

 vi) ${\bf R}_{34}=0$. This equation can be written as 
\bea
\rr_{34}=\frac{1}{2}\rr_{4343}+\sum_{c}\rr_{c3c4}=0
\eea
and the explicit expressions of $\rr_{4343}$ and $\rr_{c3c4}$ are:
\bea
&&\frac{1}{4}\rr_{4343}=\left[\dd_3\om+\dd_4\omb-4\om\omb-3|\ze|^2+|\nabb\log\oom|^2\right]\nn\\
&&\sum_{c}\rr_{c3c4}=-\left[\dd_4\tr\chib+\frac{1}{2}\tr\chi\tr\chib-2\om\tr\chib+\chih\c\chibh-2\divv\etab-2|\etab|^2\right]\nn
\eea 
so that, finally, we obtain
\bea
&&\dd_3\om+\dd_4\omb-4\om\omb-3|\ze|^2+|\nabb\log\oom|^2-\frac{1}{2}\bigg[\dd_4\tr\chib+\frac{1}{2}\tr\chi\tr\chib-2\om\tr\chib\nn\\
&&+\chih\c\chibh-2\divv\etab-2|\etab|^2\bigg]=0\ .\eql{1.20z}
\eea
We also list without derivations the remaining structure equations which depend on the conformal part of the Riemann tensor. They
will be used when we require the appropriate regularity and smallness conditions for the initial data. To do it
we first define the various conformal Riemann components with respect to the null orthonormal frame, 
\bea
&&\a({\bf C})(X,Y)={\bf C}(X,e_4,Y,e_4)\nn\\
&&\b({\bf C})(X)=\frac{1}{2}{\bf C}(X,e_4,e_3,e_4)\nn\\
&&\ro({\bf C})=\frac{1}{4}{\bf C}(e_3,e_4,e_3,e_4)\eql{3.1.19zaa}\\
&&\si({\bf C})=\frac{1}{4}\ro(^\star {\bf C})=\frac{1}{4}{^\star{\bf C}}(e_3,e_4,e_3,e_4)\nn\\
&&\bb({\bf C})(X)=\frac{1}{2}{\bf C}(X,e_3,e_3,e_4)\nn\\
&&\aa({\bf C})(X,Y)={\bf C}(X,e_3,Y,e_3)\nn
\eea
with $X,Y$ arbitrary vectors tangent to $S$ at $p$ and the left Hodge dual of ${\bf C}$:
\bea
^\star C_{\alpha\beta\gamma\delta}=\frac{1}{2}\ep_{\alpha\beta\mu\nu}
C^{\mu\nu}_{\ \ \gamma\delta}\ ,
\eea
where $\in^{\alpha\beta\gamma\delta}$ are the components of the volume element in $\M$. 
\subsubsection{The remaining structure equations.}\label{SS1.2.2}
The remaining structure equations, due to their dependance on the conformal part of the Riemann tensor,
do not play the role of constraint equations for the initial data. They are automatically satisfied as they are just the
explicit expression of the conformal part of the Riemann tensor in terms of the connection coefficients and their first derivatives.
We list all of them.
\bea
\dddd_4\chih+\tr\chi\chih-(\dd_4\log\oom)\chih=-\a\eql{1.21w}
\eea
\bea
\dddd_4\zeta+2\chi\c\zeta+\ddb_4\nabb\log\oom=-\b\eql{1.22w}
\eea
\bea
\nabb\tr\chi-\divv\chi-\zeta\c\chi+\zeta\tr\chi=\b\eql{1.23w}
\eea
\bea
&&\dd_4\tr\chib+\frac{1}{2}\tr\chi\tr\chib+(\dd_4\log\oom)\tr\chib+\chih\c\chibh+2\divv\zeta-
2\lapp\log\oom\nn\\
&&\ -2|\zeta|^2-4\zeta\c\nabb\log\oom-2|\nabb\log\oom|^2=2\ro\eql{1.24w}
\eea
\bea
&&\dd_3\tr\chi+\frac{1}{2}\tr\chib{\tr\chi}+(\dd_3\log\oom)\tr\chi+\chibh\c\chih-2\divv\zeta-2|\zeta|^2\nn\\
&&\ -2\lapp\log\oom-4\zeta\c\nabb\log\oom-2|\nabb\log\oom|^2=2\ro\eql{1.25w}
\eea
\bea
{\bf K}+\frac{1}{4}\tr\chi\tr\chib-\frac{1}{2}\chibh\c\chih=-\ro\eql{1.26w}
\eea
\bea
\curll\zeta-\frac{1}{2}\chibh\wedge\chih=\si\eql{1.27w}
\eea
\bea
\dddd_3\zeta+2\chib\c\zeta-\ddb_3\nabb\log\oom=-\bb\eql{1.28w}
\eea
\bea
\nabb\tr\chib-\divv\chib+\zeta\c\chib-\zeta\tr\chib=-\bb\eql{1.29w}
\eea
\bea
\dddd_3\chibh+\tr\chib\ \!\chibh-(\dd_3\log\oom)\chibh=-\aa\eql{1.30w}
\eea
It is important to observe that while $\b,\ro,\si,\bb$ can be expressed in two different ways in terms of the connection coefficients and their
derivatives, this is not true in the case of $\a$ and $\aa$.
\subsection{The initial data for the characteristic problem}\label{SS2.1}
In this subsection we present a definition of the initial data for the Einstein vacuum characteristic Cauchy problem from the point of view discussed
before, without prescribing a specific choice of coordinates. This will be done in three steps, first we specify a foliation on $\cal C$ and we
define a degenerate metric ``adapted" to this foliation, second we introduce as initial data some tensor fields on $\cal C$ which later have to
be interpreted as the restriction of the connection coefficients on the initial hypersurface and, third, we define the costraints they have to
satisfy. Finally we discuss the relation between these initial data and the more usual ones expressed in terms of partial derivatives of the
components of the metric tensor.
 
\subsubsection{Foliation of the initial data hypersurface}

 Let us consider two three-dimensional manifolds with an edge, we denote
$(\Cb_0,\underline{g}_0)$ and
$(C_0,g_0)$, endowed with two metric tensors
$\underline{g}_0$ and $g_0$. Let $(\Cb_0,\underline{g}_0)$ and
$(C_0,g_0)$ have the following properties:
\smallskip

i) the metrics $\underline{g}_0$ and $g_0$ are degenerate.\footnote{$\underline{g}_0$ and $g_0$ are
sufficiently regular to make all next properties meaningful.} This means the following: for any
$p\in
\Cb_0$ there exist on $T_p\Cb_0$ a vector $N_0$ such that
\bea
\underline{g}_0(N_0,Y)=0\ \ \forall\  Y\in T_p\Cb_0\ .
\eea
and the same definition for the metric $g_0$ of $C_0$. 
\smallskip

ii) The intersection of $C_0$ and $\Cb_0$ be a two-dimensional surface diffeomorphic to
$S^2$,
$S_0=C_0\cap\Cb_0$ and on $S_0$ the restrictions of $g_0$ and of $\underline{g}_0$ coincide.

We will call these two manifolds the ``initial (null) outgoing cone" $C_0$ and the ``initial (null) incoming cone" $\Cb_0$.\footnote{More precisely, as
it will be discussed in detail later on, $C_0$ and $\Cb_0$ when immersed in the Einstein spacetime, are truncated portions of an outgoing null
hypersurface and an incoming null hypersurface, the analogue of null cones of the Minkowski spacetime, see also \cite{Kl-Ni:book}.}
\smallskip

iii) The null geodesics on $\Cb_0$ and $C_0$ have past end points on $S_0$.
\smallskip

 Their union defines the initial hypersurface, ${\cal C}=C_0\cup\Cb_0$, of our characteristic problem. Once
$\cal C$ is given, to prescribe the initial data for the characteristic problem we have to specify some other properties of our initial outgoing and
incoming (truncated) ``cones". 
\smallskip

{\bf  i) Properties of $C_0$:}
\smallskip

a) $C_0$ is foliated by a family of two dimensional surfaces $S_0(\nu)$ diffeomorphic to $S^2$ with
$\nu\in[\nu_0,\infty)$, defined as the level surfaces of a scalar function $\ub(p)$ defined on $C_0$:
\bea
S_0(\nu)=\{p\in C_0|\ub(p)=\nu\}\eql{1.2}\ .
\eea
with $S_0(\nu_0)=S_0$.
The function $\ub(p)$, whose level surfaces are $S_0(\nu)$, is defined in the following way: let $\ga(\vb)$ be the null
geodesic on $C_0$ the affine parameter $\vb$ being such that $L$, the null geodesic tangent vector field, satisfies $L=2\frac{\partial}{\partial
\vb}$; let $\ga(\vb)$ starting on $S_0$ with affine parameter $\vb=0$ and passing through $p$ when $\vb=\vb(p)$, then we define
\bea
\ub(p)=\nu_0+\int_0^{\vb(p)}(4\oom)^{-2}(\ga(\vb'))d\vb'\eql{1.3}
\eea
and the scalar function $\oom$ will be specified later on.\footnote{The value of $\ub$ at $S_0$ is somewhat arbitrary, but in the following we
will choose to connect it to the ``radius" of $S_0$,
${\tilde r}_0\equiv\sqrt{4\pi^{-1}|S_0|_{g_0}}$, posing $\nu_0\in[c_1{\tilde r}_0, c_2{\tilde r}_0]$ with $c_1,c_2$ approximately equal to one.
The precise bounds on $c_1,c_2$ will be given in Section 2.}
\smallskip

 Once defined the leaves $S_0(\nu)$ of the foliation we introduce on $C_0$ a null orthonormal frame, $\{e_4,e_1,e_2\}$, 
choosing $e_4=2\oom L$, with $L$ the null geodesic vector field on $C_0$, and $\{e_1,e_2\}$ an orthonormal frame tangent to $S_0(\nu)$. 

 We define $(\ub,\theta,\phi)$ as adapted coordinates on $C_0$,
$\theta(p),\phi(p)$ being the angular coordinates of the point $p_0\in S_0=S_0(\nu_0)$ such that the null
geodesic starting from $p_0$ reaches the point $p$. With this definition an arbitrary point $p$ on $S_0(\nu)$ has coordinates
$(\nu,\theta(p),\phi(p))$.
\smallskip

b) On each $S_0(\nu)$ we define  the metric
tensor $\ga_{ab}=\ga(\nu)_{ab}$, restriction to $S_0(\nu)$ of the metric tensor ${g_0}$, $\ga(\nu)_{ab}=({{g_0}}|_{S_0(\nu)})_{ab}$. The
second null fundamental form $\chi$ is defined through the Lie derivative of $g_0$ with respect to
$N\!=\!2\oom^2\!L$,\footnote{The vector field $N\!=\!2\oom^2\!L$
defines a diffeomorphism $\Phi_N$ such that $\Phi_{\!N}(\de)[S_0(\nu)]\!=\!S_0(\nu\!+\!\de)$.}
\bea
2\oom\chi=\Lie_Ng_0
\eea
which, in the $(\nu,\theta,\phi)$ coordinates, takes the form
\bea
\chi_{ab}=\frac{1}{2\oom}\frac{\partial\ga_{ab}}{\partial\nu}\ ,
\eea
and we require that $\ga_{ab}$ be such that $\tr\chi>0$ on the whole $C_0$.
We define next, starting from the function $\oom$, the scalar function $\om$,
\bea
\om=-\frac{1}{2\oom}\frac{\partial\log\oom}{\partial\nu}\ .
\eea
We introduce on $C_0$ some other quantities: a one form $\ze$ tangent to the leaves $S_0(\nu)$ and two more quantities, a scalar function,
denoted by $\omb$, and a symmetric tensor, $\chib$ whose trace and traceless parts we denote $\tr\chib$ and $\chibh$. 

$\ze$, $\omb$ and $\chib$ do not have a direct geometrical meaning, but they will acquire it when $C_0$ becomes a null hypersurface embedded in
the vacuum Einstein spacetime $(\M,\ggg)$.

 Finally we assume that the quantities we have introduced are sufficiently regular so that on $C_0$, together with
$\chi,\chib,\om,\omb,\ze$, the derivatives\footnote{The assumptions on $\dddd_4\nabb^{k-1}\hat\chi$ are used in subsection \ref{SS2.4}. }
\bea
{\nabb^k\chi}\ ,\ {\nabb^k\chib}\ ,\  \nabb^k\om ,\  \nabb^k\omb\ ,\ \nabb^k\ze\ , \nabb^{k-1}\dddd_4\hat\chi
\eea
are well defined with $k\in [1,q]$ and $\nabb$ the covariant derivative associated to the metric $\ga_{ab}$. The choice of $q$ will be specified
in the final version of the existence theorem in subsection \ref{SS.4.3}.

 As we said these quantities have to satisfy some costraints. More precisely we require that the tensor fields
$\chi,\om,\ze,\chib,\omb$ satisfy the following (costraint) equations:\footnote{The last equation has been written in a different way using the
expression of the commutator $[\dd_4,\dd_3]$.}
\bea
&&\dd_4\tr\chi+\frac{1}{2}(\tr\chi)^2+2\om \tr\chi+|\chih|^2=0\eql{1.11}\\
&&\dddd_4\zeta+\zeta\chi+\tr\chi\zeta-\divv\chi+\nabb\tr\chi+\ddb_4\nabb\log\oom=0\eql{1.12}\\
&&\dddd_4\tr\chib+\tr\chi\tr\chib -2\om\tr\chib-2(\divv\etab)-2|\etab|^2+2{\bf K}=0\\
&&\dddd_4\chibh+\frac{1}{2}\tr\chi\chibh+\frac{1}{2}\tr\chib\chih+(\dd_4\log\oom)\chibh+\nabb\hot(\ze-\nabb\log\oom)\nn\\
&&-(\ze-\nabb\log\oom)\hot(\ze-\nabb\log\oom)=0\eql{1.13c}\\
&&\dd_4\omb\!-\!2\om\omb\!-\!\ze\c\nabb\log\oom\!-\!\frac{3}{2}|\ze|^2\!+\!\frac{1}{2}|\nabb\log\oom|^2\!-\!\big({\bf
K}\!+\!\frac{1}{4}\tr\chi\tr\chib\!-\!\frac{1}{2}\chih\c\chibh\big)\!=\!0\ \ \ \ \ \ \ \ \ \eql{1.41wz}
\eea
The nature of these equations is clear when the hypersurface $\cal C$ is considered embedded in
the Einstein spacetime $(\M,\ggg)$. In fact they are, restricted to $\cal C$, part of the vacuum Einstein equations previously written, see
\ref{1.73c},..., \ref{1.20z}.\footnote{In fact the covariant derivative $\dddd_4$ is associated to the four dimensional metric $\ggg$.} $\ze$, 
$\omb$ and $\chib$ are connection coefficients and their costraint equations are some of the structure equations defined in $(\M,\ggg)$. On the
other side observe that we can define the constraints that the initial data have to satisfy in a formulation ``intrinsic" to $\cal C$, which
does not require to consider $\cal C$ embedded in a Lorentzian manifold.\footnote{In principle both formulations are possible, the intrinsic one
is more general, but subsequently, the explicit construction of the initial data will be obtained following the second formulation.} To do it we
observe that equations
\ref{1.11}...\ref{1.41wz} can be rewritten as transport equations for scalar quantities along $C_0$, requiring that $\{e_A\}$ be a Fermi transported
orthonormal frame.\footnote{$\{e_A\}$ is Fermi transported if $\dddd_4e_a=0$. This implies that the components of $e_A$ satisfy the equations
${\partial_{\nu} e_A^a}+\oom\chi^a_{\ b}e_A^b=0$, see \cite{Kl-Ni:book}, appendix to Chapter 3, which are intrinsecally defined on $C_0$.
Other choice for transporting the $e_A$'s along $C_0$ could also be adopted.} With this frame $\{e_A\}$ the previous equations can be rewritten as,
see
\cite{Kl-Ni:book} Chapter 4,
\bea
&&\frac{\partial\tr\chi}{\partial\nu}+\frac{\oom\tr\chi}{2}\tr\chi+2\oom\om \tr\chi+\oom|\chih|^2=0\nn\\
&&\frac{\partial\zeta_A}{\partial\nu}+\oom\chi_{AB}\ze_B+\oom\tr\chi\ze_A-\oom(\divv\chi)_A+\oom\nabb_A\tr\chi
+\frac{\partial\nabb_A\log\oom}{\partial\nu}=0\nn\\
&&\frac{\partial\tr\chib}{\partial\nu}+\oom\tr\chi\tr\chib -2\oom\om\tr\chib-2\oom(\divv\etab)-2\oom|\etab|^2+2\oom{\bf K}=0\nn\\
&&\frac{\partial\chibh_{AB}}{\partial\nu}+\frac{\oom\tr\chi}{2}\chibh_{AB}+\frac{\oom\tr\chib}{2}\chih_{AB}
+(\frac{\partial\log\oom}{\partial\nu})\chibh_{AB}+\oom\nabb_A\hot(\ze-\nabb\log\oom)_B\nn\\
&&-\oom(\ze_A-\nabb_A\log\oom)\hot(\ze-\nabb\log\oom)_B=0\eql{1.41wq}\\
&&\frac{\partial\om}{\partial\nu}-2\oom\om\omb-\oom\ze_C\nabb_C\log\oom+\frac{1}{2}\left[-3\oom|\ze|^2+\oom|\nabb\log\oom|^2\right.\nn\\
&&\left.+\oom\bigg({\bf K}+\frac{1}{4}\tr\chi\tr\chib-\frac{1}{2}\chih\c\chibh\bigg)\right]=0\ .\nn
\eea
Written in this way these equations appear as ordinary differential equations on $C_0$ for some scalars functions and, therefore, can be
interpreted as the intrinsecally defined (costraint) equations for the initial data. 
\smallskip

{\bf ii) Properties of $\Cb_0$:}

 The properties we require for $\Cb_0$ are of the same type as those for $C_0$.
\smallskip

a) $\Cb_0$ is foliated by a family of two dimensional surfaces ${S}_0(\la)$ diffeomorphic to $S^2$ with
$\la\in[\la_1,\la_0]$, $\la_0<0$,\footnote{It is enough to require that $|\la_0|$ is upper bounded.} defined as the level surfaces of a scalar function
$u(p)$ defined on $\Cb_0$:
\bea
{\underline S}_0(\la)=\{p\in \Cb_0|u(p)=\la\}\ .
\eea

b) We require that \[{\underline S}_0(\la_1)=S_0=S_0(\nu_0)\ .\] 

The construction of the foliation proceeds as in the previous case and we do not repeat it here. We only observe that a function
$\underline{\oom}$ is assigned on
$\Cb_0$, analogous to the function
$\oom$ defined on the ``outgoing cone",  and
$u(p)$ is defined as
\footnote{When immersed in the spacetime the null hypersurface $\Cb_0$ is an ``incoming
null hypersurface" and its foliation is defined starting from the two dimensional surface $S_0(\la_1)$. $v=0$ is associated to the points on
$S_0(\la_1)$ and the negative values of $v$ correspond to two dimensional surfaces of smaller radius.}
\bea
u(p)=\la_1+\int_0^{v(p)}(4\underline{\oom})^{-2}(\ga(v'))dv'\ .\eql{1.47yt}
\eea
where $\ga$ is the null geodesic starting at the point $p_0\in S_0$ which reaches the point $p$.\footnote{As for $\ub$, the choice of the value for $u$
at $S_0$ is somewhat arbitrary. Nevertheless in Section 2 we will connect it to the ``radius" of $S_0$,
${\tilde r}_0=\sqrt{4\pi^{-1}|S_0|_{{\underline g}_0}}$, posing $|\la_1|\in[c_1{\tilde r}_0, c_2{\tilde r}_0]$ with
$c_1,c_2$ approximately equal to one.}

Once we have defined the leaves ${\underline S}_0(\la)$ of this foliation we introduce, as before, a null
orthonormal frame
$\{e_3,e_1,e_2\}$ on $\Cb_0$ choosing $e_3=2{\underline\oom}\ \!\Lb$, with $\Lb$ the null geodesic vector field on $\Cb_0$, and
$\{e_1,e_2\}$ an orthonormal frame tangent to ${\underline S}_0(\la)$.
 
 As done before on $C_0$ we choose $(u,\theta,\phi)$ as adapted coordinates on $\Cb_0$,
$\theta(p),\phi(p)$ being defined as the angular coordinates of the point $p_0\in{\underline S}_0(\la_1)$ such that the null
geodesic starting from it reaches the point $p$. Therefore an arbitrary point $p$ on
${\underline S}_0(\la)$ has coordinates $(\la,\theta(p),\phi(p))$.
\smallskip

b)  As before we define on each ${\underline S}_0(\la)$ the metric
tensor $\ga_{ab}$, restriction of the metric tensor $\underline{g}_0$, $\ga(\la)_{ab}=({\underline{g}_0|_{S_0(\la)}})_{ab}$. To
the metric tensor $\underline{g}_0$ is associated the second null fundamental form relative to $e_3$,
\bea
2\oomb\chib_{ab}=\Lie_{\Nb}\ga
\eea
where $\Nb=2\oomb^2\Lb$,\footnote{The vector field $\Nb\!=\!2\oomb^2\Lb$ defines a diffeomorphism $\Phi_{\Nb}$ such that
$\Phi_{\Nb}(\de)[{\underline S}_0(\la)]\!=\!{\underline S}_0(\la\!+\!\de)$.} which in the $(\la,\theta,\phi)$ coordinates has the form
\bea
\chib_{ab}=\frac{1}{2\oomb}\frac{\partial\ga_{ab}}{\partial\la}\ .
\eea
Here we require $\tr\chib|_{{\underline S}_0(\la_1)}<0$ (which justifies the name ``incoming cone" for $\Cb_0$).

 We define on $\Cb_0$, starting from the function $\oomb$, a scalar function $\omb$:
\bea
\omb=-\frac{1}{2\oomb}\frac{\partial\log\oomb}{\partial\la}
\eea
which again will acquire a geometrical meaning once the characteristic initial value problem is stated.
As in the $C_0$ case, we define on $\Cb_0$ the ``torsion" one form $\ze$, a symmetric tensor, $\chi$ on $\Cb_0$ whose
trace and traceless parts we denote $\tr\chi$ and $\chih$ and a scalar function $\om$.  Again on $\Cb_0$ $\chi$ and $\om$ do not have  a direct geometrical
meaning, but will acquire it when
$\Cb_0$ becomes a null hypersurface embedded in the vacuum Einstein spacetime $(\M,\ggg)$.

Also on $\Cb_0$ the introduced quantities have to satisfy some costraints. More precisely we require that the tensor fields
$\chi,\om,\ze,\chib,\omb$ satisfy on $\Cb_0$ the following costraint equations, again written in a way intrinsic to $\Cb_0$:
\bea
&&\frac{\partial\tr\chib}{\partial\la}+\frac{\oomb\tr\chib}{2}\tr\chib+2\oomb\om \tr\chib+\oomb|\chibh|^2=0\nn\\
&&\frac{\partial\zeta_A}{\partial\la}+\oomb\chib_{AB}\ze_B+\oomb\tr\chib\ze_A+\oomb(\divv\chib)_A-\oomb\nabb_A\tr\chib
-\frac{\partial\nabb_A\log\oomb}{\partial\la}=0\nn\\
&&\frac{\partial\tr\chi}{\partial\la}+\oomb\tr\chib\tr\chi -2\oomb\omb\tr\chi-2\oomb(\divv\eta)-2\oomb|\eta|^2+2\oomb{\bf K}=0\nn\\
&&\frac{\partial\chih_{AB}}{\partial\la}+\frac{\oomb\tr\chib}{2}\chih_{AB}+\frac{\oomb\tr\chi}{2}\chibh_{AB}
+\frac{\partial\log\oomb}{\partial\la}\chih_{AB}-\oomb\nabb_A\hot(\ze+\nabb\log\oomb)_B\nn\\
&&-\oomb(\ze_A+\nabb_A\log\oomb)\hot(\ze+\nabb\log\oomb)_B=0\eql{1.41wp}\\
&&\frac{\partial\om}{\partial\la}-2\oomb\om\omb+\oomb\ze_C\nabb_C\log\oomb+\frac{1}{2}\left[-3\oomb|\ze|^2+\oomb|\nabb\log\oomb|^2\right.\nn\\
&&\left.+\oomb\big({\bf K}+\frac{1}{4}\tr\chi\tr\chib-\frac{1}{2}\chih\c\chibh\big)\right]=0\ .\nn
\eea
Finally we assume that these quantities are sufficiently regular so that on $\Cb_0$, together with
$\chi,\chib,\om,\omb,\ze$, the derivatives
\bea
{\nabb^k\chi}\ ,\ {\nabb^k\chib}\ ,\  \nabb^k\om ,\  \nabb^k\omb\ ,\ \nabb^k\ze\ , \nabb^{k-1}\dddd_3\hat{\chib}
\eea
are well defined with $k\in [1,q]$ and $\nabb$ the covariant derivative associated to the metric $\ga_{ab}$.
\smallskip

{\bf Remarks:}
 
{a) It is appropriate to observe that the definitions and intrinsic properties we have given for $C_0$ and $\Cb_0$ do not specify completely their
relative position when they ``become" embedded hypersurfaces in the Einstein spacetime. Nevertheless the way we define explicitely the hypersurface $\cal
C$, to solve the costraint equations, as an embedded hypersurface in a $R^4$ manifold implies that $C_0$ and $\Cb_0$ have to be interpreted as
approximate null cones with their vertices lying on the vertical axis passing through the origin. See also the remark at the beginning of subsection
\ref{SS4.4}.

b) We have written the costraint equations in such a way that there is an exact symmetry between the equations on $C_0$ and those
on $\Cb_0$. To pass from the first to the second group it is enough to interchange the underlined quantities with the non
underlined ones, $\nu$ with $\la$ and finally $\ze$ with $-\ze$, a detailed discussion is in \cite{Kl-Ni:book}, Chapter 3.

\NI This is, nevertheless, a somewhat delicate point: one could be tempted to believe that the coordinates $(\nu, \theta,\phi)$ on $C_0$ and
$(\la,
\theta,\phi)$ on $\Cb_0$ are  the restriction, on
$C_0$ and $\Cb_0$ respectively, of a set of coordinates for the Einstein vacuum spacetime.\footnote{This remark holds even if the initial
hypersurface $\cal C$ is embedded in a Lorentz spacetime.} In fact, as discussed in \cite{Kl-Ni:book}, Chapter 3 for the non characteristic
case, when the existence result is achieved we have a vacuum Einstein spacetime $(\M,{\ggg})$ and $C_0$, $\Cb_0$ are embedded null
hypersurfaces part of a double null foliation made by family of null hypersurfaces $\{C(\la)\}$, $\{\Cb(\nu)\}$, namely $i(C_0)=C(\la_1)\ ,\
i(\Cb_0)=\Cb(\nu_0)$. Also in this case we can use
$u,\ub,\theta,\phi$ as coordinates, but it is \underline{not} true that in this set of coordinates we have:
\[N=\frac{\partial}{\partial\ub}\ ,\ \Nb=\frac{\partial}{\partial u}\ ,\]
as this will imply $[N,\Nb]=0$, which is false in a curved spacetime. The conclusion is that the introduction of a  unique set of
coordinates destroys the symmetry of the previous coordinate independent formulation. This
will be even more clear when, to prove the existence of initial data satisfying the constraints, we will consider $\cal C$ immersed in a Lorentz
spacetime and we will assigne to it a specific set of coordinates. The same loss of symmetry will appear when, at the end of this section, we
show how to connect our approach to a more standard one and express the connection coefficients in terms of partial derivatives of the metric
components.}

\begin{Def}\label{D1.1}
The initial data set of the characteristic problem consists in an initial data set relative to the ``null outgoing cone"
and an initial data set relative to the ``null incoming cone",
\bea
\big\{C_0\ ;\ga_{ab},\oom,\ze,\chib,\omb\big\}\cup\left\{\Cb_0\ ;\ga_{ab},\oomb,\ze,\chi,\om\right\}\ .
\eea
We denote $\{S(\nu)\}$ and $\{{\underline S}(\la)\}$ the leaves of the foliations of $C_0$ and $\Cb_0$, defined through the
functions $\oom$ and $\oomb$, with $\nu\in[\nu_0,\infty)$, $\la\in[\la_1,\la_0<0]$ and such that ${\underline
S}(\la_1)=S(\nu_0)=S_0=C_0\cap\Cb_0$. On $C_0$, assuming $\{e_A\}$ a Fermi transported orthonormal frame,
{\em
\bea
\chib \ ,\ \ \omb \ ,\ \ \ze,\ \ \chi_{ab}=\frac{1}{2\oom}\frac{\partial\ga_{ab}}{\partial\nu}\ \ \eea}
satisfy the following ``costraint equations":
{\em
\bea
&&\frac{\partial\tr\chi}{\partial\nu}+\frac{\oom\tr\chi}{2}\tr\chi+2\oom\om \tr\chi+\oom|\chih|^2=0\nn\\
&&\frac{\partial\zeta_A}{\partial\nu}+\oom\chi_{AB}\ze_B+\oom\tr\chi\ze_A-\oom(\divv\chi)_A+\oom\nabb_A\tr\chi
+\frac{\partial\nabb_A\log\oom}{\partial\nu}=0\nn\\
&&\frac{\partial\tr\chib}{\partial\nu}+\oom\tr\chi\tr\chib -2\oom\om\tr\chib-2\oom(\divv\etab)-2\oom|\etab|^2+2\oom{\bf K}=0\nn\\
&&\frac{\partial\chibh_{AB}}{\partial\nu}+\frac{\oom\tr\chi}{2}\chibh_{AB}+\frac{\oom\tr\chib}{2}\chih_{AB}
+(\frac{\partial\log\oom}{\partial\nu})\chibh_{AB}+\oom\nabb_A\hot(\ze-\nabb\log\oom)_B\nn\\
&&-\oom(\ze_A-\nabb_A\log\oom)\hot(\ze-\nabb\log\oom)_B=0\eql{1.41wqz}\\
&&\frac{\partial\om}{\partial\nu}\!-2\oom\om\omb-\oom\ze_C\nabb_C\log\oom\!+\!\frac{1}{2}\!\left[-3\oom|\ze|^2\!+\oom|\nabb\log\oom|^2\right.\nn\\
&&\left.+\big({\bf K}+\frac{1}{4}\tr\chi\tr\chib\!-\!\frac{1}{2}\chih\c\chibh\big)\!\right]\!=0\ .\nn
\eea}
 Moreover on $\Cb_0$, assuming $\{e_A\}$ a Fermi transported orthonormal frame,
\bea
&&\chi\ ,\ \ \om\ ,\ \ze\ ,\ \chib_{ab}=\frac{1}{2\oom}\frac{\partial\ga_{ab}}{\partial\la}\nn
\eea
satisfy the costraint equations:
{\em
\bea
&&\frac{\partial\tr\chib}{\partial\la}+\frac{\oom\tr\chib}{2}\tr\chib+2\oom\om \tr\chib+\oom|\chibh|^2=0\nn\\
&&\frac{\partial\zeta_A}{\partial\la}+\oom\chib_{AB}\ze_B+\oom\tr\chib\ze_A+\oom(\divv\chib)_A-\oom\nabb_A\tr\chib
-\frac{\partial\nabb_A\log\oom}{\partial\la}=0\nn\\
&&\frac{\partial\tr\chi}{\partial\la}+\oom\tr\chib\tr\chi -2\oom\omb\tr\chi-2\oom(\divv\eta)-2\oom|\eta|^2+2\oom{\bf K}=0\nn\\
&&\frac{\partial\chih_{AB}}{\partial\la}+\frac{\oom\tr\chib}{2}\chih_{AB}+\frac{\oom\tr\chi}{2}\chibh_{AB}
+\frac{\partial\log\oom}{\partial\la}\chih_{AB}-\oom\nabb_A\hot(\ze+\nabb\log\oom)_B\nn\\
&&-\oom(\ze_A+\nabb_A\log\oom)\hot(\ze+\nabb\log\oom)_B=0\eql{1.41wpz}\\
&&\frac{\partial\om}{\partial\la}-2\oom\om\omb+\oom\ze_C\nabb_C\log\oom+\frac{1}{2}\left[-3\oom|\ze|^2+\oom|\nabb\log\oom|^2\right.\nn\\
&&\left.+\oom\big({\bf K}+\frac{1}{4}\tr\chi\tr\chib-\frac{1}{2}\chih\c\chibh\big)\right]=0\ .\nn
\eea}
\end{Def}
{\bf Remark:}
When assigning the initial data to prove  a global existence result we have to require on ${\cal C}$, the regularity of 
$\chi, \chib,\om, \omb,\ze$, together with their tangential derivatives,
\bea
{\nabb^k\chi}\ ,\  \nabb^k\om\ ,\ \nabb^k\ze\ ,\ {\nabb^k\chib}\ ,\ {\nabb^k\omb}\ ,
\eea
and of
\bea
\nabb^{k-1}\dddd_4\hat\chi\ ,\ \nabb^{k-1}\dddd_3\hat{\chib}\ \ 
\eea
on $C_0$ and $\Cb_0$ respectively.
$k\in [1,q]$ and $\nabb$ is the covariant derivative associated to the metric $\ga_{ab}$. The amount of regularity needed for the existence proof
will be specified later on. 

\subsection{The characteristic initial value problem.}\label{SS1.4}
Starting with the previous definitions of the ``initial data" we state the characteristic initial value problem in the
following way:

\begin{Def}[The characteristic initial value problem]To solve the characteristic initial value problem for the
vacuum Einstein equations with initial data set
\bea
\left\{C_0\ ;\overline{\ga}_{ab},\overline{\oom},\overline{\ze}_a,\overline{\chib}_{ab},\overline{\omb}\right\}\cup\left\{\Cb_0\
;\overline{\ga}_{ab},\overline{\oomb};\overline{\ze}_a,\overline{\chi}_{ab},\overline{\om}\right\}
\eea 
means to find a four dimensional manifold $\M$, a Lorentz metric $\ggg$ satisfying the vacuum Einstein equations as well as an
imbedding
\bea
i:\left\{C_0\ ;\overline{\ga}_{ab},\overline{\oom},\overline{\ze}_a,\overline{\chib}_{ab},\overline{\omb}\right\}\cup\left\{\Cb_0\
;\overline{\ga}_{ab},\overline{\oomb};\overline{\ze}_a,\overline{\chi}_{ab},\overline{\om}\right\}\rightarrow
\M
\eea
such that: 
\smallskip

 a) $i(C_0)=C(\la_1)\ ,\ i(\Cb_0)=\Cb(\nu_0)$, $i(S_0(\nu_0))=i(\underline{S}_0(\la_1))=C(\la_1)\cap\Cb(\nu_0)$, where $C(\la_1)$ and $\Cb(\nu_0)$
are two null hypersurfaces embedded in $\M$. 
\smallskip

 b) $\M$ is the maximal future development of $C(\la_1)\cup\Cb(\nu_0)$.
\smallskip

 c) On $C(\la_1)$ with respect to the initial data foliation, we have
\bea
&&i^*(\ga)=\overline{\ga}\ ,\ i^*(\oom)=\overline{\oom}\ ,\ i^*(\ze)=\overline{\ze}\nn\\
&&i^*(\chi)=\overline{\chi}\ ,\ i^*(\chib)=\overline{\chib}\ ,i^*(\om)=\overline{\om}\ ,i^*(\omb)=\overline{\omb}\ .
\eea
 d) On $\Cb(\nu_0)$ with respect to the initial data foliation, we have
\bea
&&i^*(\ga)=\overline{\ga}\ ,\ i^*(\oom)=\overline{\oomb}\ ,\ i^*(\ze)=\overline{\ze}\nn\\
&&i^*(\chib)=\overline{\chib}\ ,\ i^*(\chi)=\overline{\chi}\ ,i^*(\omb)=\overline{\omb}\ ,i^*(\om)=\overline{\om}\ .
\eea
where $\chi,\chib,\om,\omb,\ze$ are the restriction to $C(\la_1)$ and $\Cb(\nu_0)$ of the connection coefficients of the spacetime $\M$,
\bea
\chi_{ab}&=&\ggg(\dd_{e_a}e_4,e_b)\ \ ,\ \chib_{ab}=\ggg(\dd_{e_a}e_3,e_b)\nn\\
\omb&=&\frac{1}{4}\ggg(\dd_{e_3}e_3,e_4)=-\frac{1}{2}\dd_3\log\oom\nn\\
\om&=&\frac{1}{4}\ggg(\dd_{e_4}e_4,e_3)=-\frac{1}{2}\dd_4\log\oom\eql{1.42}\\
\ze_{a}&=&\frac{1}{2}\ggg(\dd_{e_{a}}e_4,e_3)\ .\nn
\eea 
with $\{e_4,e_3.e_a\}$ a null orthonormal frame of $\M$. $\oom$ is the ``null" lapse function restricted to $C(\la_1)\cup\Cb(\nu_0)$, $\ga$ is the
metric tensor restricted to the leaves $S_0(\nu)$ and $S_0(\la)$ of $C(\la_1)$, $\Cb(\nu_0)$ respectively.
\smallskip
 
 e) The costraint equations \ref{1.41wqz}, \ref{1.41wpz} are the pull back of (some of) the structure equations of $\M$
restricted to $C(\la_1)$ and $\Cb(\nu_0)$.
\end{Def}
{\bf Remarks:}{

a) The previous definition of the characteristic initial value problem allows to visualize $C(\la_1)$ as a portion of an
outgoing truncated (before its lower vertex) cone ``starting" from $S_0(\nu_0)$ and going up indefinitely and $\Cb(\nu_0)$ as a portion of an incoming
cone truncated (before its upper vertex) at the two dimensional surface $\underline{S}_0(\la_0)$ and, going ``backward" in time, up to the two
dimensional surface $\underline{S}_0(\la_1)=S_0(\nu_0)$.

b) The reason why we consider $\Cb_0$ truncated with $\underline{S}_0(\la_0)$ is a technical one associated to the need of avoiding the
problems connected to the ``vertices" of the null cones. We specify in the next sections in which sense this characteristic
problem can be interpreted as a global existence problem.}
\medskip

\subsubsection{The initial data set in terms of the connection coefficients}\label{SS1.3.2}
The initial data set given in Definition \ref{D1.1} is equivalent to the more standard definition of initial data set made by the metric components
and their partial derivatives. To prove it we show that all the first order partial derivatives of the metric tensor
can be expressed in terms of the connection coefficients. To do it  we consider $\cal C$ as an hypersurface immersed in a four dimensional
manifold $R^4$ endowed with a Lorentzian metric $\tilde{\ggg}$.\footnote{This immersion is not the diffeomorphism $i$ introduced in the definition of the
characteristic problem, see subsection \ref{SS1.4}, and $\tilde{\ggg}$ is not a solution of the Einstein equations. This definition is a way
of defining explicitely the initial hypersurfaces $C_0$ and $\Cb_0$.} 
As we want to express the initial data as the metric components and their first partial derivatives we need to choose a coordinate set. We
introduce, therefore, the coordinates $\{\ub,u,\om^a\}$ and require that $\tilde{\ggg}$ has the following expression:
\bea
\tilde{\ggg}=|\tilde{X}|^2du^2\!-\!2\tilde{\oom}^2(dud\ub\!+\!d\ub du)\!-\!\tilde{X}_a(dud\om^a\!+\!d\om^adu)\!+\!\tilde{\ga}_{ab}d\om^a d\om^b\ .\ \
\eql{1.36b}
\eea
With this choice we define $C_0$ as (a portion of)  a level set of the function $u(p)$ and $\Cb_0$ as (a portion of)  a level set of the function
$\ub(p)$, namely:
\bea
&&C_0\equiv C(\la_1)=\{p\in R^4|u(p)=\la_1, \nu\in[\nu_0,\infty)\}\nn\\
&&\Cb_0\equiv C(\nu_0)=\{p\in R^4|\ub(p)=\nu_0, \la\in[\la_1,\la_0]\}\ .\eql{1.33d}
\eea
From the explicit form of the metric tensor $\tilde{\ggg}$ it is immediate to recognize that
$C(\la_1)$ is a portion of a truncated outgoing null cone in $R^4$ and, analogously, $\Cb(\nu_0)$, is a portion of a truncated incoming null cone.
Moreover the null geodesics generating $C(\la_1)$ and $\Cb(\nu_0)$ have tangent vector fields 
\bea
L=\frac{1}{2\tilde{\oom}^2}\frac{\partial}{\partial\ub}\ \ ,\ \ \Lb=\frac{1}{2\tilde{\oom}^2}\left(\frac{\partial}{\partial
u}+\tilde{X}\right)\eql{2.5ew}
\eea
with $\tilde{X}=\tilde{X}^a\frac{\partial}{\partial\om^a}$.\footnote{This follows immediately observing that the inverse metric $g^{-1}$ has the
following components: $g^{u\ub}=(-2\oom^2)^{-1}\ ,\ g^{a\ub}=(-2\oom^2)^{-1}\!X^a\ ,\ g^{ab}=\ga^{ab}\ ,\ g^{uu}=g^{\ub\ \!\ub}=g^{ua}=0$. Therefore
both the functions $u(p)$ and $\ub(p)$ satisfy the eikonal equation $g^{\mu\nu}\partial_{\mu}w\partial_{\nu}w=0$.}

 We require that the restriction of the various components of the metric on $\cal C$, be equal to the quantities defined before in a more abstract
way,
\bea
&&{\tilde{\ga}|_{C_0}}_{ab}=\ga_{ab},\ \ {\tilde{X}|_{C_0}}_a=X_a,\ \ \tilde{\oom}|_{C_0}=\oom\nn\\
&&{\tilde{\ga}|_{\Cb_0}}_{ab}=\ga_{ab},\ \ {\tilde{X}|_{\Cb_0}}_a=\underline{X}_a,\ \ \tilde{\oom}|_{\Cb_0}=\oomb\eql{1.51qw}
\eea
From equations \ref{2.5ew} it follows immediately that the functions $\ub(p)$ and $u(p)$ restricted to $C_0$ and $\Cb_0$ satisfy, respectively,
\bea
&&\ub(p)=\nu_0+\int_0^{\vb(p)}(4\oom)^{-2}(\ga(\vb'))d\vb'\nn\\
&&u(p)=\la_1+\int_0^{v(p)}(4\underline{\oom})^{-2}(\ga'(v'))dv'\ .\eql{1.41bw}
\eea
where, denoted $(\ub,\theta,\phi)$ the coordinates of the point $p$, $\ga$ is the null geodesic along $C_0$ starting on $S_0=C_0\cap\Cb_0$ at the point $q$ of
angular coordinates $(\theta,\phi)$. $\ga'$ is the null geodesic along $\Cb_0$ starting on $S_0=C_0\cap\Cb_0$ and reaching the point $p$ when its affine
parameter has the value $v=v(p)$. It is important to observe that from the explicit expression of $\Lb$, \ref{2.5ew}, it follows that moving along $\ga'$
the angular coordinates $\om^a$ vary differently from what happens along $\ga$. This is, again, the effect of the definition of the $\om^a$ coordinates
associated to the choice of the metric expression \ref{1.36b}.\footnote{In \cite{Kl-Ni:book} the different ways of defining the $\om^a$ coordinates
are carefully discussed, see equations (3.1.58), (3.1.59), the second choice would correspond to a metric expression of the following kind:
$\tilde{\ggg}=|\tilde{Y}|^2d\ub^2\!-\!2\tilde{\oom}^2(dud\ub\!+\!d\ub du)\!-\!\tilde{Y}_a(d\ub d\om'^a\!+\!d\om'^ad\ub)\!+\!\tilde{\ga}_{ab}d\om'^a
d\om'^b$.} Moreover due to the fact that, as discussed before, the vector fields $e_3$ and $e_4$ do not commute we have here from the explicit expression of
the commutator that on $\cal C$ the following relation holds\footnote{Of course it holds in the whole $\{R^4,\tilde{\ggg}\}$. A detailed discussion is in
\cite{Kl-Ni:book}, Chapter 3.}
\bea
\ze_a=-\frac{1}{4\oom^2}\ga_{ab}\frac{\partial X^b}{\partial\nu}\ .\eql{1.69ex}
\eea
The null frames $\{e_4,e_A\}$ defined on $C_0$ and $\{e_3,e_A\}$ on $\Cb_0$ can now be extended adding a vector field $e_3$ on $C_0$ and
a vector field $e_4$ on $\Cb_0$ such that $\tilde{\ggg}(e_4,e_3)=-2\ ,\ \tilde{\ggg}(e_4,e_a)=0\ ,\ \tilde{\ggg}(e_3,e_a)=0$. Therefore we have
defined on $\cal C$ a null frame $\{e_4,e_3, e_A\}$.

Finally we require that the tensor fields defined on $\cal C$, $\chib$, $\ze$ and $\omb$ for $C_0$, $\chi$, $\ze$ and $\om$ for $\Cb_0$) are the
corresponding connection coefficients of $(R^4,\tilde{\ggg})$ when restricted to $\cal C$.

 We can now express all the connection coefficients in
terms of the metric components and its Christoffel symbols and look for the reciprocal relations. One easily proves the following relations:
\bea
{\chi}_{AB}\!\!&=&\!e_A^ae_B^b\left[\oom^{-1}\left(-\Ga^u_{a\ub}X_b+\Ga^c_{a\ub}\ga_{cb}\right)\right]\nn\\
\chib_{AB}\!\!&=&\!e_A^ae_B^b\left[
\oom^{-1}\!\left(-\Ga^{u}_{au}X_b+\Ga^c_{au}\ga_{cb}+\partial_aX^c\ga_{cb}+X^d\Ga^c_{ad}\ga_{cb}\right)\right]\nn\\
\eta_A\!&=&\!e_A^a\left(\Ga^{u}_{ua}+X^b\Ga^{u}_{ba}\right)\ \ \ ,\ \ \ 
\etab_A=e_A^a \Ga^{\ub}_{\ub a}\nn\\
\ze_A
\!&=&\!e_A^a\left(\partial_a\log\oom-\Ga^{\ub}_{a\ub}-(2\oom^2)^{-1}|X|^2\Ga^u_{a\ub}\right)\nn\\
\om\!&=&\!-\frac{1}{2\oom}\partial_{\ub}\log\oom+\frac{1}{2}\oom^{-1}\left(\Ga^u_{\ub u}+X^c\Ga^u_{c\ub}\right)\\
\omb\!&=&\!-\frac{1}{2\oom}\partial_u\log\oom-\frac{1}{2\oom^3}X^a\partial_a\log\oom-
\frac{1}{4\oom^3}\left(2\oom^2\Ga^{\ub}_{u\ub}+X_a\Ga^{a}_{u\ub}+X^a\Ga^{u}_{a\ub}-|X|^2\Ga^{u}_{u\ub}\right)\ .\ \ \ \ \nn
\eea

 In the chosen metric the Christoffel symbols have the following expressions:
\bea
&&\Ga^{u}_{uu}=2\partial_u\log\oom+{(4\oom^2)}^{-1}\partial_{\ub}|X|^2\nn\\
&&\Ga^{u}_{ua}=\partial_a\log\oom-{(4\oom^2)}^{-1}\partial_{\ub}X_a\nn\\
&&\Ga^{u}_{ab}={(4\oom^2)}^{-1}\partial_{\ub}\ga_{ab}\nn\\
&&\Ga^{\ub}_{\ub\ub}=2\partial_{\ub}\log\oom\nn\\
&&\Ga^{\ub}_{uu}={(4\oom^2)}^{-1}X^aX^b\partial_u\ga_{ab}+{(4\oom^2)}^{-1}X^a\partial_a|X|^2\eql{1.58w}\\
&&\Ga^{\ub}_{u\ub}=-X^a\partial_a\log\oom-{(4\oom^2)}^{-1}X_a\partial_{\ub}X^a\nn\\
&&\Ga^{\ub}_{au}=-{(4\oom^2)}^{-1}\partial_a|X|^2-{(4\oom^2)}^{-1}X^b\partial_u\ga_{ab}+
{(4\oom^2)}^{-1}X^b(\partial_aX_b-\partial_bX_a)\nn\\
&&\Ga^{\ub}_{a\ub}=\partial_a\log\oom+{(4\oom^2)}^{-1}\ga_{ab}\partial_{\ub}X^b\nn\\
&&\Ga^{\ub}_{ab}={(4\oom^2)}^{-1}(\partial_aX_b+\partial_bX_a+\partial_u\ga_{ab})
-{(4\oom^2)}^{-1}X^c(\partial_b\ga_{ca}+\partial_a\ga_{cb}-\partial_c\ga_{ab})\nn\\
&&   \nn\\
&&\Ga^{a}_{uu}=-\ga^{ab}\partial_uX_b-\frac{1}{2}\ga^{ab}\partial_b|X|^2+2X^a\partial_u\log\oom
+{(4\oom^2)}^{-1}X^a\partial_{\ub}|X|^2\nn\\
&&\Ga^{a}_{u\ub}=-\frac{1}{2}\ga^{ab}\partial_{\ub}X_b+2\oom^2\ga^{ab}\partial_b\log\oom\nn\\
&&\Ga^{a}_{ub}=\frac{1}{2}\ga^{ac}\partial_u\ga_{cb}+\frac{1}{2}\ga^{ac}(\partial_cX_b-\partial_bX_c)
+X^a\partial_b\log\oom-{(4\oom^2)}^{-1}X^a\partial_{\ub}X_b\nn\\
&&\Ga^{a}_{\ub
b}=\frac{1}{2}\ga^{ac}\partial_{\ub}\ga_{cb}\ \ ,\
\ \Ga^{a}_{bc}={^{(2)}\Ga}^{a}_{bc}+{(4\oom^2)}^{-1}X^a\partial_{\ub}\ga_{bc}\nn\\
&&\Ga^{u}_{u\ub}=\Ga^{u}_{\ub\ub}=\Ga^{u}_{a\ub}=\Ga^{a}_{\ub\ub}=0\nn
\eea 
where ${^{(2)}\Ga}^{a}_{bc}$ are the Christoffel symbols associated to the metric $\ga_{ab}$\ .

 Using these relations the connection coefficients in terms of the metric components and their derivatives are:
\bea
&&\eta_a=\partial_a\log\oom-{(4\oom^2)}^{-1}\ga_{ab}\partial_{\ub}X^b\nn\\
&&\etab_a=\partial_a\log\oom+{(4\oom^2)}^{-1}\ga_{ab}\partial_{\ub}X^b\nn\\
&&\ze_a=-{(4\oom^2)}^{-1}\ga_{ab}\partial_{\ub}X^b\nn\\
&&\chi_{ab}=(2\oom)^{-1}\partial_{\ub}\ga_{ab}\eql{1.59w}\\
&&\chib_{ab}=(2\oom)^{-1}\partial_{u}\ga_{ab}+(2\oom)^{-1}({^{(2)}\nabb}_bX_a+{^{(2)}\nabb}_aX_b)\nn\\
&&\om=-(2\oom)^{-1}\partial_{\ub}\log\oom\nn\\
&&\omb=-(2\oom)^{-1}(\partial_u\log\oom+X^a\partial_a\log\oom)\nn
\eea
where ${^{(2)}\!}\nabb$ is the covariant derivative associated to the $\ga$ metric tensor of the leaves $S(\la_1,\nu)$ and $S(\la,\nu_0)$.
Through relations \ref{1.58w} and \ref{1.59w} it is possible to write all the Christoffel symbols in terms of the
connection coefficients, the metric components and their derivatives. The Christoffel symbols, different from zero, have
the following expressions:
\bea
&&\Ga^u_{uu}=-4\oom\omb+(2\oom)^{-1}\chi_{ab}X^aX^b-2X^a(\partial_a\log\oom+\ze_a)\nn\\
&&\Ga^u_{ua}=-(2\oom)^{-1}\chi_{ab}X^b+(\partial_a\log\oom+\ze_a)\nn\\
&&\Ga^u_{ab}=(2\oom)^{-1}\chi_{ab}\nn\\
&&\Ga^{\ub}_{\ub\ \!\ub}=-4\oom\om\nn\\
&&\Ga^{\ub}_{uu}=(2\oom)^{-1}\chib_{ab}X^aX^b\nn\\
&&\Ga^{\ub}_{u\ub}=-X^a(\partial_a\log\oom-\ze_a)\eql{1.57wtw}\\
&&\Ga^{\ub}_{au}=-(2\oom)^{-1}\chib_{ab}X^b\nn\\
&&\Ga^{\ub}_{a\ub}=(\partial_a\log\oom-\ze_a)\nn\\
&&\Ga^{\ub}_{ab}=(2\oom)^{-1}\chib_{ab}\nn\\
&&\Ga^{a}_{u\ub}=-\oom\ga^{ab}\chi_{bc}X^c+\ga^{ab}(\partial_b\log\oom+\ze_b)\nn\\
&&\Ga^{a}_{ub}=\oom\ga^{ac}\chib_{cb}-{^{(2)}\!\nabb}_bX^a-(2\oom)^{-1}\chi_{bc}X^cX^a+X^a(\partial_b\log\oom+\ze_b)\nn\\
&&\Ga^{a}_{\ub b}=\oom\ga^{ac}\chi_{cb}\nn\\
&&\Ga^{a}_{bc}={^{(2)}\Ga}^{a}_{bc}+(2\oom)^{-1}\chi_{bc}X^a\nn\\
&&\Ga^{a}_{uu}=-\ga^{ab}\partial_uX_b\!-\!\frac{1}{2}\ga^{ab}\partial_b|X|^2\!-\!X^a\!\left[4\oom\omb\!-\!(2\oom)^{-1}\chi_{bc}X^bX^c
\!+\!2X^b(\partial_b\log\oom+\ze_b)\!\right]\nn
\eea
All the Christoffel symbols, with the exception of the last one, $\Ga^{a}_{uu}$,  can be written in terms of the connection coefficients
the metric components and their tangential derivatives. $\Ga^{a}_{uu}$ is somewhat different for the presence of the term $\partial_uX_b$.\footnote{This is
connected to the fact that the vector field $X$ is tied to the choice of the coordinates.} Nevertheless, we will show later on, see subsection
\ref{SS2.3}, that also this term can be obtained once we prescribe on $\cal C$ the metric components and the connection coefficients. 

 All this justifies the statement that the initial data are specified once we give on $\cal C$ the connection coefficients together with the
metric components $\oom$ and $\ga_{ab}$.

\section{The construction of the initial data}\label{S2}

Once we have defined the initial data set of the characteristic Cauchy problem and the costraint equations they have to satisfy, we have to prove
that these initial data do exist. This problem is somewhat analogous to the problem of solving the costraint equations
\bea
&&\nab^j {k}_{ij}-\nab_i\tr{k}=0 \nn\\
&&{R}-|{k}|^2+(\tr{k})^2=0\eql{2.1q}
\eea
of the non characteristic problem. This section is devoted to the explicit construction of the initial data. Moreover, as done for the problem
\ref{2.1q}, see \cite{C-K:book}, \cite{Kl-Ni:book}, while trying to solve the characteristic costraint problem we will also prescribe the initial data
decay along the null hypersurfaces
$C_0$ and $\Cb_0$ when their coordinate $\nu$ or $|\la|$ tend to $\infty$. The decay rate of some initial data quantities is basically an
immediate consequence of their geometric meaning (and some smallness assumptions). This is the case for $\tr\chi$ and $\tr\chib$. For the other
quantities, $\chih,\chibh,\om,\omb,\ze$, we require a decay such that some ``energy type" integral norms defined on $C_0$ and $\Cb_0$ are
bounded.\footnote{From their decay also the decay of the metric components follows.} In fact to apply the techniques used
in \cite{Kl-Ni:book} to prove our global existence result, we need that some ``flux integrals", $\cal Q$, are bounded. In the non
characteristic case this implies that these integral norms,  $\cal Q$, see their definition in \cite{Kl-Ni:book}, Chapter 3, are bounded in terms of
the corresponding (finite) norms on the initial hypersurface $\Si_0$.

 In the characteristic case the $\cal Q$ flux norms have to be bounded in terms of the same flux norms defined on the initial hypersurface $\cal
C$. The boundedness of these norms requires that the various null Riemann components involved decay sufficiently fast along the initial cones,
$C_0$ and $\Cb_0$. In particular if the $\cal Q$ norms we use are those defined in \cite{Kl-Ni:book} it follows
immediately that on $C_0$ and $\Cb_0$ we need the following decay, with $\de>0$:
\bea
&&\a=O(r^{-(\frac{7}{2}+\de)})\ ,\ \b=O(r^{-(\frac{7}{2}+\de)})\ ,\ (\ro-{\overline\ro},\si)=O(r^{-3}|\la|^{-(\frac{1}{2}+\de)})\nn\\
&&\bb=O(r^{-2}|\la|^{-(\frac{3}{2}+\de)})\ ,\ \aa=O(r^{-1}|\la|^{-(\frac{5}{2}+\de)})\ . \eql{1.35a}
\eea
In the next subsections we prove the existence of characteristic initial data such that the decays \ref{1.35a} are satisfied. 

\subsection{The solution of the characteristic costraint problem for the initial data on the outgoing cone, with a prescribed decay
rate}\label{S2.1}

 We start, as described in the previous subsection, considering $\cal C$ as an hypersurface immersed in the four dimensional manifold $R^4$
endowed with the Lorentzian metric \ref{1.36b},
\beaa
\tilde{\ggg}=|\tilde{X}|^2du^2\!-\!2\tilde{\oom}^2(dud\ub\!+\!d\ub du)\!-\!\tilde{X}_a(dud\om^a\!+\!d\om^a du)\!+\!\tilde{\ga}_{ab}d\om^a d\om^b\ .\ \ \
\eql{1.36bbz}
\eeaa
With this choice $C_0$ is (a portion of)  a level set of the function $u(p)$, see \ref{1.33d}.
We choose the function $\tilde{\oom}$ of the metric $\tilde{\ggg}$ in such a way that its restriction on $C_0$ is an assigned function $\oom$,
\bea
\tilde{\oom}|_{C_0}=\oom\ .\eql{1.39b}
\eea
Once $\oom$ is given we define the ``$\oom$-foliation" of $C_0=C(\la_1)$ as done before. The leaves of the foliation are the surfaces
$S_0(\nu)=\{p\in C(\la_1)|\ub(p)=\nu\}$ we denote also $S(\la_1,\nu_0)$\footnote{To recall that
$S_0(\nu)=S(\la_1,\nu_0)=C(\la_1)\cap\Cb(\nu_0)$.} and $\ub(p)$ has been defined in \ref{1.3}.  
On each leave $S_0(\nu)=S(\la_1,\nu)$ we can define two orthonormal \footnote{Orthonormal with respect to the $\tilde{\ga}$ metric.} vector
fields $\tilde{e}_A$ tangent to it and Fermi transported along the cone,
\[\tilde{e}_A=\tilde{e}_A^a\frac{\partial}{\partial\om^a}\ .\]
It is important to remark that the metric $\tilde{\ga}$ introduced in \ref{1.36b}  is not the metric which we want to
prescribe as initial data; therefore it does not satisfy the ``characteristic" costraints. We use it as a background metric from which to construct
the final one satisfying the costraints together with its derivatives.\footnote{In the appendix we show how to construct explicitely a
two dimensional metric $\tilde{\ga}$.}

Given $\oom$ we define $\om$ as 
\bea
\om=-\frac{1}{2\oom}\frac{\partial}{\partial{\nu}}\log\oom\ ,
\eea
then we assigne on $C_0$ a symmetric traceless tensor $\chih$ tangent to each $S_0(\nu)$, $\nu\in [\nu_0,\infty)$, which we will consider as the
traceless part of a tensor $\chi$.

 The function $\tr\chi$ and the ``initial data metric" $\ga_{ab}$ are determined requiring that they satisfy the following
system of differential equations:
\bea
&&\frac{\partial\ga_{ab}}{\partial\nu}-\oom{\tr\chi}\ga_{ab}-2\oom\chih_{ab}=0\nn\\
&&\frac{\partial\tr\chi}{\partial\nu}+\frac{\oom\tr\chi}{2}\tr\chi+2\oom\om\tr\chi+|\chih|^2_{{\ga}}=0\eql{1.43b}
\eea
 The solution of system \ref{1.43b} is the main part of the following lemma. The estimates appearing there are pointwise estimates or
$|\c|_{p,S}$ norms, where 
\footnote{$|f|=|f|_{\ga}=f_{a_1...a_n}f_{c_1...c_n}{\ga}^{a_1c_1}...{\ga}^{a_nc_n}$} 
\[|f|_{p,S}\equiv\bigg(\int_{S}|f|^p_{\ga}d\mu_{\ga}\bigg)^{\frac{1}{p}}.\]
\begin{Le}\label{L2.1}
Assigne on the null hypersurface $C_0$ an $\oom$-foliation with leaves $S_0(\nu)$, a metric tensor $\tilde{\ga}$, tangent to the leaves
$S_0(\nu)$ satisfying, together with $\tr\chi$, the trace of the null outgoing second fundamental form associated to $\tilde{\ga}$, the following
equations {\em\bea &&\frac{\partial\tilde{\ga}_{ab}}{\partial\nu}-\oom{\tr{\tilde\chi}}\tilde{\ga}_{ab}=0\eql{2.10t}\\
&&\frac{\partial\tr\tilde{\chi}}{\partial\nu}+\frac{\oom\tr\tilde{\chi}}{2}\tr\tilde{\chi}+2\oom\om\tr\tilde{\chi}=0\ .\nn
\eea}
Assume that, with $\varepsilon>0$,
\bea\nu_0\in[c_1{\tilde r}_0, c_2{\tilde r}_0]\ \ \mbox{with}\ \ |c_{1,2}-1|=O(\varepsilon)\ . \eea   
 Given a  symmetric tensor $\chih$ on $C_0$, assume that, with $\varepsilon>0$, the following decays hold,
\bea
&&|\oom-\frac{1}{2}|=O(\varepsilon\tilde{r}^{-1})\ \ ,\ \ |\tilde{\nabb}\log\oom|=O(\varepsilon\tilde{r}^{-(2+\de)})\nn\\
&&|\om|=O(\varepsilon\tilde{r}^{-2})\ \ ,\ \ |\chih|_{\tilde{\ga}}=O(\varepsilon\tilde{r}^{-(\frac{5}{2}+\de)}), \eql{2.15zz}\\
&&|\dddd_4\chih|_{\tilde{\ga}}=O(\varepsilon\tilde{r}^{-(\frac{7}{2}+\de)})\ ,\
|\tilde{\nabb}\chih|_{\tilde{\ga}}=O(\varepsilon\tilde{r}^{-(\frac{7}{2}+\de)})\nn\\
&& |\dddd_4\tilde{\nabb}\log\oom|_{\tilde{\ga}}=O(\varepsilon\tilde{r}^{-(3+\de)})\ ,\nn
\eea
with $\de>0$,\footnote{The reason of this requirement will be clear when we discuss the costraints to impose to the one form $\ze$.} 
$\tilde{r}$ defined from the relation $4\pi\tilde{r}^2(\nu)=|S_0(\nu)|_{\tilde{\ga}}$ and $\tilde{\nabb}$ the covariant derivative associated to the
metric $\tilde{\ga}$. Assume that on $S_0(\nu_0)$,
\bea
|{\tilde r}_0^{3-\frac{2}{p}}\tilde{\nabb}\tr\chi|_{p,S}=O(\varepsilon)\ ,
\eea
then there exists on $C_0$ a metric tensor $\ga$ tangent to the leaves of the same foliation such that, denoted $\chi$ the second fundamental form of the
$S_0(\nu)$ with respect to this metric,
we have: 
\bea
\frac{1}{2\oom}\frac{\partial\ga_{ab}}{\partial\nu}=\chi_{ab}=\chih_{ab}+\frac{\ga_{ab}}{2}\tr\chi\ ,
\eea
with $\ga|_{S_0}=\tilde{\ga}|_{S_0}$ and $\tr\chi>0$ satisfies the constraint equation
\bea
\frac{\partial\tr\chi}{\partial\nu}+\frac{\oom\tr\chi}{2}\tr\chi+2\oom\om\tr\chi+|\chih|^2_{\ga}=0\ ,
\eea
with $\tr\chi|_{S_0}=\tr\tilde{\chi}|_{S_0}$. Moreover the following estimates hold, with $p\in[2,4]$,
\bea
&&|\oom-\frac{1}{2}|=O(\varepsilon{r}^{-1})\ \ ,\ \ |\nabb\log\oom|=O(\varepsilon{r}^{-(2+\de)}),\ \
|\om|=O(\varepsilon{r}^{-2})\nn\\
&&|\tr\chi-\frac{2}{r}|=O(\varepsilon r^{-2}\log r)\ \ ,\ \ |\chih|_{{\ga}}=O(\varepsilon{r}^{-(\frac{5}{2}+\de)})\eql{1.48c}\\
&&|r^{(\frac{7}{2}+\de)-\frac{2}{p}}\dddd_4\chih|_{p,S}=O(\varepsilon)\ ,\ |r^{(\frac{7}{2}+\de)-\frac{2}{p}}\nabb\chih|_{p,S}=O(\varepsilon)\nn\\
&&|r^{(3+\de)-\frac{2}{p}}\dddd_4\nabb\log\oom|_{p,S}=O(\varepsilon)\ ,\ |r^{3-\frac{2}{p}}\nabb\tr\chi|_{p,S}=O(\varepsilon)\ ,\nn
\eea
with ${r}$  defined by the relation
\bea 4\pi{r}^2(\nu)=|S_0(\nu)|_{{\ga}}\ .\eql{2.13qw}
\eea
\end{Le}
{\bf Proof:} The proof is in the appendix.\footnote{Assumptions \ref{2.15zz} are needed to prove Lemma \ref{L2.1}, other more stringent conditions on
the regularity of these quantities will be needed to prove the existence of the initial data.}
\smallskip

 Once we have solved equations \ref{1.43b} we have assigned on the whole $C_0$, $\ga$, $\oom$, $\om$ and $\chi$. $\ga$ is the (initial data)
metric relative to the leaves $S_0(\nu)$, $\chi$ is the second (null outgoing) fundamental form of the leaves $S_0(\nu)$ relative to the same metric.  
In the explicit expression of $\tilde{\ggg}$,
$\tilde{\oom}$ is the extension of the quantity previously defined on $C_0$
\bea
\oom=\tilde{\oom}|_{C_0}\eql{1.61da}
\eea 
and $\tilde\ga$ is a new extension to $R^4$ of the metric $\ga$ obtained in Lemma \ref{L2.1}, \footnote{Remark that the choice of
the null hypersurfaces $C_0$ and $\Cb_0$ does not depend on the choice of the metric $\ga_{ab}$.} 
\bea
\tilde{\ga}_{ab}|_{C_0}=\ga_{ab}\ .\eql{1.61d}
\eea
The restriction to $C_0$ of the metric components
$|\tilde{X}|^2$ and $\tilde{X}_a$ is obtained later on, after we introduce the one form $\ze$. \footnote{Observe that the leaves of the
$\oom$-foliation are not changed. This follows from the fact that the scalar function $\oom$ on
$C_0$ is not changed. Moreover the null geodesics on $C_0$ with respect the metric $\ga$ starting from the
point $(\nu_0,\theta,\phi)$ of $S_0=S_0(\nu_0)$ are the same as those relative to the metric $\tilde{\ga}$. } 

 To obtain $\ze$ along $C_0$ we have to use its evolution equation \ref{1.12} and its value on $S_0(\nu_0)$.\footnote{The way to obtain the one
form $\ze$ on $S_0$ is discussed when we obtain the initial data on the incoming cone, see later on.} We want 
$\ze$ defined on the whole $C_0$ decaying as $O(r^{-2})$. This is the result of the following lemma.
\begin{Le}\label{L1.3}
Assume that, on $C_0$, \footnote{Assumptions \ref{2.15saa}  is a condition we impose on $\oom$ on $C_0$.} 
{\em\bea
|r^{(3+\de)-\frac{2}{p}}\dddd_4\nabb\log\oom|_{p,S}=O(\varepsilon)\ ,\ |r^{3-\frac{2}{p}}\nabb\tr\chi|_{p,S}=O(\varepsilon)\eql{2.15saa}
\eea}
assume that on $S_0(\nu_0)=\underline{S}_0(\la_1)$,
{\em\bea
|r^{2-\frac{2}{p}}\ze|_{p,S}(\la_1,\nu_0)\leq c\varepsilon\ .\eql{2.15kk}
\eea}
Let $\ze$  be the solution along $C_0$ of the costraint equation \ref{1.12},
{\em\beaa
\dddd_4\zeta+\zeta\chi+\tr\chi\zeta=\divv\chi-\nabb\tr\chi-\ddb_4\nabb\log\oom\ ,
\eeaa}
then, on $C_0$,
\bea
|r^{2-\frac{2}{p}}\ze|_{p,S}(\la_1,\nu)\leq c\varepsilon\ .
\eql{1.43czz}
\eea
\end{Le}
{\bf Proof:} 
The evolution equation for $\ze$ can be written as
\bea
\dddd_4\zeta+\frac{3}{2}\tr\chi\zeta=-\chih\zeta+F(\nabb\chi,\nabb\log\oom)
\eea
\beaa
\mbox{where}\ \ \ \ \ \ F(\nabb\chi,\nabb\log\oom)=[\divv\chih-\frac{1}{2}\nabb\tr\chi-\ddb_4\nabb\log\oom]\ .
\eeaa
Using the assumptions, whose validity follows from the previous lemma,  we control the right hand side $F(\nabb\chi,\nabb\log\oom)$, choosing $\{e_A\}$ Fermi
transported and using Gronwall's Lemma, see Chapter 4 of \cite{Kl-Ni:book}, we obtain
\bea
&&|r^{3-\frac{2}{p}}\ze|_{p,S}(\la_1,\nu)\leq  \eql{1.41at}\\
&&c\!\left(|r^{3-\frac{2}{p}}\ze|_{p,S}(\la_1,\nu_0)\!+\!\int_{\nu_0}^{\nu}\!|r^{3-\frac{2}{p}}F(\nabb\chi,\nabb\log\oom)|_{p,S}(\la_1,\nu')d\nu'\!\right)\nn
\eea
Dividing the left hand side by $r(\la_1,\nu)$ and observing that on $C_0$,
\bea
r^{-1}(\la_1,\nu)\leq r^{-1}(\la_1,\nu')\leq r^{-1}(\la_1,\nu_0)\ , \eql{1.42bt}
\eea
we rewrite \ref{1.41at} as
\bea
&&|r^{2-\frac{2}{p}}\ze|_{p,S}(\la_1,\nu)\leq\eql{1.42at}\\
&& c\left(|r^{2-\frac{2}{p}}\ze|_{p,S}(\la_1,\nu_0)+
\frac{1}{r(\la_1,\nu)}\int_{\nu_0}^{\nu}|r^{3-\frac{2}{p}}F(\nabb\chi,\nabb\log\oom)|_{p,S}(\la_1,\nu')d\nu'\right)\ .\nn
\eea
The integrand on the right hand side behaves, due to the assumptions on the terms $\ddb_4\nabb\log\oom$ and $\nabb\tr\chi$, as
$O(\varepsilon)$. This implies that the integral can be bounded by $c(\la_1)r(\la_1,\nu)\varepsilon$ and the
$r(\la_1,\nu)$ factor is exactly compensated by the denominator. This allows us to conclude that
\bea
|r^{2-\frac{2}{p}}\ze|_{p,S}(\la_1,\nu)\leq c\left(|r^{2-\frac{2}{p}}\ze|_{p,S}(\la_1,\nu_0)+\varepsilon\right)\ 
\eql{1.43a}
\eea
and using assumption \ref{2.15kk} the result follows.

 
 Finally we have to specify $\chib$ and $\omb$ on $C_0$. Again we decompose $\chib$ in its trace part and its traceless part,
$\tr\chib$ and $\chibh$; also these quantities cannot be assigned freely, but have to satisfy some constraint equations. Moreover we have to
control their decay along $C_0$. Let us start considering $\tr\chi$. As discussed in subsection \ref{SS1.2}, it has to satisfy the costraint
equation
\bea 
\dddd_4\tr\chib+\tr\chi\tr\chib -2\om\tr\chib-2(\divv\etab)-2|\etab|^2+2{\bf K}=0\ ,\ \ \ \eql{1.70d}
\eea
where $\etab$ has to be read  everywhere as $(-\ze+\nabb\log\oom)$ and  $\tr\chib$ can be assigned freely on $S_0$.

 To obtain $\chibh$ and its decay we anticipate that $\chibh$ will be given freely on $\Cb_0$ and it has to satisfy
along $C_0$ the constraint equation,
\bea
\dddd_4\chibh+\frac{1}{2}\tr\chi\chibh+\frac{1}{2}\tr\chib\chih-2\om\chibh
-\nabb\hot\etab-\etab\hot\etab=0\ .\eql{1.71d}
\eea
We use evolution equations \ref{1.70d} and \ref{1.71d} to obtain, on $C_0$, $\tr\chib$ and $\chibh$ and to specify their decay in $r$. We
collect these results in the following lemma:
\begin{Le}\label{L2.3}
Assuming the results of Lemma \ref{L2.1} and Lemma \ref{L1.3} there exists a simmetric tensor $\chib$ whose trace part and traceless part
satisfy the evolution equations \ref{1.70d}, \ref{1.71d}. Moreover they satisfy the following estimates:
\bea
|r^{1-\frac{2}{p}}\tr\chib|_{p,S}\leq c \ ,\ |r^{1-\frac{2}{p}}\chibh|_{p,S}\leq c\varepsilon\ .
\eea
\end{Le}
{\bf Remark:} Again these estimates can be improved to pointwise ones, assuming enough regularity for the initial data.\footnote{The decay
estimate for
$\chibh$ on $\Cb_0$ will be stronger in $|\la|$.}

 {\bf Proof:} First we determine $\tr\chib$ solving equation \ref{1.70d}. This is easy as all the coefficients and the inhomogeneous terms
appearing in the equation have been already obtained together with their decay. To prove that $\tr\chib$ decays as $O\left({r}^{-1}\right)$
we proceed as in the case of the one form $\ze$. Equation \ref{1.70d} can be written as
\bea
\dddd_4\tr\chib+\tr\chi\tr\chib-2\om\tr\chib=H(\etab,\nabb\etab,{\bf K})\ , \eql{1.73d}
\eea
where
\[H(\etab,\nabb\etab,{\bf K})\equiv -2{\bf K}+2(\divv\etab)+2|\etab|^2\ .\]
Proceeding as done before for $\ze$, we obtain
\bea
&&|r^{2-\frac{2}{p}}\tr\chib|_{p,S}(\la_1,\nu)\leq\eql{1.74d}\\
&&c\!\left(\!|r^{2-\frac{2}{p}}\tr\chib|_{p,S}(\la_1,\nu_0)\!+\!\int_{\nu_0}^{\nu}|r^{2-\frac{2}{p}}
H(\etab,\nabb\etab,{\bf K})|_{p,S}(\la_1,\nu')d\nu'\!\right)\nn
\eea
Dividing the left hand side by $r(\la_1,\nu)$ and recalling \ref{1.42bt} we obtain
\bea
|r^{1-\frac{2}{p}}\tr\chib|_{p,S}(\la_1,\nu)\!&\leq&\! c\left(|r^{1-\frac{2}{p}}\tr\chib|_{p,S}(\la_1,\nu_0)\right.\eql{1.76c}\\
 &+&\!\left.\frac{1}{r(\la_1,\nu)}\int_{\nu_0}^{\nu}|r^{2-\frac{2}{p}}H(\etab,\nabb\etab,{\bf K})|_{p,S}(\la_1,\nu')d\nu'\right)\ .\nn
\eea
As, due to the presence of the Gauss curvature ${\bf K}$, $H(\etab,\nabb\etab,{\bf K})=O(r^{-2})$, the second term in the right hand side can
be bounded by a constant so that, finally, 
\bea 
|r^{1-\frac{2}{p}}\tr\chib|_{p,S}(\la_1,\nu)\leq c\left(|r^{1-\frac{2}{p}}\tr\chib|_{p,S}(\la_1,\nu_0)+1\right)\ .\eql{1.76d}
\eea
Once we have obtained $\tr\chib$, $\chibh$ is obtained immediately applying Gronwall's Lemma to  equation \ref{1.71d}. \footnote{Observe that due to
the fact that we are integrating forward in time along $C_0$ we cannot improve the decay for $\chibh$.}

{\bf Remark:} More precisely, to prove Lemma \ref{L2.3} we need an extended version of Lemma \ref{L2.1} as we have to control $\nabb\ze$. This
is achieved writing the evolution equation for $\nabb\ze$:
\bea
&&\dddd_4\nabb_a\ze_b
=-2\tr\chi\nabb_a\ze_b-(\chih_{ac}\nabb_c\ze_b+\chih_{bc}\nabb_a\zeta_c)-\eta_b\chi_{ac}\ze_c+\chi_{ab}(\etab\c\ze)\nn\\
&&\ \ \ \ \ -\frac{3}{2}\zeta_b\nabb_a\tr\chi-\zeta_b\nabb_a\chih+\nabb_a(\divv\chih)_b-\frac{1}{2}\nabb_a\nabb_b\tr\chi-\nabb_a\ddb_4\nabb_b\log\oom\ .\nn
\eea
 From it the control of $\nabb\ze$ proceeds in the same way as the one for $\ze$, but it requires the control of $\nabb^2\chi$ and of
$\nabb^2\dddd_4\log\oom$. The terms $\nabb_a\nabb_b\tr\chi$, $\nabb_a\nabb_b\chih$ and $\nabb^2\dddd_4\log\oom$ have to satisfy the estimates
\bea
&&|r^{4-\frac{2}{p}}\nabb_a\nabb_b\tr\chi|_{p,S}\leq c\varepsilon\ \ ,\ \ |r^{(\frac{9}{2}+\de)-\frac{2}{p}}\nabb^2\chih|_{p,S}\leq
c\varepsilon\nn\\
&&|r^{(4+\de)-\frac{2}{p}}\nabb^2\dddd_4\log\oom|_{p,S}\leq c\varepsilon\ .\eql{2.32pr}
\eea
Therefore estimates \ref{2.32pr} have to be included in the initial assumptions of Lemma \ref{L2.3}. In other words they must be part of the
results of an easily extended version of Lemma \ref{L2.1} and Lemma \ref{L1.3}.
\smallskip

 We are left with determining the function $\omb$. In the Einstein vacuum spacetime it will be related to the function $\oom$ by the relation
\bea
\omb=-\frac{1}{2\oom}\frac{\partial\log\oom}{\partial\la}\ .
\eea
Again the Einstein equations imply a constraint equation for it on $C_0$. 
In the introduction we have shown that $\rr_{34}=0$ implies for $\om$ and $\omb$ the following equation:
\bea
\dd_3\om+\dd_4\omb-4\om\omb-3|\ze|^2+|\nabb\log\oom|^2=\ro(\chi,\chib,\eta,\etab)\eql{1.83c}
\eea
where, see \ref{1.26w},
\bea\ro(\chi,\chib,\eta,\etab)=-\left[{\bf K}+\frac{1}{4}\tr\chi\tr\chib-\frac{1}{2}\chih\c\chibh\right]\ .\eea  
This equation can be written in a slightly different way in the Einstein vacuum spacetime, $(\M,\ggg)$. In fact, there, the following relation
holds:\footnote{Equation \ref{1.84c}  holds in the Einstein spacetime, but cannot be derived intrinsecally on $C_0$.
On the other side on $C_0$ we can choose as costraint equation either \ref{1.83c} or \ref{1.86c}.}
\bea
\dd_3\om=\dd_4\omb+(\etab-\eta)\c\nabb\log\oom\ \ \ \ \ \eql{1.84c} 
\eea
and \ref{1.83c} can be rewritten as
\bea
\dd_4\omb\!-\!2\om\omb\!=\!\bigg[\ze\c\nabb\log\oom\!+\!\frac{3}{2}|\ze|^2\!-\!\frac{1}{2}|\nabb\log\oom|^2
\!+\!\frac{1}{2}\ro(\chi,\chib,\eta,\etab)\bigg]\eql{1.86c}
\eea
This evolution equation is the costraint equation that the scalar quantity $\omb$ has to satisfy on $C_0$.
All the terms in the right hand side of \ref{1.86c} have already been obtained, therefore from this equation we determine $\omb$ on $C_0$, once
we specify it on $S_0$.
 
{\bf Remark:}
We can choose the initial data for $\omb$ on $S_0$ in such a way that $\lim_{r\rightarrow\infty}\omb\rightarrow 0$ along $C_0$. This requires that on
$S_0$,\footnote{Specifying the value of $\omb$ on $S_0$ amounts to
assigne $-\frac{1}{2}\frac{\partial\log\oomb}{\partial\la}$ on $S_0$. Recall that on
$\Cb_0$ we have: $u(v,\theta,\phi)=\int_0^v\frac{1}{\oomb^2}(v',\theta,\phi)dv'$. From the change of variable $u=u(v,\theta,\phi)$ it follows that
$\frac{du}{dv}=(\frac{dv}{du})^{-1}$ which implies $\frac{\partial\oomb}{\partial u}=
\frac{\partial\oomb}{\partial v}\frac{du}{dv}^{-1}=\oomb^2\frac{\partial\oomb}{\partial v}$. Therefore imposing $\omb|_{S_0}=k_0$ is
equivalent to require that $\oomb$ satisfies on $S_0$: $-\frac{1}{2}\oomb\frac{\partial\oomb}{\partial v}=k_0$.}
\bea
\omb(\la_1,\nu_0)=-\!\int_{\nu_0}^{\infty}\!\left[\ze\c\nabb\log\oom+\frac{3}{2}|\ze|^2\!
-\!\frac{1}{2}|\nabb\log\oom|^2+\!\frac{1}{2}\ro(\chi,\chib,\eta,\etab)\right]\ .\ \ \ \eql{1.94d}
\eea
 The choice
\ref{1.94d} for $\omb(\la_1,\nu_0)$ implies that $\omb$ decays as $O(r^{-2})$ along $C_0$, a condition which will be used
in the proof of the global existence result.\footnote{The same procedure is not needed for obtaining $\om$ along $\Cb_0$ due to the
fact that $\Cb_0$ is an incoming cone.}

\subsubsection{The determination of the vector field $X$}

The subject of this subsection is somewhat separated from the previous ones. In fact the choice of $X$ is connected to the choice of the $\om$ coordinates.
From the knowledge of $\ze$ on $C_0$, using the relation \ref{1.69ex}
\bea
\ze_a=-\frac{1}{4\oom^2}\ga_{ab}\frac{\partial X^b}{\partial\nu}\ ,\nn\eql{1.82ex}
\eea
and requiring that $X^a$ takes a well defined value on $S_0$, we can obtain
$X^a$ on the whole $C_0$ such that \footnote{If $|r^{2-\frac{2}{p}}\ze|_{p,S}(\la_1,\nu)$ is bounded it follows that
$|\ze^a|=O(r^{-3})$ and $|X^a|=O(r^{-2})$.}
\bea
|X|=O(r^{-1})\ .
\eea
To prove this last estimate we have to assume that $X^b$ goes to zero, as $r\rightarrow\infty$, on $C_0$. In this case from \ref{1.69ex} we have:
\bea
X^a|_{S_0}=X^a(\la_1;\nu_0,\theta,\phi)=\int_{\nu_0}^{\infty}4\oom^2\ga^{ac}\ze_c(\la_1;\nu',\theta,\phi)d\nu'\ \eql{1.84e}
\eea
and from it
\bea
X^a(\la_1;\nu,\theta,\phi)=\int_{\nu}^{\infty}4\oom^2\ga^{ac}\ze_c(\la_1;\nu',\theta,\phi)d\nu'\ .\eql{1.84ef}
\eea
As the integrand decays as $O(\frac{1}{r^3})$ it follows that $X^b=O(r^{-2})$ and $|X|=\sqrt{\ga_{ab}X^aX^b}=O(r^{-1})$. Observe that the right
hand side of \ref{1.84e} does not depend on $X$, therefore, once we have obtained $\ze$, choosing $X^a$ on $S_0$ as in \ref{1.84e} implies
the result.

\subsubsection{The control of $\frac{\partial X^a}{\partial u}$ on $C_0$.}\label{SS2.3}
 In subsubsection \ref{SS1.3.2} it was said that, knowing the connection coefficients on $C_0\cup\Cb_0$ and the orthonormal moving frame, we
can obtain, with respect to the $\{u,\ub,\om^a\}$ coordinates, the Christoffel symbols and the first derivatives of the metric tensor
 \ref{1.36b} restricted to the initial hypersurface ${\cal C}=C_0\cup\Cb_0$. Nevertheless the issue was not complete as one Christoffel symbol,
$\Ga^a_{uu}$, could not be expressed directly as a combination of the connection coefficients and the metric components. This is due to the
presence, in its explicit expression, of the term $\frac{\partial X_a}{\partial u}$. To prove our statement we have, therefore, also to express
this quantity in  terms of the connection coefficients and the moving frame. This will be obtained observing that
$\frac{\partial X^a}{\partial u}$ cannot be assigned freely along $C_0$, but has to satisfy a transport equation.
\begin{Le}
Given on $C_0$  the components of the metric $\tilde{\ggg}$, the connection coefficients $\ze, \chib,\omb$  and $\partial_{u}X_{b}$ on $S_0$, we
can obtain $\partial_{u}X_{b}$ on the whole $C_0$. 
\end{Le}
{\bf Proof:} the result is obtained writing a transport equation for $\partial_{u}X^{a}$ which specifies it
on the whole $C_0$. This transport equation is derived starting from the relation:
\bea
\frac{\partial}{\partial\ub}\left(\frac{\partial X^a}{\partial u}\right)=\frac{\partial}{\partial u}\left(\frac{\partial X^a}{\partial \ub}\right)
=\frac{\partial}{\partial u}\left(4\oom^2\ze^a\right)=8\oom^2\frac{\partial\log\oom}{\partial u}\ze^a+4\oom^2\frac{\partial}{\partial u}\ze^a\nn
\eea
The first term in the right hand side is known as it is expressed in terms of $\oom$, $\omb$ and $\ze$; the second term can be rewritten as
\bea
\frac{1}{\oom}\frac{\partial}{\partial u}\ze^a\!=\!\partial_{e_3}\ze^a\!-\!\oom^{-1}\partial_{X}\ze^a
\!=\!(\dddd_{e_3}\ze)^a\!-\!\oom^{-1}(\dddd_{X}\ze)^a\!-\!e_3^{\mu}\Ga^a_{\mu\ro}\ze^{\ro}\!+\!\oom^{-1}X^c\Ga^a_{c\ro}\ze^{\ro}\ \ .
\ \ \eql{2.111t}
\eea
Observe that the components $\ze^{\ro}$ different from zero are $(\ze^b,\ze^{\ub})$ therefore all the Christoffel symbols written in the right hand
side of \ref{2.111t} are different from $\Ga^a_{uu}$ and can be expressed in terms of the connection coefficients, see \ref{1.57wtw}. The term
$(\dddd_{X}\ze)^a$ is also known in terms of the previous quantities, therefore the only thing left is $(\dddd_{e_3}\ze)^a$ which is known once we
know $(\dddd_{e_3}\ze)_a$. The knowledge of this last quantity is, nevertheless a consequence of the structure equations. In fact equation
$\rr_{3a}=0$, which must be satisfied on $C_0$, is
\bea
\dddd_3\zeta=-2\chib\c\zeta+\ddb_3\nabb\log\oom+\nabb\tr\chib-\divv\chib+\zeta\c\chib-\zeta\tr\chib\ .
\eea
therefore we know explicitely $\dddd_3\zeta$ as the right hand side is expressed in terms of quantities already obtained. Substituting this
expression in \ref{2.111t} we end with a linear transport equation, where all terms in the right hand side are known,
\bea
\frac{\partial}{\partial\ub}\left(\frac{\partial X^a}{\partial
u}\right)=H\left(\oom,\nabb\oom,\ze,\nabb\ze,\chi,\nabb\chi,\chib,\nabb\chib,\omb,\nabb\omb\right)\ .
\eea

\subsection{The solution of the characteristic costraint problem for the initial data on the incoming cone, with a prescribed decay
rate.}\label{SS2.2}

The construction of the initial data on $\Cb_0$ is very similar to what has been done on $C_0$, basically the role of the underlined and not
underlined quantities is interchanged, but there are also some other relevant differences and in this subsection we will focus mainly on them.

 The first remark is that, as already discussed, the symmetry between the structure equations along the outgoing or the incoming cone
is true when we consider the equations written in a coordinate independent way, see for instance subsection \ref{SS1.2}. This symmetry is lost if, viceversa,
we write the equations in a coordinate dependant way, choosing, for instance, the coordinates $\{u,\ub,\om^a\}$ such that the metric has the
expression \ref{1.36b},
\beaa
\tilde{\ggg}=|\tilde{X}|^2du^2\!-\!2\tilde{\oom}^2(dud\ub\!+\!d\ub du)\!-\!\tilde{X}_a(dud\om^a\!+\!d\om^a du)\!+\!\tilde{\ga}_{ab}d\om^a d\om^b\ .\ \ \
\eeaa
In this case, in fact, the expression for the geodesic vector fields $L$ and $\Lb$ are different, see \ref{2.5ew}. Therefore, while all the
estimates for the connection coefficients along $\Cb_0$ can be obtained exactly as before, just interchanging the
underlined and the not underlined coefficients and taking into account that we are considering here a portion of the incoming ``cone", the
estimates relative to the metric components
$\ga_{ab}$ depend necessarily on the coordinates choice. It follows that the proof of Lemma \ref{L2.4}, the analogue of
Lemma \ref{L2.1}, if done in the same chart is slightly different, while it will be completely equivalent if we choose
a different set of coordinates such that $\Lb=\partial/\partial u$ and $L=\partial/\partial\ub+Y$, see the discussion in
subsection \ref{SS1.3.2} and footnote 20. It is easy to realize that in most of our work the choice of a set of coordinates is not
relevant and that both strategies to prove Lemma \ref{L2.4} are equivalent. The only part of this work where we have to use a specific set
of coordinates is when we look for a connection with harmonic coordinates, which is needed only to use a local existence proof needed to start
our argument. This will be discussed in more detail in subsections \ref{SS3.1} and \ref{SS4.4}.
\medskip

 We have defined $\Cb_0$ as the level surface, see \ref{1.33d}, 
\[\Cb_0\equiv \Cb(\nu_0)=\{p\in R^4|\ub(p)=\nu_0, \la\in[\la_1,\la_0]\}\]
in the manifold $(R^4,\tilde{\ggg})$. Choosing the previous coordinates the metric $\tilde{\ggg}$ has the expression \ref{1.36b},
\beaa
{\tilde{\ggg}}=|\tilde{X}|^2du^2\!-\!2\tilde{\oom}^2(dud\ub\!+\!d\ub du)\!-\!\tilde{X}_a(dud\om^a\!+\!d\om^a du)\!+\!\tilde{\ga}_{ab}d\om^a d\om^b
\eeaa
where $\tilde{\oom},\tilde{X}$ and $\tilde\ga$ can be thought as extensions to $R^4$ of the quantities we have obtained on $C_0$,
\beaa
X_a=\tilde{X}_a|_{C_0}\ ,\ \oom=\tilde{\oom}|_{C_0}\ ,\ \ga_{ab}=\tilde{\ga}_{ab}|_{C_0}\ .
\eeaa
As we said these extended components are arbitrary in the interior region between $C_0$ and $\Cb_0$. The construction of the appropriate initial
data on $\Cb_0$ will determine, as before, the restriction of these metric components on $\Cb_0$.

 We start specifying the restriction of the function $\tilde\oom$ on $\Cb_0$, ${\tilde\oom}|_{\Cb_0}=\underline{\oom}$.

$\underline{\oom}$ is assigned freely except for the constraint on $S_0$ following from the condition \ref{1.94d} required to have the
correct decay of $\omb$ on $C_0$,
\bea
\dd_3\log{\underline{\oom}}=-2\omb|_{S_0}\ .
\eea
 Once $\underline{\oom}$ is given we define, as done before for $C_0$, the ``$\underline{\oom}$-foliation" of $\Cb_0$. The
leaves of the foliation are the surfaces ${\underline S}_0(\la)=S(\la,\nu_0)=\{p\in \Cb(\nu_0)|u(p)=\la\}$. 
Denoting $v$ the affine parameter of the null geodesics on $\Cb_0$ we have: \footnote{In this case differently from $C_0$, denoted $(u,\theta,\phi)$ the
coordinates of the point $p$, $\ga$ is the null geodesic starting on $S_0$ at the point $q$, whose angular coordinates are different
from $(\theta,\phi)$, such that $\ga(v(p))=p$.}
\bea
u(p)=\la_1+\int_0^{v(p)}(4\oom)^{-2}(\ga(v))dv\ .\eql{1.41bq}
\eea
 As in the $C_0$ case the vector field $\Nb$ is the equivariant vector field associated to this foliation; denoting $\underline{\phi}_{\de}$ the
diffeomorphisms generated by $\Nb$ on $\Cb(\nu_0)$ it follows
\[\underline{\phi}_{\de}[S(\la,\nu_0)]=S(\la+\de,\nu_0)\ .\]
From now on the estimates on $\Cb_0$ are performed assuming on $(R^4,\tilde{\ggg})$ a different set of coordinates such that $\tilde{\ggg}$ has the
following expression, see the discussion at the beginning of subsection \ref{SS1.3.2} and footnote 19,
\bea
\tilde{\ggg}=|\tilde{Y}|^2d\ub^2\!-\!2\tilde{\oom}^2(dud\ub\!+\!d\ub du)\!-\!\tilde{Y}_a(d\ub
d\om'^a\!+\!d\om'^ad\ub)\!+\!\tilde{\ga}'_{ab}d\om'^ad\om'^b\ \ \eql{1.79d} 
\eea
On each leave ${\underline S}_0(\la)=S(\la,\nu_0)$ we define two orthonormal \footnote{Orthonormal with respect to the $\tilde{\ga}$ metric.}
vector fields
$\tilde{e}_A$ tangent to it,
\[\tilde{e}_A=\tilde{e}_A^a\frac{\partial}{\partial\om'^a}\ \]
and require that the orthonormal frame $\{e_A\}$ be Fermi transported along $\Cb_0$.
Again the metric $\tilde{\ga}'$  in \ref{1.79d}  is not the metric we want to give as initial data on $\Cb_0$, but we use it as a background
metric from which to construct the final one satisfying, with its derivatives, the ``characteristic" costraints.

 To perform the construction of the initial data, proceeding as in the previous section, we start defining  $\omb$,
\bea
\omb=-\frac{1}{2\oom}\frac{\partial}{\partial{\la}}\log{\underline\oom}\ .
\eea
Then we assigne a symmetric traceless tensor $\chibh$ on ${\underline S}_0(\la)$, $\la\in [\la_1,\la_0]$, which we will consider as the
traceless part of a tensor
$\chib$. We require that the scalar functions ${\underline\oom}$, $\omb$ and the tensor $\chibh$ have the following asymptotic behaviours:
\bea
&&|{\underline\oom}-\frac{1}{2}|=O(\varepsilon\tilde{r}^{-1})\ \ ,\ \ |\tilde{\nabb}\log{\underline\oom}|=O(\varepsilon\tilde{r}^{-(2+\de)})\nn\\
&&|\omb|=O(\varepsilon\tilde{r}^{-2})\ \ ,\ \ |\chibh|_{\tilde{\ga}}=O(\varepsilon\tilde{r}^{-1}|\la|^{-(\frac{3}{2}+\de)}), \eql{2.15z}\\
&&|\tilde{\nabb}\chibh|_{\tilde{\ga}}=O(\varepsilon\tilde{r}^{-2}|\la|^{-(\frac{3}{2}+\de)})\ ,\
|\dddd_3\chibh|_{\tilde{\ga}}=O(\varepsilon\tilde{r}^{-1}|\la|^{-(\frac{5}{2}+\de)})\nn\\
&&|\dddd_3\tilde{\nabb}\log\oom|=O(\varepsilon|\la|^{-1}\tilde{r}^{-(2+\de)})\nn
\eea
with $\de>0$ and $\tilde{r}$ is defined from the relation $4\pi \tilde{r}^2(\la)=|{\underline S}_0(\la)|_{\tilde{\ga}'}$.

 We require that the function $\tr\chib$ and the initial metric components $\ga'_{ab}$, we denote hereafter again  $\ga_{ab}$, assigned on
$S_0={\underline S}_0(\la_1)=S_0(\nu_0)$,  satisfy the following system of differential equations:
\bea
&&\frac{\partial\ga_{ab}}{\partial\la}-{{\underline\oom}\tr\chib}\ga_{ab}-2{\underline\oom}\chibh_{ab}=0\eql{1.84d}\\
&&\frac{\partial\tr\chib}{\partial\la}+\frac{{\underline\oom}\tr\chib}{2}\tr\chib+2\oom\omb\tr\chib+|\chibh|^2_{{\ga}}=0\nn
\eea
\begin{Le}\label{L2.4}
Assigne on the null hypersurface $\Cb_0$ a ${\underline\oom}$-foliation with leaves ${\underline S}_0(\la)$, a metric tensor $\tilde{\ga}$,
tangent to the leaves ${\underline S}_0(\la)$ satisfying, together with $\tr\chib$, the trace of the null outgoing second fundamental form
associated to $\tilde{\ga}$, the equations
{\em\bea
&&\frac{\partial\tilde{\ga}_{ab}}{\partial\la}-{\underline\oom}{\tr\tilde{\chib}}\tilde{\ga}_{ab}=0\eql{2.10tt}\\
&&\frac{\partial\tr\tilde{\chib}}{\partial\la}+\frac{{\underline\oom}\tr\tilde{\chib}}{2}\tr\tilde{\chib}+2{\underline\oom}\ \!\omb\tr\tilde{\chib}=0\ .\nn
\eea}
Assume that, given $\varepsilon>0$,
\bea|\la_1|\in[c_1{\tilde r}_0, c_2{\tilde r}_0]\ \ \mbox{with}\ \ |c_{1,2}-1|=O(\varepsilon)\ .\eea  
Given a tensor $\chibh$ on $\Cb_0$, assume that, with $\varepsilon>0$, the following decays hold
\bea
&&|{\underline\oom}-\frac{1}{2}|=O(\varepsilon\tilde{r}^{-1})\ \ ,\ \ |\nabb\log\underline\oom|=O(\varepsilon\tilde{r}^{-(2+\de)})\nn\\
&&|\omb|=O(\varepsilon\tilde{r}^{-1}|\la|^{-1})\ \ ,\ \ |\chibh|_{\tilde{\ga}}=O(\varepsilon\tilde{r}^{-1}|\la|^{-(\frac{3}{2}+\de)})\\
&&|\tilde{\nabb}\chibh|_{\tilde{\ga}}=O(\varepsilon\tilde{r}^{-2}|\la|^{-(\frac{3}{2}+\de)})\ ,\
|\dddd_3\chibh|_{\tilde{\ga}}=O(\varepsilon\tilde{r}^{-1}|\la|^{-(\frac{5}{2}+\de)})\nn\\
&&|\dddd_3\tilde{\nabb}\log\underline\oom|=O(\varepsilon\tilde{r}^{-(3+\de)})\nn
\eea
with $\de>0$, and $\tilde{r}$ defined through the relation $4\pi\tilde{r}^2(\la,\nu_0)=|{\underline S}_0(\la)|_{\tilde{\ga}}$. Assume that on
${\underline S}_0(\la_1)$,
\bea
|{\tilde r}_0^{3-\frac{2}{p}}\tilde{\nabb}\tr\chib|_{p,S}=O(\varepsilon)\ ,
\eea then there exists on
$\Cb_0$ a metric tensor $\ga$ tangent to the leaves of the same foliation such that, denoted $\chib$ the second fundamental form of the
${\underline S}_0(\la)$ with respect to this metric, $e_A$ an orthonormal frame tangent to the ${\underline S}_0(\la)$ with respect to $\ga$ we
have:
\bea
\chib_{ab}=\frac{1}{2\oomb}\frac{\partial\ga_{ab}}{\partial\la}=\chibh_{ab}+\frac{\ga_{ab}}{2}\tr\chib\ ,
\eea
with $\ga|_{S_0}=\tilde{\ga}|_{S_0}$ and $\tr\chib<0$ satisfies the constraint equation
\bea
\frac{\partial\tr\chib}{\partial\la}+\frac{\oomb\tr\chib}{2}\tr\chib+2\oomb\ \!\omb\tr\chib+|\chibh|^2_{\ga}=0
\eea
with $\tr\chib|_{S_0}=\tr{\tilde{\chib}}|_{S_0}$. Moreover the following estimates hold, with $p\in[2,4]$, 
\bea
&&|{\underline\oom}-\frac{1}{2}|=O(\varepsilon{r}^{-1})\ \ ,\ \ |\nabb\log{\underline\oom}|=O(\varepsilon{r}^{-(2+\de)}),\ \
|\omb|=O(\varepsilon{r}^{-1}|\la|^{-1})\nn\\
&&|\tr\chib+\frac{2}{r}|=O(\varepsilon r^{-2}\log r)\ \ ,\ \ |\chibh|_{{\ga}}=O(\varepsilon{r}^{-1}|\la|^{-(\frac{3}{2}+\de)})\ ,\eql{1.48d}\\
&&|r^{2-\frac{2}{p}}|\la|^{(\frac{3}{2}+\de)}\nabb\chibh|_{p,S}=O(\varepsilon)\ ,
\ |r^{1-\frac{2}{p}}|\la|^{(\frac{5}{2}+\de)}\dddd_3\chibh|_{p,S}=O(\varepsilon)\nn\\
&&|r^{(2+\de)-\frac{2}{p}}|\la|\dddd_3\nabb\log\underline\oom|_{p,S}=O(\varepsilon)\ ,\ |r^{3-\frac{2}{p}}\nabb\tr\chib|_{p,S}=O(\varepsilon)\ ,\nn
\eea
and ${r}$  defined by the relation: $4\pi{r}^2(\la,\nu_0)=|{\underline S}_0(\la)|_{{\ga}}$\ .
\end{Le}

{\bf Proof:} We do not report the proof of Lemma \ref{L2.4} as it is obtained exactly as the one of Lemma \ref{L2.1}.  

 Once solved equations \ref{1.84d} we have assigned on the whole $\Cb_0$, $\ga$, $\underline\oom$, $\omb$ and $\chib$. $\ga$ is the
(initial data) metric relative to the leaves ${\underline S}_0(\la)$, $\chib$ is the second fundamental form of the leaves ${\underline
S}_0(\la)$ relative to this metric.  
\smallskip

 Let us discuss the way to fulfill the remaining costraint equations which have to be satisfied on $\Cb_0$.
For $\ze$ we use the evolution equation \ref{1.74c},
\bea
\dddd_3\zeta+\frac{3}{2}\tr\chib\zeta=-\chibh\c\zeta+\left[-\divv\chib+\nabb\tr\chib+\ddb_3\nabb\log\oomb\right]\ .\eql{2.89tt}
\eea
 We want that $r^2\ze$ be bounded and small, in the appropriate norm, on the whole $\Cb_0$. In this case we integrate
going down along the $\Cb_0$ cone, namely moving toward the increasing radius of the leaves, as in the case of $C_0$. The result is in the
following lemma.
\begin{Le}\label{L2.5} Assume that, on $\Cb_0$,  
{\em\bea
|r^{(2+\de)-\frac{2}{p}}|\la|\dddd_3\nabb\log\oom|_{p,S}=O(\varepsilon)\ ,\ |r^{3-\frac{2}{p}}\nabb\tr\chib|_{p,S}=O(\varepsilon)\eql{2.15sa}
\eea}
Let $\ze$  be the solution along $\Cb_0$ of the costraint equation \ref{2.89tt}
then  for $p\in[2,4]$, we have the following bound:
\bea
|r^{2-\frac{2}{p}}\ze|_{p,S}(\la,\nu_0)\leq c\left(|r^{2-\frac{2}{p}}\ze|_{p,S}(\la_0,\nu_0)+\varepsilon\right)\ .
\eql{1.43c}
\eea
\end{Le}

{\bf Proof:}
The evolution equation for $\ze$ can be written as
\bea
\dddd_3\zeta+\frac{3}{2}\tr\chib\zeta+\chibh\c\zeta=\Fb(\nabb\chib,\nabb\log{\underline\oom})\eql{2.91t}
\eea
where
\beaa
\Fb(\nabb\chi,\nabb\log\oom)=[-\divv\chibh+\frac{1}{2}\nabb\tr\chib+\ddb_3\nabb\log{\underline\oom}]\ .
\eeaa
Using standard techniques, see Chapter 4 of \cite{Kl-Ni:book}, we obtain, as we have already proved that
$|\chibh|=O(r^{-1}|\la|^{-(\frac{3}{2}+\de)})$,
\bea
|r^{3-\frac{2}{p}}\ze|_{p,S}(\la,\nu_0)\leq
c\!\left(|r^{3-\frac{2}{p}}\ze|_{p,S}(\la_0,\nu_0)\!+\!
\int_{\la_0}^{\la}|r^{3-\frac{2}{p}}\Fb(\nabb\chib,\nabb\log{\underline\oom})|_{p,S}(\la',\nu_0)d\la'\right)\ \ \
\eql{1.41a}
\eea
As on $\Cb_0$,
\bea
r^{-1}(\la,\nu_0)\leq r^{-1}(\la',\nu_0)\leq r^{-1}(\la_0,\nu_0)\ , \eql{1.42b}
\eea
from \ref{1.41a} it follows
\bea
&&|r^{2-\frac{2}{p}}\ze|_{p,S}(\la,\nu_0)\leq\eql{1.42a}\\
&& c\!\left(|r^{2-\frac{2}{p}}\ze|_{p,S}(\la_0,\nu_0)+
\frac{1}{r(\la,\nu_0)}\int_{\la_0}^{\la}|r^{3-\frac{2}{p}}\Fb(\nabb\chib,\nabb\log{\underline\oom})|_{p,S}(\la',\nu_0)d\la'\right)\ .\nn
\eea
The integrand of the right hand side behaves, due to the terms $\ddb_3\nabb\log\oomb$ and $\nabb\tr\chib$, as
$O(\varepsilon)$. Therefore the integral can be bounded by $c(\la_0)r(\la,\nu_0)$ and the
$r(\la,\nu_0)$ factor is exactly compensated by the denominator. In conclusion, if we assume
$|\tilde{r}^{2-\frac{2}{p}}\ze|_{p,S}(\la_0,\nu_0)\leq c\varepsilon$,
\bea
|r^{2-\frac{2}{p}}\ze|_{p,S}(\la,\nu_0)\leq c\left(|r^{2-\frac{2}{p}}\ze|_{p,S}(\la_0,\nu_0)+\varepsilon\right)\leq c\varepsilon \ .
\eql{1.96c}
\eea
The knowledge of $\ze$ on $S_0(\la_0)\!=\!S(\la_0,\nu_0)$, allows to obtain an estimate for $\ze$ on $S_0=S(\la_1,\nu_0)$.
Therefore the value on $S_0$ which was used to obtain and control $\ze$ on the whole $C_0$ can be still considered arbitrary. This will turn out important
when, in the next subsection, we introduce the integral norms $\cal Q$ associated to the initial data.

 The tensor quantity $\chi$ on $\Cb_0$ is obtained in the same way as we obtained $\chib$ on $C_0$. The trace and the traceless part of $\chi$,
$\tr\chi$ and $\chih$, have to satisfy on
$\Cb_0$ the costraint equations analogous to \ref{1.70d} and \ref{1.71d}, namely
\bea 
\dddd_3\tr\chi+\tr\chib\tr\chi -2\omb\tr\chi-2(\divv\eta)-2|\eta|^2+2{\bf K}=0\ ,\ \ \ \eql{2.96t}
\eea
\bea
\dddd_3\chih+\frac{1}{2}\tr\chib\chih+\frac{1}{2}\tr\chi\chibh-2\omb\chih-\nabb\hot\eta-\eta\hot\eta=0\ .\eql{2.97t}
\eea 
Proceeding as before, we prove the following lemma analogous to Lemma \ref{L2.3}.
\begin{Le}\label{L2.6}
Assuming the results of Lemma \ref{L2.4} and Lemma \ref{L2.5} there exists a simmetric tensor $\chi$ whose trace part and traceless part
satisfy the evolution equations \ref{2.96t}, \ref{2.97t}. Moreover they satisfy the following estimates:
{\em\bea
|r^{1-\frac{2}{p}}\tr\chi|_{p,S}\leq c \ ,\ |r^{2-\frac{2}{p}}\chih|_{p,S}\leq c\varepsilon\ .
\eea}
\end{Le}
{\bf Proof:} To estimate $\tr\chi$ we proceed exactly as done befor for $\tr\chib$ but integrating $\tr\chi$ on $\Cb_0$ backward, starting from
$S(\la_0,\nu_0)$.\footnote{We will choose $\tr\chi$ on ${\underline S}_0(\la_0)=S(\la_0,\nu_0)$ in a way consistent with the initial
value of $\tr\chi$ on $S_0(\nu_0)={\underline S}_0(\la_1)$, specified in Lemma \ref{L2.1}.}
 To obtain the decay of
$\chih$ we recall that
$\chih$ has been given freely on
$C_0$. Therefore we know it on $S_0$ and we can determine it along $\Cb_0$ using its evolution equation \ref{2.97t}
Applying standard techniques, see \cite{Kl-Ni:book}, Chapter 4, we obtain
\bea
|r^{1-\frac{2}{p}}\chih|_{p,S}(\la,\nu_0)\leq
c\!\left(\!|r^{1-\frac{2}{p}}\chih|_{p,S}(\la_1,\nu_0)\!+\!\int_{\la_1}^{\la}\!|r^{1-\frac{2}{p}}G(\tr\chi,\chibh,\eta,\nabb\eta)|\!\right)\ \ \ \ 
\eea
where 
\bea
G(\tr\chi,\chibh,\eta,\nabb\eta)=\left[\frac{1}{2}\tr\chi\chibh-\nabb\hot\eta-\eta\hot\eta\right]
\eea
satisfies
\bea
|r^{1-\frac{2}{p}}G(\tr\chi,\chibh,\eta,\nabb\eta)|_{p,S}(\la,\nu_0)\leq
c\!\left(\frac{1}{r(\la,\nu_0)|\la|^{\frac{3}{2}+\de}}\!+\!\frac{1}{r(\la,\nu_0)^2}\right)\!\ .\ \ \ 
\eea
This implies the following bound of $\chih$ along $\Cb_0$ (for all the $\la$ values such that $|\la|<r(\la,\nu_0)$),
\bea
|r^{2-\frac{2}{p}}\chih|_{p,S}\leq c\varepsilon\ ,\eql{2.102w}
\eea
proving the lemma. Again this estimate can be improved to a pointwise one, assuming enough regularity for the initial data.
 
 We are left with determining the function $\om$ on $\Cb_0$.\footnote{In the Einstein vacuum spacetime it will be connected to the function
$\oom$ by the relation $\om=-\frac{1}{2}\frac{\partial\log\oomb}{\partial\nu}$.}
Again this function cannot be assigned freely but has to satisfy on $\Cb_0$ the constraint equation \ref{1.83c}:
\bea
\dd_3\om+\dd_4\omb-4\om\omb-3|\ze|^2+|\nabb\log{\underline\oom}|^2=\ro(\chi,\chib,\eta,\etab)\ \ \ \eql{1.83cww}
\eea
which, proceeding as done for $\omb$ on $C_0$, can be written as:
\bea
\dd_3\om\!-\!2\omb\om=G(\ze,\nabb\log{\underline\oom},\ro)
\eea
where
\bea
G(\ze,\nabb\log{\underline\oom},\ro)=-\ze\c\nabb\log{\underline\oom}\!+\!\frac{3}{2}|\ze|^2\!-\!\frac{1}{2}|\nabb\log{\underline\oom}|^2
+\frac{1}{2}\ro(\chi,\chib,\eta,\etab)\ .\ \ \ \ \ \eql{2.106t}
\eea
 All the terms in the right hand side of the evolution equation \ref{2.106t} are known, therefore this equation can be used to determine $\om$ on
$\Cb_0$ once we know it on $S_0$.
 
{\bf Remark:} Observe that, differently from the estimate of $\omb$ on $C_0$, in this case equation \ref{2.106t} allows to obtain the decay
for $\om$ on $\Cb_0$ without requiring a specific choice of $\om$ on $S_0$. In fact observing that $\omb$ is integrable along $\Cb_0$ and
that $G(\ze,\nabb\log\oom,\ro)=O(r^{-3})$, we conclude, integrating forward in time, that, assuming on $S_0$,
$|r_0^2\dd_4\log\oom|=O(\varepsilon)$, the following bound holds on $\Cb_0$, with $\de>0$, 
\bea
||\la|^{\de}r^{(2-\de)-\frac{2}{p}}\om|_{p,S}\leq c \varepsilon\ .
\eea

\smallskip

{\bf Remark:} As discussed at the beginning of this subsection the initial data quantities on $\Cb_0$ have been obtained in a way competely
symmetric with the one used to obtain the initial data on $C_0$. This implies for the estimates of $\ga_{ab}$ in Lemma \ref{L2.4} that we used a
set of coordinates different from the coordinates $\{u,\ub,\om^a\}$ used in Lemma \ref{L2.1}. As we already said we could also use the
original set of coordinates and the difference would be that equations \ref{1.84d} have to be substituted by the equations, see \ref{1.59w},
\bea
&&\frac{\partial\ga_{ab}}{\partial\la}-{{\underline\oom}\tr\chib}\ga_{ab}+({^{(2)}\nabb}_b{\underline X}_a
+{^{(2)}\nabb}_a{\underline X}_b)-2{\underline\oom}\chibh_{ab}=0\eql{2.78zx}\\
&&\frac{\partial\tr\chib}{\partial\la}+\frac{{\underline\oom}\tr\chib}{2}\tr\chib+2\oom\omb\tr\chib+{\underline
X}^a\nabb_a\tr\chib+|\chibh|^2_{{\ga}}=0\ .\nn
\eea
It is easy to realize that Lemma \ref{L2.4} can be proved also starting from equations \ref{2.78zx}. This requires to impose some conditions
on ${\underline X}$, allowed as ${\underline X}$ can be assigned freely on $\Cb_0$.\footnote{We require some decay for
${^{(2)}\nabb}_b{\underline X}_a$ and that the condition ${\underline X}^a\nabb_a\tr\chib=0$ be satisfied.}

\subsection{The decay of the null Riemann components on $C_0$}\label{SS2.4}
In this subsection we show how to deduce from the initial data and their asymptotic behaviour the decay conditions for the various null
Riemann components. Let us explain, first of all, why we need this result. 

 In the global existence proof of the characteristic Cauchy problem we use, mimicking the strategy developed in \cite{Kl-Ni:book}, some integral
norms along the null outgoing and null incoming  ``cones" which are $L^2$ integrals of the conformal part of the Riemann tensor. In particular we
prove that these norms are bounded in terms of the same norms defined on the initial hypersurface $\cal C$. Therefore it is crucial that the initial
data be such that these norms are bounded (and small). To fulfill this request the initial data must have appropriate decays along $C_0$ and
$\Cb_0$.\footnote{The request of a decay along $\Cb_0$ has to be interpreted as we always consider a truncated finite portion of it.
Nevertheless in the existence proof there is not an upper bound on its size, this is the sense in which our result is a global one, see the
discussion in Section \ref{S.3}.} Let us start to consider the initial data on $C_0$.

 It is important to emphasize that the various null components of the Riemann tensor whose decay along $C_0$ we want to control refer to
the Riemann tensor of the Einstein vacuum spacetime (whose global existence we want to prove). Nevertheless, given the initial data satisfying the
characteristic constraints, they can be defined, all except one,\footnote{But this will not be harmful.} as specific combinations, still denoted
$\a,\b,\ro,\si,\bb$, of the initial data defined on $C_0$. Once solved the existence problem, $C_0$ becomes an embedded hypersurface,
$C(\la_1)$, in $(\M,\ggg)$ and the initial data tensor fields are the various connection coefficients restricted to $C(\la_1)$ and the various
combinations $\a,\b,...$ become the various null components of the Riemann tensor expressed in terms of the connection coefficients.

 More precisely, see for instance \cite{Kl-Ni:book}, Chapter 3, the various null Riemann components in the vacuum Einstein spacetime
$(\M,\ggg)$ are defined as \footnote{In a vacuum Einstein spacetime the Riemann tensor coincides with its conformal part.}
\bea
&&\a({\rr})({e_A},{e_B})={\rr}({e_A},e_4,{e_B},e_4)\nn\\
&&\b({\rr})({e_A})=\frac{1}{2}{\rr}({e_A},e_4,e_3,e_4)\nn\\
&&\ro({\rr})=\frac{1}{4}{\rr}(e_3,e_4,e_3,e_4)\eql{3.1.19za}\\
&&\si({\rr})=\frac{1}{4}\ro(^\star {\rr})=\frac{1}{4}{^\star{\rr}}(e_3,e_4,e_3,e_4)\nn\\
&&\bb({\rr})({e_A})=\frac{1}{2}{\rr}({e_A},e_3,e_3,e_4)\nn\\
&&\aa({\rr})({e_A},{e_B})={\rr}({e_A},e_3,{e_B},e_3)\ .\nn
\eea 
On the other side, as discussed in Section \ref{S.1} the structure equations give the explicit expression of the conformal Riemann tensor ${\bf C}$ in
terms of the connection coefficients. Therefore we can write the null components $\a,\b,\ro,\si,\bb$ (restricted to $C_0$) in terms
of the initial data, see subsubsection \ref{SS1.2.2}, as \footnote{Observe that $\aa$ cannot be expressed in terms of the initial data
and their derivatives along $C_0$, see \ref{1.30w}.}
\bea
&&\a=-[\dddd_4\chih+\tr\chi\chih-(\dd_4\log\oom)\chih]\nn\\
&&\b=\nabb\tr\chi-\divv\chi-\zeta\c\chi+\zeta\tr\chi\nn\\
&&\ro=-[{\bf K}+\frac{1}{4}\tr\chi\tr\chib-\frac{1}{2}\chibh\c\chih]\eql{1.69c}\\
&&\si=\curll\zeta-\frac{1}{2}\chibh\wedge\chih\nn\\
&&\bb=-[\nabb\tr\chib-\divv\chib+\zeta\c\chib-\zeta\tr\chib]\nn
\eea 
These expressions can also be interpreted as the restriction of the Riemann
tensor of the Lorentzian manifold $(R^4,\tilde{\ggg})$ where $\tilde{\ggg}$ is the metric \ref{1.36b} \footnote{Obviously $(R^4,\ggg)$ is not an
Einstein vacuum spacetime as the extensions of the metric components outside $C_0$ does not satisfy the Einstein equations.}
where $\tilde{\oom},\tilde{X}$ and $\tilde\ga$ are extensions to $R^4$ of the quantities we have obtained on $C_0$.
Therefore in $(R^4,\tilde{\ggg})$ the Bianchi equations, when restricted to $C_0$, have exactly the same expression as in the vacuum Einstein
manifold. In other words, as a consequence of the constraint equations, the Ricci tensor of $(R^4,\tilde{\ggg})$ restricted to $C_0$, is
identically zero.\footnote{The structure equations \ref{1.73c},...,\ref{1.20z} which on the right hand side are
equal to zero, in a general manifold would be equal to the Ricci part of the Riemann tensor. Requiring that the constraints equations are
satisfied implies, therefore, that on $C_0$ the Ricci tensor is zero.} This
allows us to use the Bianchi equations to show that the initial data imply the appropriate decay for the null components $\b,\ro,\si,\bb$. This
discussion and the decay estimates we are looking for are summarized in the following propositions:
\begin{Prop}
Let $\{{\ga}_{ab},{\oom},{\ze}_a,{\chib}_{ab},\omb\}$ be the initial data on $C_0$ satisfying the $|\c|_{p,S}$ norm bounds of Lemma \ref{L2.1} and
Lemma \ref{L1.3}, then the combinations of the (derivatives of the) initial data $\b,\ro,\si,\bb$, defined in \ref{1.69c}, satisfy on
$C_0$ the following evolution equations, with $\etab=-\ze+\nabb\log\oom$:
\bea
&&\dddd_4\b+2\tr\chi\b=\divv\a-\left[2\om \b-(2\ze+\etab)\a\right]\nn\\
&&\dd_4\ro+\frac{3}{2}\tr\chi\ro=\divv\b-\left[\frac{1}{2}\hat{\chib}\c\a-\ze\c\b-2\etab\c\b\right]\nn\\
&&\dd_4\si+\frac{3}{2}\tr\chi\si=-\divv\dual\b+\left[\frac{1}{2}\hat{\chib}\c\dual\a-\ze\c\dual\b-2\etab\c\dual\b\right]\eql{1.70cb}\\
&&\dddd_4\bb+\tr\chi\bb=-\nabb \ro+\left[2\om\bb+2\hat{\chib}\c\b+\dual\nabb\si-3(\etab\ro-\dual\etab\si)\right]\nn
\eea 
\end{Prop}
{\bf Proof:} This explicit expression of the Bianchi equations has been derived, for instance, in \cite{C-K:book}, see also \cite{Kl-Ni:book},
Chapter 3. 

\NI{\bf Remarks:} 

 a) Observe that the Bianchi equations do not provide an evolution equation along the outgoing null cones for the
null component $\a$. Therefore its decay along $C_0$  has to be obtained in a different way.

 b) We are not interested on the decay along $C_0$ of the null component $\aa$. This is due to the fact that the $\cal Q$ flux-norms we will use
along the outgoing cones do not depend on $\aa$, see their definition in \cite{Kl-Ni:book}, subsection 3.5.1.\footnote{This is consistent with our
setting as to express $\aa$ in terms of initial data on $C_0$ we would need the derivative along $u$ of the $\chib$ tensor which is not
provided by the initial data (on $C_0$).} 

\begin{Prop}\label{P1.2}
Let $\{{\ga}_{ab},{\oom},{\ze}_a,{\chib}_{ab}\}$ be the initial data on $C_0$ satisfying the $|\c|_{p,S}$ norm bounds of Lemma \ref{L2.1} and
Lemma \ref{L1.3} and analogous estimates for the derivatives of the initial data \footnote{In
the following remark we discuss how much regularity for the initial data is needed to prove Poposition \ref{P1.2}.} assume that on
$S_0=S(\la_1,\nu_0)$ the following conditions are satisfied:  {\em\bea
&&\nabb\tr\chi+\ze\tr\chi=O(\varepsilon r_0^{-(\frac{7}{2}+\de)}) \nn\\
&&{\bf K}-\overline{\bf K}+\frac{1}{4}\!\left(\tr\chi\tr\chib-\overline{\tr\chi\ \!\tr\chib}\right)=O(\varepsilon r_0^{-(\frac{7}{2}+\de)})\nn\\
&&\curll\zeta-\frac{1}{2}(\chibh\wedge\chih)=O(\varepsilon r_0^{-(\frac{7}{2}+\de)})\eql{2.113r}\\
&&\nabb\tr\chib-\ze\tr\chib=O(\varepsilon r_0^{-(\frac{7}{2}+\de)})\ ,\nn
\eea}then the combinations $\a,\b,\ro,\si,\bb$ of the (derivatives of the) initial data defined in \ref{1.69c} satisfy the following bounds,
with $\de>0$:
\bea
&&|r^{(\frac{7}{2}+\de)}\a|_{C_0}\leq c\varepsilon\ ,\ |r^{(\frac{7}{2}+\de)-\frac{2}{p}}\b|_{p,S}\leq c\varepsilon\ ,\
|r^{(3+\de)-\frac{2}{p}}(\ro-\overline{\ro},\si)|_{p,S}\leq c\varepsilon\nn\\
&&|r^{3-\frac{2}{p}}\ro|_{p,S}\leq c\varepsilon\ ,\ |r^{(2+\de)-\frac{2}{p}}\bb|_{p,S}\leq c\varepsilon\ \ \ \ \ \eql{2.85rt}
\eea
\end{Prop}
{\bf Proof:} We start looking at $\a$. From its definition in \ref{1.69c}, 
\beaa
\a=-[\dddd_4\chih+\tr\chi\chih-(\dd_4\log\oom)\chih]\ ,
\eeaa
and the results of Lemma \ref{L2.1}, see \ref{1.48c}, the correct decay for $\a$ follows immediately.

 To determine the asymptotic behaviour of $\b$, through the expression
\beaa
\nabb\tr\chi-\divv\chi-\zeta\c\chi+\zeta\tr\chi=\b\ ,
\eeaa
we use the evolution equation
\beaa
\dddd_4\b+2\tr\chi\b=2\om\b-\left[\divv\a-(\ze+\nabb\log\oom)\a\right]\ .\eql{1.70c}
\eeaa
Proceeding as we did for $\ze$ we obtain, using the Gronwall's lemma, the following estimate, with
$G(\a,\nabb\a,\ze,\nabb\log\oom)\equiv[\divv\a-(\ze+\nabb\log\oom)\a]$,
\bea
&&|r^{4-\frac{2}{p}}\b|_{p,S}(\la_1,\nu)\leq\eql{1.49as}\\
&& c\left(|r^{4-\frac{2}{p}}\b|_{p,S}(\la_1,\nu_0)+
\int_{\nu_0}^{\nu}|r^{4-\frac{2}{p}}G(\a,\nabb\a,\ze,\nabb\log\oom)|_{p,S}(\la_1,\nu')d\nu'\right)\ .\nn
\eea
Dividing both sides by ${r^{(\frac{1}{2}-\de)}(\la_1,\nu)}$ using inequality \ref{1.42b} we rewrite \ref{1.49as} as
\bea
&&|r^{(\frac{7}{2}+\de)-\frac{2}{p}}\b|_{p,S}(\la_1,\nu)\leq
c\!\left(\!|r^{(\frac{7}{2}+\de)-\frac{2}{p}}\b|_{p,S}(\la_1,\nu_0)\right.\eql{1.49aa}\\ && 
\left.+\frac{1}{r^{(\frac{1}{2}-\de)}(\la_1,\nu)}\!\int_{\nu_0}^{\nu}|r^{4-\frac{2}{p}}G(\a,\nabb\a,\ze,\nabb\log\oom)|_{p,S}(\la_1,\nu')d\nu'\right)\
.\nn
\eea
From the previous estimate on $\a$, $G=O(r^{-(\frac{9}{2}+\de)})$ and the integral in the right hand side can be estimated by
$c(\la_1)r^{(\frac{1}{2}-\de)}(\la_1,\nu)\varepsilon$, we conclude that
\bea
|r^{(\frac{7}{2}+\de)-\frac{2}{p}}\b|_{p,S}(\la_1,\nu)\leq c\left(|r^{(\frac{7}{2}+\de)-\frac{2}{p}}\b|_{p,S}(\la_1,\nu_0)+\varepsilon\right)\
.\eql{1.51az}
\eea
Finally to prove that $|r^{(\frac{7}{2}+\de)-\frac{2}{p}}\b|_{p,S}(\la_1,\nu_0)\leq c\varepsilon$,\footnote{Observe that the estimate for $\b$ can also be a pointwise estimate if we assume
enough regularity for the initial data.} uniformily in $r_0=r(\la_1,\nu_0)$ we 
need that \[\b|_{S_0}=(\nabb\tr\chi-\divv\chi-\zeta\c\chi+\zeta\tr\chi)|_{S_0}=O(\varepsilon r^{-(\frac{7}{2}+\de)})\ .\]
This follows, recalling the behaviour of $\chih$, from the assumption, see \ref{2.113r}: \footnote{This quantity plays the role of the $\Us$ form
defined in \cite{Kl-Ni:book}, equation (4.3.5).} 
\bea
\nabb\tr\chi|_{S_0}+\ze|_{S_0}\tr\chi|_{S_0}=O(\varepsilon r_0^{-(\frac{7}{2}+\de)}) \ .\eql{2.117q}
\eea
  To
determine the asymptotic decay of $\ro$ and $\ro-\overline{\ro}$ we proceed in the same way. Again the initial data on $S_0$ have to be such
that
\bea
{\bf K}-\overline{\bf K}+\frac{1}{4}\!\left(\tr\chi\tr\chib-\overline{\tr\chi\tr\chib}\right)=O(\varepsilon r_0^{-(\frac{7}{2}+\de)})\ .
\eea

 To prove that $\si$ has the correct decay we must require, after a simple application of the Gronwall's Lemma to its evolution equations, that
on $S_0$ the following relation holds:
\bea
\curll\zeta-\frac{1}{2}\chibh\wedge\chih=O(\varepsilon r_0^{-(\frac{7}{2}+\de)})\ .
\eea
Finally proceeding in the same way for $\bb$, using Gronwall's Lemma and the decay of the previous null components we obtain the expected result
provided on $S_0$ the following estimate holds:
\bea
\nabb\tr\chib|_{S_0}-\ze|_{S_0}\tr\chib|_{S_0}=O(\varepsilon r_0^{-(\frac{7}{2}+\de)})\ .
\eea
{\bf Remark:} To complete the proof we have to specify the exact regularity of the initial data required to prove Proposition \ref{P1.2}. To control $\b$ we
use the first transport equation in \ref{1.70cb}, which requires the control of $\divv\a$, to control $\ro$ and $\si$ we need to control $\nabb\b$ which at
its turn requires the control of $\nabb^2\a$ and finally to control $\bb$ we require the control of $\nabb\ro$ which finally implies the control of
$\nabb^3\a$. Recalling the explicit expression of $\a$, \ref{1.69c}, this implies that we need to control
$|r^{(\frac{13}{2}+\de)-\frac{2}{p}}\nabb^3\dddd_4\chih|_{p,S}$,
$|r^{(\frac{11}{2}+\de)-\frac{2}{p}}\nabb^3\chih|_{p,S}$ and $|r^{(4+\de)-\frac{2}{p}}\nabb^3\om|_{p,S}$. The needed regularity is also
higher if we require pontwise norms for all the Riemann components.
\footnote{In the global existence proof the boundedness of the ${\cal Q}$ norms defined on $\cal C$ 
requires the boundedness of the $L^2(C_0)$ and $L^2(\Cb_0)$ norms of the various Riemann components and their derivatives up to
second order.}

 This completes the existence proof for the initial data satisfying the characteristic constraints and the appropriate decay on
$C_0$. To achieve our result we have to implement the analogous program for $\Cb_0$. 

\subsection{The decay of the null Riemann components on the incoming null cone.}\label{SS2.5}
The discussion about the null Riemann components on $\Cb_0$ goes exactly as in the case of $C_0$ and we do not repeat it here. The various combinations of
connection coefficients which describe, once the global existence problem is solved, the restriction to $\Cb_0$ of the Riemann tensor are the
same as in \ref{1.69c}, with the expression for $\aa$, instead of the one for $\a$,
\bea
\aa=-[\dddd_3\chibh+\tr\chib\ \!\chibh-(\dd_3\log\oom)\chibh]\ .\eql{2.87aaa}
\eea
Again in $(R^4,\ggg)$ the Bianchi equations, restricted to $\Cb_0$, have exactly the same expression as in the vacuum Einstein manifold.
Therefore we prove the following propositions:
\begin{Prop}\label{P2.3}
Let $\{{\ga}_{ab},{\oom},{\underline X}^a,{\ze}_a,{\chib}_{ab}\}$ be the initial data on $\Cb_0$ satisfying the $|\c|_{p,S}$ norm bounds of Lemma
\ref{L2.4} and Lemma \ref{L2.5}, then $\b,\ro,\si,\bb$, combinations of the (derivatives of the) initial data defined in \ref{1.69c} satisfy on
$\Cb_0$ the following evolution equations, with $\eta=\ze+\nabb\log\oom$:
\bea
&&\dddd_3\bb+2\tr\chib\s\bb=-\divv\aa-\left[2\omb \bb+(-2\ze+ \eta)\c\aa\right]\nn\\
&&\dd_3\ro+\frac{3}{2}\tr \chib\ro=-\divv\bb-\left[\frac{1}{2}\hat{\chi}\c\aa
-\ze\c \bb+2\eta\c\bb\right]\nn\\
&&\dd_3\si+\frac{3}{2}\tr \chib\si=-\divv\dual\bb+\left[\frac{1}{2}\hat{\chi}\c\dual\aa
-\ze\c\dual\bb-2\eta\c\dual\bb\right]\eql{1.106c}\\
&&\dddd_3\b+\tr\chib\b=\nabb\ro+\left[2\omb\b+\dual\nabb\si+2\hat{\chi}\c\bb+
3(\eta\ro+\dual\eta\si)\right]\nn
\eea 
\end{Prop}
{\bf Remark:} 
The Bianchi equations do not provide an evolution equation along the incoming null cones for $\aa$. Therefore its decay along $\Cb_0$  has to
be obtained in a different way.
\begin{Prop}\label{P2.4}
Let $\{{\ga}_{ab},{\oomb},{\underline X}^a,{\chib}_{ab}\}$ be the initial data on $\Cb_0$ satisfying the $|\c|_{p,S}$ norm bounds of Lemma
\ref{L2.4} and \ref{L2.5} then, provided that on $S_0=S(\la_1,\nu_0)$ the following conditions are satisfied:  
{\em\bea
&&\nabb\tr\chi+\ze\tr\chi=O(\varepsilon r_0^{-(\frac{7}{2}+\de)}) \nn\\
&&{\bf K}-\overline{\bf K}+\frac{1}{4}\!\left(\tr\chi\tr\chib-\overline{\tr\chi\ \!\tr\chib}\right)=O(\varepsilon r_0^{-(\frac{7}{2}+\de)})\nn\\
&&\curll\zeta-\frac{1}{2}\chibh\wedge\chih=O(\varepsilon r_0^{-(\frac{7}{2}+\de)})\eql{2.113rs}\\
&&\nabb\tr\chib-\ze\tr\chib=O(\varepsilon r_0^{-(\frac{7}{2}+\de)})\ ,\nn
\eea}
 the combinations $\b,\ro,\si,\bb,\aa$ of the (derivatives of the) initial data defined in \ref{1.69c} satisfy the following bounds with
$\de>\de'>\de''>0$ and $p\in [2,4]$:
\bea
&&|r|\la|^{\frac{5}{2}+\de}\aa|_{C_0}\leq c\varepsilon\ ,\ |r^{2-\frac{2}{p}}|\la|^{\frac{3}{2}+\de'}\bb|_{p,S}\leq c\varepsilon
\ ,\ |r^{3-\frac{2}{p}}\ro|_{p,S}\leq c\varepsilon\nn\\
&&|r^{3-\frac{2}{p}}|\la|^{\frac{1}{2}+\de''}(\ro-{\overline\ro},\si)|_{p,S}\leq c\varepsilon\ ,\ 
|r^{3-\frac{2}{p}}|\la|^{\frac{1}{2}+\de''}\b|_{p,S}\leq c\varepsilon 
\eea
\end{Prop}

{\bf Proof:} We start looking at $\aa$. From its definition in \ref{2.87aaa}
and the decay of $\chibh$, equation \ref{1.48d}, the estimate for $\aa$ follows
immediately. The second quantity to control is $\bb$, defined through the expression
\beaa
\bb=\nabb\tr\chib-\divv\chib+\zeta\c\chib-\zeta\tr\chib\ .
\eeaa
To determine the bound of $\bb$ on $\Cb_0$ we use its evolution equation:
\beaa
\dddd_3\bb+2\tr\chib\ {\!\bb}=-2\omb\bb-\left[\divv\aa-(-\ze+\nabb\log\oomb)\aa\right]\ .\eql{1.70ca}
\eeaa
Proceeding as we did for $\b$ along $C_0$ we obtain, using the Gronwall's lemma, the following estimate, with
$\Gb(\aa,\nabb\aa,\ze,\nabb\log\oomb)=-\left[\divv\aa+(\ze-\nabb\log\oomb)\aa\right]$,
\bea
|r^{4-\frac{2}{p}}\bb|_{p,S}(\la,\nu_0)\leq c\!\left(\!|r^{4-\frac{2}{p}}\bb|_{p,S}(\la_1,\nu_0)
\!+\!\!\int_{\la_1}^{\la}\!\!|r^{4-\frac{2}{p}}\Gb(\aa,\nabb\aa,\ze,\nabb\log\oomb)|_{p,S}(\la',\nu_0)d\la'\!\right)\ .\eql{1.108c}
\eea
Consider now the function \footnote{$u(p)$ is not exactly equal to $\nu_0-2r$, but the argument holds also in that case.}
\[\frac{|\la|^{(\frac{3}{2}+\de)}}{r^2}=\frac{|\nu_0\!-\!2r|^{(\frac{3}{2}+\de)}}{r^2}=\frac{(2r\!-\!\nu_0)^{(\frac{3}{2}+\de)}}{r^2}\ ,\]
with ${\nu_0}/{2}\leq r\leq\nu_0$. This function is increasing in $r$ in the interval $[0,\nu_0]$. In fact
\bea
&&\frac{d}{dr}\frac{(2r\!-\!\nu_0)^{(\frac{3}{2}+\de)}}{r^2}=2({{3}/{2}}+\de)\frac{(2r\!-\!\nu_0)^{(\frac{1}{2}+\de)}}{r^2}
-2\frac{(2r\!-\!\nu_0)^{(\frac{3}{2}+\de)}}{r^3}\eql{1.109c}\\
&&=2\frac{(2r\!-\!\nu_0)^{(\frac{1}{2}+\de)}}{r^2}\left[({{3}/{2}}+\de)-\frac{(2r\!-\!\nu_0)}{r}\right]=
2\frac{|\la|^{(\frac{1}{2}+\de)}}{r^2}\left[({{3}/{2}}+\de)-\frac{|\la|}{r}\right]>0\nn
\eea
as $|\la|\leq r$. Therefore
\bea
\frac{|\la|^{(\frac{3}{2}+\de)}}{r^2(\la,\nu_0)}\leq \frac{|\la_1|^{(\frac{3}{2}+\de)}}{r^2(\la_1,\nu_0)}\ .
\eea
Multiplying both sides by $\frac{|\la|^{(\frac{3}{2}+\de')}}{r^2(\la,\nu_0)}$, we rewrite \ref{1.108c} as
\bea
&&|r^{2-\frac{2}{p}}|\la|^{(\frac{3}{2}+\de')}\bb|_{p,S}(\la,\nu_0)\leq\eql{1.49a}\\
&& c\left(|r^{2-\frac{2}{p}}|\la_1|^{(\frac{3}{2}+\de')}\bb|_{p,S}(\la_1,\nu_0)\!+\!
\int_{\la_1}^{\la}\!|{|\la'|^{(\frac{3}{2}+\de')}}r^{2-\frac{2}{p}}G(\aa,\nabb\aa,\ze,\nabb\log\oomb)|_{p,S}(\la',\nu_0)d\la'\!\right)\nn
\eea
and as, from the previous estimates on $\aa$, the integrand in the right hand side can be estimated by
\[c(\la')\frac{|\la'|^{(\frac{3}{2}+\de')}}{r^2(\la',\nu_0)}r^{2}(\la',\nu_0)|\la'|^{-(\frac{5}{2}+\de)}\varepsilon
\leq c(\la')\frac{1}{|\la'|^{1+(\de-\de')}}\varepsilon\ ,\] we conclude again that
\bea
|r^{2-\frac{2}{p}}|\la|^{(\frac{3}{2}+\de')}\bb|_{p,S}(\la,\nu)\leq
c\left(|r^{2-\frac{2}{p}}|\la|^{(\frac{3}{2}+\de')}\bb|_{p,S}(\la_1,\nu_0)+\varepsilon\right)\ .\eql{1.51a}
\eea
Finally to prove that $|r^{2-\frac{2}{p}}|\la|^{(\frac{3}{2}+\de')}\bb|_{p,S}(\la_1,\nu_0)\leq c\varepsilon$, uniformily in $r_0=r(\la_1,\nu_0)$
we require that
\[\bb|_{S_0}=(\nabb\tr\chib-\divv\chib+\zeta\c\chib-\zeta\tr\chib)|_{S_0}=O(r_0^{-2}|\la_1|^{-(\frac{3}{2}+\de')})=O(r_0^{-(\frac{7}{2}+\de')})\
.\]
This estimate, recalling the assumption on $\chibh$, is satisfied from the assumptions of Proposition \ref{P2.4}:
\beaa
\nabb\tr\chib|_{S_0}-\ze|_{S_0}\tr\chib|_{S_0}=O(\varepsilon r_0^{-(\frac{7}{2}+\de)})\ .
\eeaa
 Observe that the estimate for $\bb$ can also be a pointwise estimate if we assume enough regularity for the initial data.

 To prove that, given the initial data, $\ro$, $\si$ and $\b$ satisfy the  right bounds along $\Cb_0$ is a simple application of the Gronwall's
Lemma to their evolution equations, \ref{1.106c}, which can be estimated starting from $S_0$ plus the assumptions of Proposition \ref{P2.4}.

{\bf Remark:} It is important to observe that to obtain the expected bound for $\b$ we have to use, togheter with the evolution equation for $\ro$,
\ref{1.106c}, an improved estimate for $\nabb\ro$, namely $|r^{4-\frac{2}{p}}|\la|^{\frac{1}{2}+\de''}\nabb\ro|_{p,S}\leq c\varepsilon$. The same happens
when we estimate $\ro$, $\si$ and $\bb$.

Proposition \ref{P1.2} and Proposition \ref{P2.4} complete the construction of initial data satisfying the appropriate decay, provided we show that
conditions \ref{2.113rs} can be satisfied on $S_0$. This is possible as the one form $\ze$ can be assigned freely on $S_0$, as previously discussed.
Therefore we can proceed in the following way: Given $\chih$ and $\chibh$ on $S_0$ we solve the equation
\bea
\curll\zeta-\frac{1}{2}\chibh\wedge\chih=0\ .\eql{2.132q}
\eea
Then with this $\ze$ we solve the equations 
\bea
&&\nabb\tr\chi|_{S_0}+\ze|_{S_0}\tr\chi|_{S_0}=0\nn\\
&&\nabb\tr\chib|_{S_0}-\ze|_{S_0}\tr\chib|_{S_0}=0\ .\eql{2.133q}
\eea
Finally we verify that with these solutions on $S_0$ the estimate
\bea
{\bf K}|_{S_0}-\overline{\bf
K}|_{S_0}+\frac{1}{4}\!\left(\tr\chi|_{S_0}\tr\chib|_{S_0}-\overline{\tr\chi|_{S_0}{\tr\chib}|_{S_0}}\right)=O(\varepsilon
r_0^{-(\frac{7}{2}+\de)})\eql{2.134q}
\eea
can be satisfied.

 More precisely we recall that in the initial data construction we were free to assigne on $S_0$, $\tr\chi,\tr\chib,\ze$ and
$\tilde{\ga}_{ab}$. We can therefore assigne them from the beginning requiring they satisfy equations \ref{2.132q}, \ref{2.133q} and
the estimate \ref{2.134q}. Then we proceed in the construction of the initial data with the appropriate decay and, finally, we prove that, due to
the asymptotic behaviour of the restriction of the null Riemann components, the $\cal Q$ norms on $C_0$ and $\Cb_0$ are finite and bounded
by $c\varepsilon$.
\section{The global existence theorem, strategy of the proof.}\label{S.3}

In this section we state the global existence theorem we want to prove and describe the main steps of the proof. The following
sections are devoted to their implementation.
\subsection[The regularity of the initial data]{The regularity of the initial data}\label{SS4.1}
To state and prove an existence theorem we have to specify the function spaces of the initial data, in other words their regularity. Let us focus first
on the initial data on $C_0$, an analogous discussion is done later for the initial data on $\Cb_0$. 
\subsubsection{The regularity of the initial data on $C_0$}
In Section \ref{S2} we have shown how to obtain the initial data on $C_0$ satisfying the costraints and some appropriate decay rates.
We want to investigate here their regularity with respect to the tangential derivatives, $\nabb$. It is clear that the amount of regularity
we need will be fixed by the construction of the solution to the characteristic problem. Here we want to show how, due to the constraints, the
regularities of different terms of the initial data are related. To see it in more detail let us go through the various steps done in Section
\ref{S2} to obtain the initial data. Let us first introduce the following definition:
\begin{Def}let $f$ be a covariant tensor defined on $C_0$ and at each point tangent to the leaves
$S_0(\nu)$ of the $\oom$-foliation. We say $f\in C^q(S)$ if, for any component of $f$, $f_{\a_1,...,\a_n}(\nu,\om)\in
C^q(S^2)$ for any $\nu\in [\nu_0,\infty)$.\footnote{Assuming the appropriate derivability in $\nu$.}
\end{Def}

 Given this definition, following the strategy of Section \ref{S2} we prescribe the initial data starting from $\chih$ and $\oom$. Let us assume,
therefore, that $\chih,\ \oom,\ \dddd_4\chih,\ \dddd_4\oom\in  C^{q}(S)$, then from Lemma \ref{L2.1} it follows that $\tr\chi$ and $\ga$ also
belong to
$C^q(S)$. The connection coefficient
$\ze$, obtained solving equation {1.12} in Lemma \ref{L1.3} is, therefore, in $C^{q-1}(S)$ as $\nabb\log\oom$. Moreover from its definition $\om\in
C^q(S)$ and $X$, due to its relation with $\ze$, see equation \ref{1.69ex}, belong also to $C^{q-1}(S)$.

 On $C_0$ the initial data are completely assigned once we prescribe also the connection coefficients $\chib$ and $\omb$. To do it we have to solve
equations \ref{1.70d}, \ref{1.71d} and \ref{1.86c}. From their inspection one sees immediately that $\chib\in C^{q-2}(S)$. The proof that
$\omb\in C^{q-2}(S)$ has, viceversa, to be postponed after the determination of the regularity of $\ro(\chi,\chib,\eta,\etab)$. Summarizing we
have
\bea
&&\oom,\ \ga,\ \chi,\ \om,\ \dddd_4\chih\in C^q(S)\nn\\
&&X,\ \ze,\ \nabb\log\oom\in C^{q-1}(S)\eql{4.1qw}\\
&& \chib,\ \omb \in C^{q-2}(S)\ \ .\nn
\eea 
Once the regularity of the metric components and the connection coefficients is stated we can also see which regularity this implies for the
various components of the Riemann tensor. From equations \ref{1.69c} we have immediately
\bea
\a\in C^{q}(S)\ ,\ \b\in C^{q-1}(S)\ ,\ (\ro,\si)\in C^{q-2}(S)\ ,\ \bb\in C^{q-3}(S)\ ,\eql{4.2qw}
\eea
and from the regularity of $\ro$ that one of $\omb$ follows, using equation \ref{1.86c}.
\subsubsection{The regularity of the initial data on $\protect\underline{C}_0$}
We do not repeat for $\Cb_0$ the previous argument and we only report the result obtained exactly in the same way:
\bea
&&\oom,\ \ga,\ \chib,\ \om,\ \dddd_3\chibh\in C^q(S)\ ,\eql{4.1qwz}\\
&&{\underline X},\ \ze,\ \nabb\log\oom\in C^{q-1}(S)\nn\\
&&\chi,\om \in C^{q-2}(S)\nn
\eea
and 
\bea
\aa\in C^{q}(S)\ ,\ \bb\in C^{q-1}(S)\ ,\ (\ro,\si)\in C^{q-2}(S)\ ,\ \b\in C^{q-3}(S)\ .\eql{4.2qwz}
\eea

\subsubsection{The ``loss of derivatives" of the initial data}
It is well known, see for instance H.Muller Zum Hagen, \cite{Muller}, that to solve the characteristic problem we need to control on the initial null
hypersurfaces the initial data and their derivatives along the ``normal" direction. On $C_0$, for instance, we need an estimate of the derivatives along
the $e_3$ direction. On the other side, due to the costraints associated to the characteristic problem, these data cannot be given freely and,
therefore, are related to the data on the initial hypersurface and their tangential derivatives. This has the effect that to control the derivatives
along the ``normal" direction up to order, say, $k$, the order of tangential derivatives of the initial data which have to be controlled is
greater than $k$. This is what we call the ``loss of derivatives" of the initial data. The aim of this section is to see how this fact appears
in our formalism. In this way  we can specify in the subsequent paper, where we state and prove in a precise way the existence result, which is
the regularity required to the initial data. In
\cite{Muller}, see Lemma 4.3 and Lemma 4.4, it is shown that the control of the first derivative normal to a null hypersurface
requires the control of second tangential derivatives and in general that of a normal derivative of order $k$ requires $2k$ tangential derivatives. To
recognize the same phenomenon in our case we proceed in the following way.

 First of all in our formalism we do not consider the partial derivatives of the metric components, but the connection coefficients  associated to
the $\oom$ foliation. Let us consider for the moment the $C_0$ part of the initial data null hypersurface, to select the connection coefficients
associated to the ``normal" null direction $e_3$ relative to $C_0$ \footnote{We use the `` " for the adjective normal as, in fact, $\ggg(e_3,r_4)=-2$.}
we use the notion of signature introduced by D. Christodoulou and S. Klainerman in \cite{C-K:book}, see also \cite{Kl-Ni:book}, paragraph 3.1.24.
Extending this notion also to the connection coefficients it follows that the signature of the various connection coefficients is
\bea
\mbox{sig}(\chi)=+1\ ,\ \mbox{sig}(\om)=+1\ ,\ \mbox{sig}(\ze)=0\ ,\ \mbox{sig}(\omb)=-1\ ,\ \mbox{sig}(\chib)=-1\ \ \ \ 
\eea
moreover each derivative along $e_3$ decreases the signature by $1$ and any derivative along $e_4$ increases it by $1$.Any decrease of $-1$ in
the corresponds, in some sense, to a derivative along the $e_3$ direction.

Let us recall that if $\ga\in C^q(S)$ then $\chib \in C^{q-2}(S)$,
consistent with the loss of derivatives we discussed. To substantiate this result for higher order derivatives, let us consider the derivative
along $e_3$ of $\chib$, $\dd_3\chib$. To control $\dd_3\chib$ on $C_0$ we have to look at the evolution equation of it along $C_0$ which can be
obtained applying $\dd_3$ to the evolution equations
\ref{1.70d} and \ref{1.71d}. Let us consider, to have a concrete example, the second equation,
\beaa
\dddd_4\chibh+\frac{1}{2}\tr\chi\chibh+\frac{1}{2}\tr\chib\chih-2\om\chibh
-\nabb\hot\etab-\etab\hot\etab=0\ .
\eeaa
Applying $\dd_3$ to this equation, neglecting the commutator between $\dd_3$ and $\dd_4$, we obtain an evolution equation along $C_0$, formally of this
type
\bea
\dd_4(\dd_3\chibh)=\c\c\c+\nabb(\dd_3\ze)+\c\c\c\ ,
\eea
where dots denote the less relevant terms, and using for $\dd_3\ze$ the transport equation \ref{1.28w},\ 
$\dddd_3\zeta+2\chib\c\zeta-\ddb_3\nabb\log\oom=-\bb$,
we can write
\bea
\dd_4(\dd_3\chibh)=\c\c\c-\nabb\bb+\c\c\c
\eea
which implies, recalling \ref{4.2qw}, that $\dd_3\chib\in C^{q-4}(S)$, again one derivative $\dd_3$ implying the loss of two tangential derivatives.
Iterating the argument, with the same simplifications, we have immediately, using the Bianchi equations,
\bea
\dd_4(\dd_3^2\chibh)=\c\c\c-\nabb\dd_3\bb+\c\c\c=\c\c\c-\nabb^2\aa+\c\c\c\eql{4.8qww}
\eea
where
\bea
\aa=-[\dddd_3\chibh+\tr\chib\ \!\chibh-(\dd_3\log\oom)\chibh]\ ,\eql{3.9uu}
\eea
and $\nabb^2\aa\in C^{q-6}(S)$ as expected. Finally for an arbitrary order $k$ of $\dd_3$ derivatives we substitute in \ref{4.8qww} the
explicit expression \ref{3.9uu}, obtaining
\bea
\dd_3^2\chibh\ \  ``="\ \ \nabb^2\dddd_3\chibh\eql{4.9qww}
\eea
implying that each extra ``$\dd_3$ regularity" requires the regularity of two more tangential derivatives. 
These considerations will imply the well known phenomenon that the regularity of the solution will turn out to be lower than the one of (some
of) the initial data. In fact in the present case $\chibh$ will be inside the spacetime $C^{q-4}(S)$ while on $C_0$ it is $C^{q-2}(S)$.

 An analogous argument can be done for $\tr\chib$ and $\omb$ and we do not report it here.
\subsection{The integral $\cal Q$ norms on the initial hypersurface}

In Chapter 3 of \cite{Kl-Ni:book}, section 3.5.1 we have introduced a family of integral norms, denoted ``$\cal Q$-integral norms", see also
\cite{C-K:book}, $L^2$ integrals made along the incoming and outgoing cones (or portions of them). The control of these norms is a crucial
step in the global existence proof as from them it is possible to control the family of norms relative to the Riemann tensor null components.
This allows to start the bootstrap mechanism discussed in detail in \cite{Kl-Ni:book} which we do not repeat here. To control these norms means
to prove that they can be bounded by the same norms on the initial hypersurface. As the $\cal Q$ norms are $L^2$ weighted integral norms of the
various Riemann null components expressed on the initial hypersurface in terms of the initial data, \ref{1.69c}, the regularity
and the decay of the initial data have to be prescribed to satisfy the existence of these norms. From the inspection of the explicit
expression of the $\cal Q$ norms, one easily sees that the assumptions on the decay of the initial data along the ``cones", are such
that the $\cal Q$ norms are finite, provided that also the derivatives of the initial data are sufficienty regular and decay in a consistent way.
More precisely we have the following expressions for the
$\cal Q$ norms:
\bea
&&\QQ(\la,\nu)=\QQ_1(\la,\nu)+\QQ_2(\la,\nu)\nn\\
&&\QQb(\la,\nu)=\QQb_1(\la,\nu)+\QQb_2(\la,\nu)
\eea
where,\index{$\QQ_1(\la,\nu)$ integral norm}\index{$\QQ_2(\la,\nu)$ integral norm}\index{$\QQb_1(\la,\nu)$ integral norm}
\index{$\QQb_2(\la,\nu)$ integral norm} 
\bea
\QQ_1(\la,\nu)&\equiv&\int_{C(\la)\cap V(\la,\nu)}Q(\lie_{T}\rr)(\acc,\acc,\acc,e_4)\nn\\
&&+\int_{C(\la)\cap V(\la,\nu)}Q(\lie_{O}\rr)(\acc,\acc,T,e_4)\nn\\
\QQ_2(\la,\nu)&\equiv&
\int_{C(\la)\cap V(\la,\nu)}Q(\lie_{O}\lie_{T}\rr)(\acc,\acc,\acc,e_4)\nn\\
&&+\int_{C(\la)\cap V(\la,\nu)}Q(\lie^2_{O}\rr)(\acc,\acc,T,e_4)\\
&&+\int_{C(\la)\cap V(\la,\nu)}Q(\lie_{S}\lie_{T}\rr)(\acc,\acc,\acc,e_4)\nn
\eql{QQ12}
\eea 
\bea
\QQb_1(\la,\nu)&\equiv&\sup_{V(\la,\nu)\cap{\cal C}}|r^3{\overline\ro}|^2+
\int_{\Cb(\nu)\cap V(\la,\nu)}Q(\lie_{T}\rr)(\acc,\acc,\acc,e_3)\nn\\
&&+\int_{\Cb(\nu)\cap V(\la,\nu)}Q(\lie_{O}\rr)(\acc,\acc,T,e_3)\nn\\
\QQb_2(\la,\nu)&\equiv&\int_{\Cb(\nu)\cap V(\la,\nu)}Q(\lie_{O}\lie_{T}\rr)(\acc,\acc,\acc,e_3)\nn\\
&&+\int_{\Cb(\nu)\cap V(\la,\nu)}Q(\lie^2_{O}\rr)(\acc,\acc,T,e_3)\\
&&+\int_{\Cb(\nu)\cap V(\la,\nu)}Q(\lie_{S}\lie_{T}\rr)(\acc,\acc,\acc,e_3)\ ,\nn
\eql{QQb12}
\eea
with
\bea 
Q(\rr)(K_0,K_0,T,e_4)\!&=&\!\frac{1}{4}\ub^4|\a|^2
+\frac{1}{2}(\ub^4+2\ub^2u^2)|\b|^2+\frac{1}{2}(u^4+2\ub^2u^2)(\ro^2+\si^2)\nn\\
&&+\frac{1}{2}u^4|\bb|^2\nn\\
Q(\rr)(K_0,K_0,T,e_3)\!&=&\!\frac{1}{4}u^4|\aa|^2
+\frac{1}{2}(u^4+2\ub^2u^2)|\bb|^2+\frac{1}{2}(\ub^4+2\ub^2u^2)(\ro^2+\si^2)\nn\\
&&+\frac{1}{2}\ub^4|\b|^2\eql{2.4.38}\\
Q(\rr)(K_0,K_0,K_0,e_4)\!&=&\!\frac{1}{4}\ub^6|\a|^2+\frac{3}{2}\ub^4u^2|\b|^2+
\frac{3}{2}u^4\ub^2(\ro^2+\si^2)+\frac{1}{2}u^6|\bb|^2\nn\\
Q(\rr)(K_0,K_0,K_0,e_3)\!&=&\!\frac{1}{4}u^6|\aa|^2+\frac{3}{2}u^4\ub^2|\bb|^2+
\frac{3}{2}\ub^4u^2(\ro^2+\si^2)+\frac{1}{2}\ub^6|\b|^2\ \ \ \ \ \ \ \ \ \ \ \ 
\eea
and $V(\la,\nu)$ is the part of $J^{-}(S(\la,\nu))$ above $\cal C$. $\lie_{O},\lie_{T}, \lie{S}$ are some ``modified" Lie derivatives,
$T,O,S,K$ are vector fields, Killing or conformal Killing in the Minkowski spacetime, ``near" Killing or conformal Killing in the general case,
see for a detailed discussion on these definitions \cite{Kl-Ni:book} Chapter 3.  From these expressions we see that the norms depend on the
second Lie derivatives of the null Riemann components
$\a,\b,\ro,\si,\bb,\aa$. Expressing the Lie derivatives in terms of the covariant derivatives we conclude that we have to control the null
components of the Riemann tensor field and their
$\nabb,\dddd_3,\dddd_4$ derivatives up to second order. On $C_0$ the
$\dddd_3$ derivatives of the null components can be expressed in terms of the tangential $\nabb$ derivatives\footnote{Except for $\aa$ which does not
appear in the $\cal Q$ norms along the outgoing cones.} using the Bianchi equations. This, as discussed in the previous subsection,
implies the so called ``loss of derivatives". This does not happen, on $C_0$, for the $\dddd_4$ derivatives, while on $\Cb_0$ the role of
$\dddd_3$ and $\dddd_4$ are interchanged.

 The detailed examination of the $\cal Q$ norms tells us the amount of regularity in the tangential derivatives we have to require.
We do not discuss it in detail here as it will be done in the second part of this work. We only show how the argument goes in the case of the
null component $\a$ for the $\cal Q$ norms along $C_0$.

 Starting with the explicit expression of the $\cal Q$ norms it will follow that in the existence proof we have to control a weighted
$L^2(C_0)$ norm of $\dddd_3^2\a$. From the Bianchi equations, \ref{1.70cb}, \ref{1.106c} we have that $\dddd_3\a\ ``\!=\!"\nabb\b$ and
$\dddd_3^2\a\ ``\!=\!"\nabb^2\ro$. On the other side from \ref{4.2qw} it follows that if $\a\in C^{q}$ then $\dddd_3^2\a\  ``\!=\!"\nabb^2\ro\in
C^{q-4}$ showing the loss of derivatives previously discussed. From the requirement that $\dddd_3^2\a\in L^2(C_0)$ it follows, therefore, that
$\nabb^2\ro\in L^2(C_0)$. This condition implies, at its turn, that $\ro\in C^0(S)$ which requires on $C_0$, $q=2$ for the initial data defined in
\ref{4.1qw} .

\subsection{Global initial data smallness condition}\label{SS4.2}
We say that the initial data on ${\cal C}=C_0\cup\Cb_0$ are small if the following quantity, $J^{(q)}_{C_0\cup\Cb_0}$, is small:
\footnote{Observe that in Definition \ref{D1.1} of the initial data the quantities $X$ or $\underline X$ are not present. In fact although we can obtain
it on $C_0$ and prescribe on $\Cb_0$ they are not needed to build the initial data $\cal Q$ norms. Nevertheless in the local part of the existence
proof we need to write our initial data in terms of the harmonic coordinates. This requires to connect our
$\{u,\ub,\om\}$ coordinates with the harmonic coordinates and this implies the knowledge of the vector fields $X$ or $\underline X$. Motivated by this
argument we also add $\underline X$ as initial data and specify its appropriate norms.}
\bea
J^{(q)}_{C_0\cup\Cb_0}=J^{(q)}_{C_0}\left[\overline{\ga}_{ab},\overline{\oom},\overline{\ze}_a,\overline{\chib},\overline{\omb}\right]
+J^{(q)}_{\Cb_0}\left[\overline{\ga}_{ab},\overline{\oomb},\overline{\underline X}^a,\overline{\ze}_a,\overline{\chi},\overline{\om}\right]\leq
\varepsilon\ ,\eql{smallness}
\eea
where, with $p\in[2,4]$ and $\de>0$, 
\bea
&&J^{(q)}_{C_0}\left[\overline{\ga}_{ab},\overline{\oom},\overline{\ze}_a,\overline{\chib},\overline{\omb}\right]=
\sup_{C_0}\left(r\big|\overline{\oom}-\frac{1}{2}\big|+\frac{r^2}{\log r}|\tr\overline{\chi}-\frac{2}{r}|
+r^{(\frac{5}{2}+\de)}|\hat{\overline{\chi}}|\right)+\nn\\
&&\sup_{C_0}\left[\left(\sum_{l=1}^q|r^{(2+l-\frac{2}{p})}\nabb^l\tr\chi|_{p,S}
+\sum_{l=1}^q|r^{(\frac{5}{2}+l+\de-\frac{2}{p})}\nabb^l\hat{\overline{\chi}}|_{p,S}
+\sum_{l=0}^q|r^{(\frac{7}{2}+l+\de-\frac{2}{p})}\nabb^l\dddd_4\hat{\overline{\chi}}|_{p,S}\right.\right.\nn\\
&&\left.\left.|r^{(2-\frac{2}{p})}\overline{\om}|_{p,S}+\sum_{l=1}^{q}|r^{(2+l+\de-\frac{2}{p})}\nabb^{l}\overline{\om}|_{p,S}\right)
+\sum_{l=1}^{q-1}|r^{(1+l+\de-\frac{2}{p})}\nabb^l\log\overline{\oom}|_{p,S}\right.\nn\\
&&\left.+\sum_{l=0}^{q-1}|r^{(2+l-\frac{2}{p})}\nabb^l\overline{\ze}|_{p,S}+\sum_{l=0}^{q-2}|r^{(2+l-\frac{2}{p})}\nabb^{l}\overline{\omb}|_{p,S}
+\sum_{l=0}^{q-2}|r^{(1+l-\frac{2}{p})}\nabb^{l}\overline{\chib}|_{p,S}\right]\eql{Initialsmallcond1}\
.\nn
\eea 
\bea
&&J^{(q)}_{\Cb_0}\left[\overline{\ga}_{ab},\overline{\oomb},\overline{\underline X}^a,\overline{\ze}_a,\overline{\chi},\overline{\om}\right]=
\sup_{\Cb_0}\left(r\big|\overline{\oomb}-\frac{1}{2}\big|+\frac{r^2}{\log r}|\tr\overline{\chib}+\frac{2}{r}|
+r|\la|^{(\frac{3}{2}+\de)}|\hat{\overline{\chib}}|\right)+\eql{Initialsmallcond2}\nn\\
&&\sup_{\Cb_0}\left[\left(\sum_{l=1}^q|r^{(2+l-\frac{2}{p})}\nabb^l\tr\chib|_{p,S}
+\sum_{l=1}^q||\la|^{(\frac{3}{2}+\de)}r^{(1+l-\frac{2}{p})}\nabb^l\hat{\overline{\chib}}|_{p,S}
+\sum_{l=0}^q||\la|^{(\frac{5}{2}+\de)}r^{(1+l-\frac{2}{p})}\nabb^l\dddd_3\hat{\overline{\chib}}|_{p,S}\right.\right.\nn\\
&&\left.\left.+|r^{(2-\frac{2}{p})}\overline{\omb}|_{p,S}+\sum_{l=1}^{q}|r^{(2+l+\de-\frac{2}{p})}\nabb^{l}\overline{\omb}|_{p,S}\right)
+\sum_{l=1}^{q-1}|r^{(1+l+\de-\frac{2}{p})}\nabb^l\log\overline{\oomb}|_{p,S}
\right.\nn\\
&&\left.+\sum_{l=0}^{q-1}|r^{(2+l-\frac{2}{p})}\nabb^l\overline{\ze}|_{p,S}
+\sum_{l=0}^{q-2}||\la|^{\de}r^{((2-\de)+l-\frac{2}{p})}\nabb^{l}\overline{\om}|_{p,S}
+\sum_{l=0}^{q-2}|r^{(1+l-\frac{2}{p})}\nabb^{l}\tr\overline{\chi}|_{p,S}\right.\nn\\
&&\left.+\sum_{l=0}^{q-2}|r^{(2+l-\frac{2}{p})}\nabb^{l}\hat{\overline{\chi}}|_{p,S}
+\sum_{l=0}^{q-1}|r^{(1+l-\frac{2}{p})}\nabb^l\overline{\underline
X}|_{p,S}\right]\ .\nn
\eea

 Observe that, due to the costraint equations for the initial data, to satisfy the smallness condition
\ref{smallness} it will be enough to impose that the norms of the quantities which are assigned freely on $C_0$ and
$\Cb_0$ be small, together with the smallnesss of some norms relative to $S_0=\underline{S}_0=C_0\cap\Cb_0$. These restricted
conditions plus the transport equations \ref{1.41wqz}, \ref{1.41wpz} along $C_0$ and $\Cb_0$, respectively, will imply the smallness of
\ref{smallness}. The explicit form of the ``restricted condition" will be given in the forthcoming paper when we discuss the local
existence solution.

\subsection{Statement of the characteristic global existence theorem.}\label{SS.4.3}

\begin{theorem}[characteristic global existence theorem]\label{T2.1}
Let the initial data 
\bea
\left\{C_0\ ;\overline{\ga}_{ab},\overline{\oom},\overline{\ze}_a,\overline{\chib}_{ab},\overline{\omb}\right\}\cup\left\{\Cb_0\
;\overline{\ga}_{ab},\overline{\oomb},\overline{\underline X}^a\overline{\ze}_a,\overline{\chi}_{ab},\overline{\om}\right\}\eql{2.54}
\eea 
be assigned together with their partial tangential derivatives to a fixed order specified by the integer $q\geq 7$.\footnote{The minimum value of $q$
will be discussed in a forthcoming paper. The techniques in \cite{Kl-Ni:book} to obtain a ``global" solution would require $q\geq s=4$, but the local
existence requires a stronger condition, namely $q\geq 2s-1=7$, see for instance \cite{MullerSeifert}.}  Let us assume they satisfy the smallness
conditions:
\bea
J^{(q)}_{C_0\cup\Cb_0}=J^{(q)}_{C_0}\left[\overline{\ga}_{ab},\overline{\oom},\overline{\ze}^a,\overline{\chib},\overline{\omb}\right]
+J^{(q)}_{\Cb_0}\left[\overline{\ga}_{ab},\overline{\oomb},\overline{\ze}_a,\overline{\chi},\overline{\om}\right]\leq
\varepsilon
\eea
Then there exists and it is unique a vacuum Einstein spacetime $\{\M,\ggg\}$ solving the characteristic initial value
problem with initial data \ref{2.54}. $\M$ is the maximal future development of
$C(\la_1)\cup\Cb(\nu_0)$, \footnote{Sometimes to avoid any confusion we write explicitely
$C(\la_1;[\nu_0,\nu_1])\cup\Cb(\nu_0;[\la_1,\la_0])$.}
\bea
\M=\lim_{\nu_1\rightarrow\infty}J^{+}(S(\la_1,\nu_0))\cap J^{-}(S(\la_0,\nu_1))\ .
\eea
Moreover $\M$ is endowed with the following structures:

 a) $\M$ is foliated by a ``double null canonical foliation" $\{C(\la)\}$, $\{\Cb(\nu)\}$, with $\la\in [\la_1,\la_0]\
,\nu\in[\nu_0,\nu_1]$.\footnote{To avoid any misunderstanding ``double canonical foliation" refers to a foliation of the spacetime
$(\M,\ggg)$, while with canonical foliation we denote a specific foliation of the initial data on $C_0$. Of course the first is related to the
second.} Double canonical foliation means that the null hypersurfaces $C(\la)$ are the level hypersurfaces of a function
$u(p)$ solution of the eikonal equation
\[g^{\mu\nu}\partial_{\mu}w\partial_{\nu}w=0\ ,\]
with initial data  a function $u_*(p)$ defining the foliation of the ``final" incoming cone
$\Cb(\nu_1)$,\footnote{A detailed discussion of the function $u_*(p)$ when $\nu_1\rightarrow\infty$ is in \cite{Kl-Ni:book}, Chapter 8.} while
the the null hypersurfaces
$\Cb(\nu)$ are the level hypersurfaces of a function
$\ub(p)$ solution of the eikonal equation with initial data  a function $\ub_{(0)}(p)$ defining a canonical foliation of the initial outgoing cone
$C(\la_1)$.\footnote{It is important to remark that the ``canonical" foliation on the portion $C(\la_1)$ of the initial hypersurface is not the
one given when the initial data are specified. Nevertheless it can be proved, see the discussion during the global existence proof, that given
the initial data foliation of $C(\la_1)$, it is possible to build on $C(\la_1)$ a ``canonical" foliation. Its precise definition and the way for
doing it will be made clear in the course of the proof. See also \cite{Kl-Ni:book}, Chapter 3.} The family of two dimensional surfaces
$\{S(\la,\nu)\}$, where
$S(\la,\nu)\equiv C(\la)\cap\Cb(\nu)$, defines a two dimensional foliation of $\M$. 
\smallskip

 b) $i(C_0)=C(\la_1)\ ,\ i(\Cb_0)=\Cb(\nu_0)$, $i(S_0(\nu_0))=i(\underline{S}_0(\la_1))=C(\la_1)\cap\Cb(\nu_0)$\ .

 c) On $C(\la_1)$ with respect to the initial data foliation,\footnote{Observe that while $\cal M$ is globally foliated by a double null canonical
foliation, it is not possible to foliate it with a double foliation solution of the eikonal equation with as initial data the $\oom$-foliation of $C_0$
and $\Cb_0$. This is, nevertheless possible in a small neighbourhood of $C_0$ and $\Cb_0$.} we have
\bea
&&i^*(\ga')=\overline{\ga}\ ,\ i^*(\oom')=\overline{\oom}\ ,\ {i_*}^{-1}(X')=\overline{X}\nn\\
&&i^*(\chi')=\overline{\chi}\ ,\ i^*(\om')=\overline{\om}\ ,\ i^*(\ze')=\overline{\ze}
\eea
together with their tangential derivatives up to $q$. 
\smallskip

 d) On $\Cb(\nu_0)$ with respect to the initial data foliation, we have
\bea
&&i^*(\ga')=\overline{\ga}\ ,\ i^*(\oom')=\overline{\oomb},\ {i_*}^{-1}({\underline X}')=\overline{\underline X}\nn\\
&&i^*(\chib')=\overline{\chib}\ ,\ i^*(\omb')=\overline{\omb}\ ,\ i^*(\ze')=\overline{\ze}
\eea
where $\ga',\oom',X',\chi',\chib',\om',\omb',\ze'$ are the metric components and the connection coefficients in a neighbourhood
of $C(\la_1)$ and $\Cb(\nu_0)$.\footnote{These quantities denoted here $\ga',\oom',......$ were denoted in the previous sections
$\ga,\oom,\oomb,X,{\underline X}$, $\chi,\chib,\om,\omb,\ze$. In the sequel these notations are referred to the corresponding quantities relative to the
double canonical foliation of $\M$ and their restrictions to $C_0\cup\Cb_0$.}
\smallskip
 
 e) The costraint equations \ref{1.41wqz},\ref{1.41wpz} are the pull back of (some of) the structure equations of $\M$
restricted to $C(\la_1)$ and $\Cb(\nu_0)$.\footnote{With respect to the foliations of the initial data.}

 f) The double null canonical foliation and the associated two dimensional one, $\{S(\la,\nu)\}$, implies different foliations of
$C(\la_1)$ and $\Cb(\nu_0)$; We can define on the whole $\M$ a null orthonormal frame $\{e_4,e_3,e_a\}$ adapted to the double null canonical
foliation, the (Lorentzian) metric components, $\ga_{ab},\oom,X^a$ with respect to the adapted coordinates
$\{u,\ub,\theta,\phi\}$\footnote{The precise way the coordinate $\theta$ and $\phi$ are defined will be discussed elsewhere. See, anyway
\cite{Kl-Ni:book}, paragraph 3.1.6.} and the corresponding connection coefficients
\bea
\chi_{ab}&=&\ggg(\dd_{e_a}e_4,e_b)\ \ ,\ \chib_{ab}=\ggg(\dd_{e_a}e_3,e_b)\nn\\
\omb&=&\frac{1}{4}\ggg(\dd_{e_3}e_3,e_4)=-\frac{1}{2}\dd_3(\log\oom)\nn\\
\om&=&\frac{1}{4}\ggg(\dd_{e_4}e_4,e_3)=-\frac{1}{2}\dd_4(\log\oom)\eql{4.16ww}\\
\ze_{a}&=&\frac{1}{2}\ggg(\dd_{e_{a}}e_4,e_3)\ .\nn
\eea
Moreover $X^a$ is a vector field defined in $\M$ such that denoting $N$ and $\Nb$ two null vector fields equivariant with respects to
the $S(\la,\nu)$ surfaces, the following relations hold: $N=\oom e_4$, $\Nb=\oom e_3$ and in the $\{u,\ub,\theta,\phi\}$ coordinates 
\[N=2\oom^2\frac{\partial}{\partial\ub}\ ,\ \Nb=2\oom^2\left(\frac{\partial}{\partial u}+X^a\frac{\partial}{\partial\om^a}\right)\ .\]

 g) The null geodesics of $\M$ along the outgoing and incoming null direction
$e_4,e_3$ are defined for all $\nu\in[\nu_0,\infty)$ and all $\la\in[\la_1,\la_0]$ respectively.
Finally the existence result is uniform in $\la_1(<\la_0<0)$.  
\end{theorem}

\subsection{The general structure of the proof.}\label{SS4.4}
The proof is made by two different parts whose structure is, basically, the same.
\smallskip

{\bf First part:} We prove the existence of the spacetime
\[\M'=J^+(S(\la_1,\nu_0))\cap J^-(S(\la_0,\nu_1\!=\!\la_0\!+\!2\ro_0))\ ,\]
{where}$\ \ \ \ \ \ \ \ \ \ \ \ \ \ \ \ \ \ \ \ \ \ \ \ \ \ \ \ \ \ \ 2\ro_0=\nu_0-\la_1\ .$ 
\medskip

{\bf Second part:} We prove the existence of the spacetime $\M\supset\M',$
\[\M=\lim_{\nu_1\rightarrow\infty}J^+(S(\la_1,\nu_0))\cap J^-(S(\la_0,\nu_1))\ .\]

{\bf Remark:} {\em It is appropriate now to give an intuitive picture of the spacetime we are building. As our result is a small data result we
expect that our spacetime stays near to (a portion of) the Minkowski spacetime. This implies that the functions $\ub(p)$ and $u(p)$ do not differ
much, written in spherical Minkowski coordinates, from $t+r$ and $t-r$, respectively. Moreover the radius $r(\la,\nu)$ defined as proportial to the
square root of the area of $S(\la,\nu)$, see \ref{2.13qw}, will stay near to the $r$ spherical coordinate. The initial cones $C_0$ and $\Cb_0$
approximate two minkowskian cones, one outgoing and one incoming, with their vertices on the origin vertical axis. The assumption that
$\nu_0$ and $|\la_1|$ are approximately equal to $r_0$ just pictures the two dimensional surface $S_0$ as lying (approximately) on the hyperplane $t=0$
where $t$ is ``near" to $\frac{1}{2}(\ub+u)$.}\footnote{Although the picture defined here is the more natural, with a redefinition of the functions
$\ub,u,r$ in terms of the standard coordinates one could also treat in a similar way the case where the two ``initial" cones $C_0$ and
$\Cb_0$ are shifted one respect to the other and their vertices do not lie even approximatley on the same vertical axis.}
\medskip

 {\bf Proof of the first part:}
We denote $({\cal K}(\tau),\ggg)$ a solution of the ``characteristic Cauchy problem" with the following properties:
\smallskip

{\bf i)} $({\cal K}(\tau),\ggg)$ is foliated by a double null canonical foliation $\{C(\la)\}$, $\{\Cb(\nu)\}$
with $\la\in [\la_1,{\overline\la}]\ ,\ \nu\in [\nu_0,{\overline\nu}]$. Moreover
\[i(C_0)\cap{\cal K}(\tau)=C(\la_1;[\nu_0,{\overline\nu}])\ ;\  i(\Cb_0)\cap{\cal K}(\tau)=\Cb(\nu_0;[\la_1,{\overline\la}])\]
where \footnote{As the spacetime we obtain is ``near" to a portion of the Minkowski spacetime,
condition \ref{3.22gg} can be visualized as the requirement that $S({\overline\la},{\overline\nu})$ coincides with a vertical (time) translation of
$S_0$.} 
\bea
{\overline\nu}+{\overline\la}=2\tau\ \ \ \ \ ;\ \ \ \ \ {\overline\nu}-{\overline\la}=2\ro_0\ . \eql{3.22gg}
\eea

{\bf ii)}
Denoted $\{e_3,e_4,e_a\}$ the null orthonormal frame adapted to the double null canonical foliation. We introduce a family of norms ${\cal R} , {\cal
O}$ for the null components of the Riemann curvature tensor and for the connection coefficients respectively, as done in \cite{Kl-Ni:book},
Chapter 3. 
\smallskip

{\bf iii)} Given $\ep_0>0$ sufficiently small, but larger than $\varepsilon$, the norms ${\cal R} , {\cal O}$ satisfy the
following inequalities
\bea
{\cal R}\leq \ep_0\ \ ,\ \ {\cal O}\leq \ep_0\ .
\eea

{\bf iv)} Denoted by $\cal T$ the set of all values $\tau$ for which the spacetime
${\cal K}(\tau)$ does exist, we define $\tau_*$ as the $\sup$ over all the values of $\tau\in{\cal T}$:
\bea
\tau_* =\sup \{\tau\in {\cal T}\}\eql{2.58}\ .
\eea
There are, now, two possibilities:
\[\tau_*=\la_0+\ro_0 \ \ \ \ \mbox{or}\ \ \ \ \tau_*<\la_0+\ro_0\]
If $\tau_*=\la_0+\ro_0$ then 
\bea
{\cal K}(\tau_*)=J^+(S(\la_1,\nu_0))\cap J^-(S(\la_0,\la_0\!+\!2\ro_0))
\eea 
and the first part is proved. If, viceversa, $\tau_*<\la_0+\ro_0$,
we show that in ${\cal K}(\tau_*)$ the norms $\cal R$ and $\cal O$ satisfy the following estimates
\bea
{\cal R}\leq \frac{1}{2}\ep_0\ \ ;\ \ {\cal O}\leq \frac{1}{2}\ep_0\ .\eql{2.59}
\eea

\smallskip

{\bf v)} Using the inequalities \ref{2.59} restricted to $\Cb(\nu_*)$ it is possible to prove that we can extend the
spacetime
${\cal K}(\tau_*)$ to a spacetime ${\cal K}(\tau_*+\de)$. This contradicts the fact that $\tau_*$ is the supremum defined in
\ref{2.58}, unless $\tau_*=\la_0+\ro_0$. In fact in this case ${\cal K}(\tau_*)$ describes the maximal region ${\cal
K}(\tau)$ which can be determined from the initial conditions assigned on $\Cb(\nu_0;[\la_1,\la_0])\cup
C(\la_1;[\nu_0,\nu_1])$ even if ${\cal R}\leq \frac{1}{2}\ep_0 ;\ {\cal O}\leq \frac{1}{2}\ep_0$.



\smallskip

{\bf vi)} Observe that, due to the fact, which will be evident in the course of the proof, that the estimates for
$\cal R$ and $\cal O$ do not depend on the magnitude of the interval $[\la_1,\la_0]$, also the proof of the first part can be
considered a global existence result. In other words fixed $\la_0$, we can choose $\ro_0=\frac{1}{2}(\nu_0-\la_1)$
arbitrarily large implying that $\tau_*=\la_0+\ro_0$ can be chosen arbitrarily large.

 In conclusion the result of the first part is achieved if we can prove that:
\smallskip

i) The set ${\cal T}$ is not empty.
\smallskip

ii) in ${\cal K}(\tau_*)$ the norms $\cal R$ and $\cal O$ satisfy the following estimates
\bea
{\cal R}\leq \frac{1}{2}\ep_0\ \ ;\ \ {\cal O}\leq \frac{1}{2}\ep_0\ .\eql{2.59a}
\eea

iii) If $\tau_*<\la_0+\ro_0$ we can extend ${\cal K}(\tau_*)$ to ${\cal K}(\tau_*+\de)$ with $\de>0$.
\bigskip

 The proofs of i), ii) and iii) are the technical parts of the result. Their main lines are sketched in the next
subsection. The detailed proofs are in a forthcoming paper.
\smallskip

 {\bf Proof of the second part:} It is, basically, a corollary of the proof of the first part.
We consider in this case the spacetimes $\M({\overline\nu})$ solutions of the ``characteristic Cauchy problem" with the following properties:

{\bf i)} $({\M}(\nu),\ggg)$ is foliated by a double null canonical foliation $\{C(\la)\}$, $\{\Cb(\nu)\}$
with
$\la\in [\la_1,\la_0]\ ,\ \nu\in [\nu_0,{\overline\nu}]$. Moreover
\[i(C_0)\cap{\M}({\overline\nu})=C(\la;[\nu_0,{\overline\nu}])\ ;\ 
i(\Cb_0)\cap{\M}({\overline\nu})=\Cb(\nu_0;[\la_1,\la_0])\] 

{\bf ii)}
Denoted $\{e_3,e_4,e_a\}$ the null orthonormal frame adapted to the double null canonical foliation, we introduce a family of norms ${\cal O}, {\cal R}$
for the connection coefficients and for the null components of the Riemann curvature tensor respectively, as done in \cite{Kl-Ni:book}, Chapter 3. 
\smallskip

{\bf iii)} Given $\ep_0>0$ sufficiently small, but larger than $\varepsilon$, the norms ${\cal R} , {\cal O}$ satisfy the
following inequalities
\bea
{\cal R}\leq \ep_0\ \ ,\ \ {\cal O}\leq \ep_0\ .
\eea

{\bf iv)} Denoted by $\cal N$ the set of all values $\overline\nu$ for which the spacetime
$\M(\overline\nu)$ does exist, we define $\nu_*$ as the $\sup$ over all the values of $\overline\nu\in{\cal N}$:
\bea
\nu_* =\sup\{\overline\nu\in {\cal N}\}\eql{2.58a}\ .
\eea
Again there are two possibilities: if $\nu_*=\infty$ the result is achieved, therefore we are left to show that the
second possibility, $\nu_*<\infty$, leads to a contradiction. In fact we prove that in $\M(\nu_*)$ the norms $\cal R$
and $\cal O$ satisfy the following estimates
\bea
{\cal R}\leq \frac{1}{2}\ep_0\ \ ;\ \ {\cal O}\leq \frac{1}{2}\ep_0\ .\eql{2.59aa}
\eea
\smallskip

{\bf v)} Using the inequalities \ref{2.59aa} restricted to $\Cb(\nu_*;[\la_1,\la_0])$ it is possible to prove that we can
extend the spacetime $\M(\nu_*)$ to a spacetime ${\M}(\nu_*+\de)$ which contradicts the definition of $\nu_*$ unless
$\nu_*=\infty$.

{\bf Remark:} Observe that in the second part of the result we do not have to prove that the set $\cal N$ is not empty as, due
to the first part of the proof, it contains at least the element $\overline\nu=\la_0+2\ro_0$.

\subsection{Thechnical parts of the proof, a broad sketch.}

{\bf First part:}
\smallskip

{\bf i)} To prove that the set $\cal T$ is not empty we have to show that a spacetime ${\cal K}(\tau)$ exists, possibly with
small $\tau$. This requires a local existence theorem and we can use the results of A.Rendall,
\cite{rendall:charact} or of H.Muller Zum Hagen, H.Muller Zum Hagen and H.J.Seifert, \cite{Muller}, \cite{MullerSeifert}. The main difficulty in adapting
these results to our case is that they are proved using harmonic coordinates while the ``gauge" we use for ${\cal K}(\tau)$ is
the one associated to the double null canonical foliation.
This requires a precise connection between the initial data written in the two different gauges in such a way
that we can reexpress their results in our formalism. This will be achieved in the next section.
\smallskip

{\bf ii)} Once we have proved that $\cal T$ is not empty we can define ${\cal K}(\tau_*)$ and prove that in this spacetime
inequalities \ref{2.59} hold. This requires, see \cite{Kl-Ni:book}, Chapter 3 for a  detailed discussion, that the double
null foliation of ${\cal K}(\tau_*)$ be canonical which, at its turn, implies proving the existence of specific foliations, denoted again
``canonical" on $C(\la_1)$ and on $\Cb(\nu_*)$. These ``canonical" foliations will play the role of the ``initial data" for the solutions of the
eikonal equations
\[g^{\mu\nu}\partial_{\mu}w\partial_{\nu}w=0\]
whose level hypersurfaces define the double null canonical foliation $\{C(\la)\}$, $\{\Cb(\nu)\}$ of ${\cal K}(\tau_*)$, see also
\cite{Niclast}.\footnote{The existence of this foliation has to be proved also for the (local) spacetime of i). The proof one gives for ${\cal
K}(\tau_*)$ can be easily adapted to this case. The canonical foliation is crucial to obtain on on $\Cb(\nu_*)$ all the connection coefficients
to be used again as initial data, without any loss of derivatives.}
\medskip

{\bf iii)} The central part of the proof consists in showing that we can extend ${\cal K}(\tau_*)$ to ${\cal K}(\tau_*+\de)$ with
$\de>0$. To achieve it we have to implement the following steps:
\smallskip

 1) First of all we have to prove an existence theorem in the strips $\Delta_1{\cal K}\ ,\ \Delta_2{\cal K}$
\bea
&&\Delta_1{\cal K}=J^+(S(\la_1,\nu_*)\cap J^-(S({\overline\la}(\nu_*),\nu_*+\de))\eql{3.31wq}\\
&&\Delta_2{\cal K}=J^+(S({\overline\la}(\nu_*),\nu_0)\cap J^-(S({\overline\la}(\nu_*+\de),\nu_*))\nn
\eea 
whose boundaries are:
\bea
\partial\Delta_1{\cal K}\!&=&\!C(\la_1;[\nu_*,\nu_*+\de])\cup\Cb(\nu_*+\de;[\la_1,{\overline\la}(\nu_*)])\eql{aa}\\
&&\cup\,C({\overline\la}(\nu_*);[\nu_*,\nu_*+\de])\cup\Cb(\nu_*;[\la_1,{\overline\la}(\nu_*)])\nn
\eea
\bea
\partial\Delta_2{\cal K}\!&=&\!\Cb(\nu_0;[{\overline\la}(\nu_*),{\overline\la}(\nu_*+\de)])\cup
C({\overline\la}(\nu_*+\de);[\nu_0,\nu_*])\\
&&\cup\,\Cb(\nu_*;[{\overline\la}(\nu_*),{\overline\la}(\nu_*+\de)])\cup C({\overline\la}(\nu_*);[\nu_0,\nu_*])\ ,\nn
\eea
where \[{\overline\la}(\nu)=-2\ro_0+\nu\ .\]
\smallskip

 2) We are then left with proving the (local) existence of a diamond shaped spacetime $\Delta_3{\cal K}$,
\bea
\Delta_3{\cal K}=J^{+}(S({\overline\la}(\nu_*),\nu_*)\cap J^{-}(S({\overline\la}(\nu_*+\de),\nu_*+\de))\ ,
\eea
specified by the initial data
\bea
C({\overline\la}(\nu_*);[\nu_*,\nu_*+\de])\cup\Cb(\nu_*;[{\overline\la}(\nu_*),{\overline\la}(\nu_*+\de)])\ .
\eea

 3) Finally on the portion of the boundary of
\[{\cal K}(\tau_*+\de)={\cal K}(\tau_*)\cup\Delta_1{\cal K}\cup\Delta_2{\cal K}\cup\Delta_3{\cal K}\ ,\]
made by $\Cb(\nu_*+\de;[\la_1,{\overline\la}(\nu_*+\de))])$ \footnote{ The canonical foliation on $C(\la_1;[\nu_0,\nu_*+\de])$
was already proved and it does not change.} a new canonical foliation has to be constructed which stays near to the one
obtained extending the double null canonical foliation of ${\cal K}(\tau_*)$ up to
$\Cb(\nu_*+\de;[\la_1,{\overline\la}(\nu_*+\de))])$ which will be considered, in this case, as the background foliation.

 This completes the description of the steps needed to prove our result. The detailed proof will be given in a subsequent paper.
\smallskip

{\bf Second part:}
\smallskip

In this case, as we said, we already know that the set $\cal N$ is not empty. To prove that $\nu_*=\infty$ we have to show
that, if $\nu_*$ were finite, the spacetime
${\M}(\nu_*)$ could be extended to a spacetime ${\M}(\nu_*+\de)$ with the same properties. To prove it we need an
existence theorem for the strip
\[J^+(S(\la_1,\nu_*))\cap J^{-}(S(\la_0,\nu_*+\de))\]
Again, as in the case of $\Delta_1{\cal K}$ and $\Delta_2{\cal K}$, this existence theorem is a non local result as the
``length" of the strip is not required to be small in the $e_3$ direction. Therefore as in previous cases the proof requires a bootstrap
mechanism. 

 Finally as at the end of the proof of the first part we have to show that $\M(\nu_*+\de)$ can be endowed with a double
canonical foliation whose existence can be proved once we prove the existence of a canonical foliation on the ``last slice",
see \cite{Kl-Ni:book} Chapter 3, on $\Cb(\nu_*+\de;[\la_1,\la_0])$.
\section{The ``harmonic" initial data}\label{S.4}
In this section we use the results of A.Rendall, \cite{rendall:charact} and of H.Muller Zum Hagen \cite{Muller} to provide the local existence
result that we need to start the bootstrap mechanism for our problem.

 The main difficulty in adapting these results to our case is that they have been proved using harmonic coordinates while we want that the
``gauge" for ${\cal K}(\tau)$ be the one associated to the double null canonical foliation. This requires stating the connection
between the initial data written in the two different gauges so that we can reexpress their results in our formalism. 
\subsection{The relation between the harmonic and the $\oom$-foliation gauges}\label{SS3.1} 

\subsubsection[The reduction of the Einstein equations to hyperbolic equations in the characteristic case]{The reduction of
the Einstein equations to hyperbolic equations in the characteristic case}\label{SS4}

As it is well known the standard procedure to solve locally the Einstein equations has been to find a ``gauge" which reduces
them to hyperbolic P.D.E. equations.\footnote{Different approaches have been used to solve the equations globally, see
\cite{C-K:book}, \cite{Kl-Ni:book}, \cite{Rod-lin}.} The ``reduction" mechanism for the Einstein equations is based on the fact that if
the initial data satisfy, besides the constraint equations, the conditions
\[\Ga^{\mu}\equiv g^{\ro\si}\Ga^{\mu}_{\ro\si}=0\ ,\]
then it can be proved that the equality $\Ga^{\mu}=0$ holds in the whole spacetime $(\M,\ggg)$ where $\ggg$ is solution of the
``reduced" Einstein equations, (the Einstein equations where $\Ga^{\mu}$ is posed, ab initio, equal to zero). This fact is relevant as the
reduced Einstein equations are of (quasilinear) hyperbolic type.

 The strategy to find a ``gauge", that is a change of coordinates, such that on the initial hypersurface
$\Ga^{\mu}=0$, is well known in the non characteristic case. This ``gauge" is usually called ``harmonic" gauge. A similar approach can be also used in
the characteristic case.  In particular this has been investigated in great detail by A.Rendall,  \cite{rendall:charact}.

 Here we show how to express the initial data in a different family of gauges in terms of data expressed in the
harmonic gauge and viceversa.

 More precisely, specified our initial data null hypersurface, we define on it the ``$\oom$-foliation", described in the previous sections.
Then we introduce another foliation of the same null hypersurface, whose leaves are different from those of the $\oom$-foliation, but still
diffeomorphic to $S^2$ surfaces, which has the property of being the foliation associated to the ``harmonic gauge". This foliation will be called the
``harmonic foliation".

 The advantage of this approach is that it keeps separated in a clear way the two main difficulties connected with the choice of the initial data
for the characteristic problem.

 The first one, discussed at length in the previous sections, is that not all the initial data can be freely assigned
on the initial hypersurface; some of them have to satisfy ``transport equations" and therefore depend, through
these equations, on the initial data assigned on the specific leaf $S_0$. This problem is not related to the harmonic gauge
and to the reduction problem; whatever ``gauge" we choose, not all the initial data can be assigned freely. In particular the ``normal" derivatives
with respect to the null hypersurfaces of the components of the metric tensor have to satisfy some transport equations.\footnote{See the previous
sections. This has also been discussed in detail by A.Rendall, \cite{rendall:charact} in a more coordinate dependent way.}

 The second difficulty is associated to the ``reduction problem". Namely to solve the Einstein equations in the ``reduced form", the ``reduction
problem" one has to implement the ``harmonic" condition $\Ga^{\mu}=0$. 

 Once we are able to compare in a precise way the ``$\oom$-foliation" and
the ``harmonic" foliation,  we can express the norms of data for the ``harmonic" situation in terms of the norms of
the ``$\oom$-foliation" data and viceversa. This is important for us as we will choose the initial data in the
``$\oom$-foliation" setting such that the corresponding data expressed in the harmonic gauge are suitable for using the
local existence theorems for the characteristic problem in the harmonic gauge, see H.Muller Zum Hagen, \cite{Muller} and M.Dossa,
\cite{Dossa}.

\subsubsection{The ``harmonic" null frame.}

Let us recall that the null initial hypersurface $\cal C$ has been thought as immersed in the four dimensional manifold $R^4$ endowed with a
Lorentzian metric $\tilde{\ggg}$. We introduced in $R^4$ the coordinates $\{\ub,u,\om^a\}$ and we required that $\tilde{\ggg}$ has, in these
coordinates, the following expression, see \ref{1.36b}:
\bea
\tilde{\ggg}=|\tilde{X}|^2du^2\!-\!2\tilde{\oom}^2(dud\ub\!+\!d\ub du)\!-\!\tilde{X}_a(dud\om^a\!+\!d\om^a
du)\!+\!\tilde{\ga}_{ab}d\om^a d\om^b\ .\ \
\eql{1.36bz}
\eea
With this choice of coordinates the restriction on ${\cal C}=C_0\cup\Cb_0$ of the various components of the metric $\tilde{\ggg}$ has been specified
in the previous sections during the initial data construction, see \ref{1.51qw}. Moreover adapted to the foliation specified by $\oom$ and
$\oomb$, see \ref{1.41bw} a null orthonormal frame is specified $\{e_4,e_3,e_a\}$,
\[e_4=2\oom L\ ,\ e_3=2\oomb\ \!\Lb\ ,\ e_A=e^a_A\frac{\partial}{\partial\om^a}\ .\]
where\footnote{Here $X$ is $\underline X$ when defined on $\Cb_0$.}
\bea
L=\frac{1}{2\oom^2}\frac{\partial}{\partial\ub}\ \ ,\ \ \Lb=\frac{1}{2\oom^2}\left(\frac{\partial}{\partial u}+X\right)
\eea
and $X=X^a\frac{\partial}{\partial\om^a}$. Therefore
\bea
&&e_4={\oom}^{-1}N={\oom}^{-1}\frac{\partial}{\partial\ub}\nn\\
&&e_3={\oom}^{-1}\Nb={\oom}^{-1}\left(\frac{\partial}{\partial u}+X\right)\eql{3.4aaq}\\
&&e_A=e_A^a\frac{\partial}{\partial\om^a}\nn
\eea
We introduce now on $\cal C$ a different foliation, we call ``harmonic foliation", tied to the harmonic coordinates, 
denoted $\{x^1,x^2,x^a\}$. The harmonic coordinates as functions of the $\{u,\ub,\om^a\}$ coordinates,
$x^{\mu}(u,\ub,\om^a)$, satisfy the wave equations
\bea
\square_{\!\!\tilde{g}}x^{\mu}=0\eql{3.3.aa}
\eea
where the D'alembertian $\square_{\!\!\tilde{g}}$ is written in the coordinates $\{u,\ub,\om^a\}$ and $\tilde g$ is the metric \ref{1.36bz}.
Explicitely, equation
\ref{3.3.aa} has the following form:
\bea
\tilde{g}^{\ro\si}\partial_{\ro}\partial_{\si}x^{\mu}-\tilde{g}^{\ro\si}\Ga^{\la}_{\ro\si}\partial_{\la}x^{\mu}=0\ .\eql{3.4aa}
\eea 
To define the ``harmonic foliation" we need that the coordinates $\{x^1,x^2,x^a\}$ satisfy, on ${\cal C}=C(\la_1)\cup\Cb(\nu_0)$, equations
\ref{3.4aa}. 
To do it we start considering the hypersurface $C_0$, the argument for the hypersurface $\Cb_0$ is analogous and will be done later on. The new
harmonic coordinates, (they will be made to satisfy \ref{3.4aa}) are denoted 
$\{x^1,x^2,x^{3,4}\}$ following Rendall, \cite{rendall:charact}.
We assume that the null outgoing ``cone" $C_0$, in these coordinates, is the level hypersurface $\{x^2=0\}$ ($\Cb_0$, the level
hypersurface $\{x^1=0\}$) and we require that the metric tensor has, restricted to $C_0$, the following expression:\footnote{It is important to
recognize that, differently from the case of the $\{u,\ub,\om^a\}$ coordinates, the other metric components in the harmonic coordinates are
different from zero when not restricted to $C_0$, for instance the component $g'_{11}$ is different from zero. Moreover the expression of the
metric when restricted to $\Cb_0$ will be different. This will be discussed in more detail later on.}
\bea
\tilde{\ggg}|_{C_0}
\!=\!g'_{22}d{x^2}d{x^2}\!+\!g'_{12}(dx^1dx^2\!+\!dx^2dx^1)\!+\!g'_{2a}(d{x^2}d{x^a}\!+\!d{x^a}d{x^2})\!+\!\ga'_{ab}d{x^a}d{x^b}\ .\ \ \eql{3.16a}
\eea
We specify the coordinate $x^1$ on $C_0$ as the affine parameter of the null geodesics
generating $C_0$. Therefore, denoting $L$ their tangent vector fields, it follows that $L=\frac{\partial}{\partial x^1}$.

 We introduce a foliation on $C_0$, we call ``harmonic-foliation", defining the leaves of the foliation, $\{S'_0(\nu')\}$, in the following
way:
\bea
S'_0(\nu')=\{p\in C_0|x^1(p)=\nu'\}\ .
\eea
We define the coordinates $x^a$, $a\in \{3,4\}$, as adapted to these leaves (the analogues of the $\{\om^a\}$
in the $\oom$-foliation) and an orthonormal frame tangent to each leave $S'_0(\nu')$,
\bea
e'_A={e'_A}^a\frac{\partial}{\partial x^a}\ .
\eea
We consider the null frame $\{L,e'_A\}$ relative to the outgoing cone $C_0$ and we extend it to a null frame $\{L,\Lb^*,e'_A\}$, we call
``harmonic null frame", satisfying the conditions
\bea
g(L,\Lb^*)=-1\ ,\ g(\Lb^*,\Lb^*)=0\ ,\ g(\Lb^*,e'_A)=0\ .\eql{3.19a}
\eea
The expression of the metric \ref{3.16a}, with $g'_{2a}\neq 0$, and the condition $g(L,\Lb^*)=-1$ imply that $\Lb^*$ has to be written as
\bea
\Lb^*=\si\frac{\partial}{\partial x^2}+X'
\eea
with $X'=X'_Ae'_A$. The remaining conditions \ref{3.19a} determine the relations between $\si$, $X'_a=\ga'_{ab}X'^b$ and the various components of the
metric
\ref{3.16a}:
\bea
&&2\si g'_{2a}X'^a=-\left(\si^2g'_{22}+|X'|^2\right)\ ,\ \si g'_{12}=-1\nn\\
&&\si g'_{2a}=-X'_a \ ,\ \si^2g'_{22}=|X'|^2\ .
\eea
The metric tensor $\tilde{\ggg}|_{C_0}$, in the harmonic coordinates, can be rewritten as
\bea
\ggg|_{C_0}\!&=&\!\!\si^{-2}|X'|^2d{x^2}d{x^2}-\si^{-1}(dx^1dx^2+dx^2dx^1)\nn\\
&-&\!\!\si^{-1}X'_a(d{x^2}d{x^a}+d{x^a}d{x^2})+\ga'_{ab}d{x^a}d{x^b}\ \ \ \eql{3.22a}
\eea
where $\si$ and $X'_a$ can be thought as extensions in $R^4$ of the quantities defined on $C_0$.

 Summarizing, the ``harmonic null frame" $\{L,\Lb^*,e'_A\}$, on $C_0$, written in the ``harmonic" coordinates, has the
following expression:
\bea
&&L=\frac{\partial}{\partial x^1}\nn\\
&&\Lb^*=\si\frac{\partial}{\partial x^2}+X'\eql{3.23a}\\
&&e'_A={e'_A}^a\frac{\partial}{\partial x^a}\ .\nn
\eea
To justify the name of the harmonic null frame for $\{L,\Lb^*,e'_A\}$ we have to impose that the $\{x^1,x^2,x^{3,4}\}$ be harmonic coordinates.
To relate these coordinates to the $\{u,\ub,\om^a\}$ ones it turns out convenient to look first at the relation between the $\oom$-null
frame and the harmonic null frame. Therefore we express the ``harmonic null frame" in terms of the $\oom$-null frame
$\{e_4,e_3,e_A\}$ writing
\bea
&&L=(2\oom)^{-1}e_4\nn\\
&&\Lb^*=\a e_3+\b e_4+\de_Ae_A\\
&&e'_A=\ga e_A+\ro_Ae_4+\si_Ae_3\ .\nn
\eea
where all the coefficients are functions defined on $\cal C$.
As both frames are null orthonormal the following relations hold:
\bea
\a=\oom\ ,\ \b=\frac{|\de|^2}{4\oom}\ ,\ \ga=1\ ,\  \ro_A=\frac{\de_A}{2\oom}\ ,\ \si_A=0
\eea
and, therefore,
\bea
&&L=(2\oom)^{-1}e_4\nn\\
&&\Lb^*=\oom e_3+\frac{|\de|^2}{4\oom}e_4+\de_Ae_A\eql{3.26a}\\
&&e'_A=e_A+\frac{\de_A}{2\oom}e_4\nn
\eea
From \ref{3.4aa}, \ref{3.23a}, \ref{3.26a} we obtain immediately the explicit expressions of the first derivatives of the functions
$x^{\mu}(\ub,u,\om^a)$. The relation between $L$ and $e_4$ gives immediately
\bea
\frac{\partial}{\partial\ub}=2\oom^2\frac{\partial}{\partial x_1}\ .
\eea
Let us consider the relation
\bea
e_3={\oom}^{-1}\!\left(\frac{\partial}{\partial u}+X\right)
={\oom}^{-1}\!\left(\frac{|\de|^2}{2}\frac{\partial}{\partial x_1}+\si\frac{\partial}{\partial x^2}
+(X'^a-\de^a)\frac{\partial}{\partial x^a}\right)\ \ \ \ 
\eea
where $\de^a\equiv \de_Ae_A^a$. From it follows
\bea
\frac{\partial}{\partial u}=
\si\frac{\partial}{\partial x^2}+\left(\frac{|\de|^2}{2}+\de\c X\right)\frac{\partial}{\partial x_1}
+\left(\lap_A-\de_A\right){e'_A}^a\frac{\partial}{\partial x^a}\ .
\eea
where $\lap_A\equiv X'_A-X_A$. Finally the relation
\bea
e_A^a\frac{\partial}{\partial\om^a}=-{\de_A}\frac{\partial}{\partial x^1}+{e'_A}^a\frac{\partial}{\partial x^a}
\eea
implies
\bea
\frac{\partial}{\partial\om^a}=-\theta^A_a{\de_A}\frac{\partial}{\partial x^1}+\theta^A_a{e'_A}^c\frac{\partial}{\partial x^c}
\eea
where $\theta^A=\theta^A_ad\om^a$ is the one form associated to the vector $e_A$, $\theta^A(e_B)=\de^A_B$.
Collecting all these results we have
\bea
&&\frac{\partial}{\partial\ub}=2\oom^2\frac{\partial}{\partial x_1}\nn\\
&&\frac{\partial}{\partial u}=\left(\frac{|\de|^2}{2}+\de\c\!X\right)\frac{\partial}{\partial
x_1}+\si\frac{\partial}{\partial x^2}+\left(\lap_A-\de_A\right){e'_A}^a\frac{\partial}{\partial x^a}\eql{3.22qa}\\
&&\frac{\partial}{\partial\om^a}=-\theta^A_a{\de_A}\frac{\partial}{\partial x^1}
+\theta^A_a{e'_A}^c\frac{\partial}{\partial x^c}\nn
\eea
and, immediately,
\bea
&&\frac{\partial x^1}{\partial\ub}=2\oom^2\ \ ,\ \ \frac{\partial x^2}{\partial\ub}=0\ \ ,\ \ \frac{\partial
x^a}{\partial\ub}=0\nn\\
&&\frac{\partial x^1}{\partial u}=\left(\frac{|\de|^2}{2}+\de\c\!X\right)\ \ ,\ \ \frac{\partial x^2}{\partial u}=\si\ ,\
\frac{\partial x^a}{\partial u}=\left(\lap_A-\de_A\right){e'_A}^a\nn\\
&&\frac{\partial x^1}{\partial\om^b}=-\theta^A_b{\de_A}\ \ ,\ \
\frac{\partial x^2}{\partial\om^b}=0\ \ ,\ \ \frac{\partial x^a}{\partial\om^b}=\theta^A_b{e'_A}^a\ .\eql{3.25sa}
\eea
The null orthonormal frame $\{L,\Lb^*,e'_A\}$ is harmonic if the coordinates
$\{x^1,x^2,x^a\}$ are harmonic, that is if the functions $x^{\mu}(u,\ub,\om^a)$ satisfy the wave equation \ref{3.3.aa},
\bea
{\tilde g}^{\ro\si}\partial_{\ro}\partial_{\si}x^{\mu}-{\tilde g}^{\ro\si}\Ga^{\la}_{\ro\si}\partial_{\la}x^{\mu}=0\ .\eql{3.27a}
\eea
The components of the inverse metric $g^{\ro\si}$ different from zero are
\bea
g^{u\ub}=-(2\oom^2)^{-1}\ ,\ g^{a\ub}=-\frac{X^a}{2\oom^2}\ ,\ g^{ab}=\ga^{ab}
\eea
and, therefore, \ref{3.27a} takes the form
\bea
&&-\frac{1}{2\oom^2}\partial_{\ub}\partial_ux^{\mu}-\frac{X^a}{2\oom^2}\partial_a\partial_{\ub}x^{\mu}
+\frac{1}{2}\ga^{ab}\partial_a\partial_bx^{\mu}\nn\\
&&+\frac{1}{2\oom^2}\left(\Ga^{\ub}_{u\ub}\partial_{\ub}x^{\mu}+\Ga^{u}_{u\ub}\partial_{u}x^{\mu}
+\Ga^{a}_{u\ub}\partial_{a}x^{\mu}\right)\nn\\
&&+\frac{X^a}{2\oom^2}\left(\Ga^{\ub}_{a\ub}\partial_{\ub}x^{\mu}+\Ga^{u}_{a\ub}\partial_{u}x^{\mu}
+\Ga^{b}_{a\ub}\partial_{b}x^{\mu}\right)\eql{3.29}\\
&&-\frac{1}{2}\ga^{ab}\left(\Ga^{\ub}_{ab}\partial_{\ub}x^{\mu}+\Ga^{u}_{ab}\partial_{u}x^{\mu}
+\Ga^{c}_{ab}\partial_{c}x^{\mu}\right)=0\nn
\eea
 The Christoffel symbols $\Ga^{\la}_{\ro\si}$ relative to the coordinates $u,\ub,\om^a$ have been previously expressed in
terms of the connection coefficients, see subsection \ref{SS1.3.2}.  Therefore, once the initial data in the ``$\oom$-foliation" are assigned, 
all the factors which, in equation \ref{3.29}, multiply the first and second partial derivatives are known (on $C_0$).\footnote{Observe that the
only Christoffel symbol which cannot be directly expressed in terms of the connection coefficients, namely $\Ga^a_{uu}$, see subsections
\ref{SS1.3.2}, \ref{SS2.3}, does not appear.}

 Writing the explicit expressions for equations \ref{3.29} for each value of $\mu$, using \ref{3.25sa} and the expressions of the connection
coefficients given in \ref{1.57wtw}, we obtain:
\bea
&&\frac{\partial}{\partial\ub}T+\frac{\oom\tr\chi}{2}T-\eta(\de)+\oom^2(\nabb_A\de)_A+\frac{\oom\tr\chi}{2}X\!\c\de
\!-\!\left[\nabb_X\log\oom-\oom^2(\oom\tr\chib)\right]=0\nn\\
&&\frac{\partial}{\partial\ub}\si+\frac{\oom\tr\chi}{2}\si=0\eql{3.28wq}\\
&&\frac{\partial}{\partial\ub}W^a+\frac{\oom\tr\chi}{2}W^a-\left[\oom^2\lapp X^a-\left(\frac{\oom\tr\chi}{2}
X^c+\eta^c\right)\!\partial_cX^a\right]=0\ ,\nn
\eea
where
\bea
&&T=\bigg(\!\frac{|\de|^2}{2}\!+\!\de\c\!X\!\bigg)\ \ ,\ \ W^a=(\Delta_A-\de_A){e'}_A^a\ \ \nn\\
&&\de=\de^a\frac{\partial}{\partial\om^a}\ , \ X=X^a\frac{\partial}{\partial\om^a}\ , \ \ X'={X'}^a\frac{\partial}{\partial x^a}\eql{3.29wq}\\
&&\Delta_A=X'_A-X_A\ .\nn
\eea
A simplification of \ref{3.28wq} is obtained if we choose ${e'}^a_A=e^a_A$. In this case $W^a=({X'}^a-\de^a)-X^a$. The terms in brackets of the
right hand side of \ref{3.28wq} are known terms as they depend on the initial data associated to the $\oom$-foliation. Therefore their decay is
known. More precisely we have
\bea
&&\left[\nabb_X\log\oom-\oom^2(\oom\tr\chib)\right]=O(\frac{1}{r})\nn\\
&&\left[\oom^2\lapp X^a-\left(\frac{\oom\tr\chi}{2}X^c+\eta^c\right)\!\partial_cX^a\right]=O(\frac{\varepsilon}{r^3})\ .\eql{3.30qw}
\eea
Using these inequalities we prove the following lemma,

\begin{Le}\label{L3.1} Assume we control the norm $|\c|_{p,S}$, $p\in [2,4]$, for the connection coefficients relative to the $\oom$-foliation and
their first derivatives, then, assuming $\de|_{S_0}=0,\ \si|_{S_0}=2\oom^2,\ X'^a|_{S_0}=0$, equations \ref{3.28wq} have a solution along the
whole
$C_0$ such that the following estimates hold
\bea
\si=O(\frac{c}{r})\ ,\ \de^a=O(\frac{c}{r})\ ,\ W^a=O(\frac{\varepsilon}{r^{2}})\ , (X'^a-\de^a)=O(\frac{\varepsilon}{r^{2}})\ \ \ \ 
\eea
Moreover\ \ $\de\c\!X=O(\frac{\ep}{r})$.
\end{Le}
{\bf Proof:}
Observe that $\de\c X$ satisfies $|\de\c X|\leq c {\ep}{r^{-1}}|\de|$.
An application of Gronwall's Lemma to the first equation in \ref{3.28wq} gives
\bea
|r^{1-\frac{2}{p}}T|_{p,S}(\la_1,\nu)&\leq&
c|r^{1-\frac{2}{p}}T|_{p,S}(\la_1,\nu_0)+\int_{\nu_0}^{\nu}|r^{1-\frac{2}{p}}(\eta(\de)-(\nabb_A\de)_A-\frac{\oom\tr\chi}{2}X\!\c\de)|_{p,S}\nn\\
&&+\int_{\nu_0}^{\nu}|r^{1-\frac{2}{p}}\left[\nabb_X\log\oom-\oom^2(\oom\tr\chib)\right]|_{p,S}\ .
\eea
As \[\left[\nabb_X\log\oom-\oom^2(\oom\tr\chib)\right]=O\left(\frac{1}{r}\right)\ ,\]
the last integral is bounded by $cr(\la_1,\nu)$. Dividing both members by $r(\la_1,\nu)$ and observing that
\[r^{-1}(\la_1,\nu)\leq r^{-1}(\la_1,\nu')\leq r^{-1}(\la_1,\nu_0)\]
we can write
\bea
|T|_{p,S}(\la_1,\nu)&\leq& c\!\left(\frac{|r^{1-\frac{2}{p}}T|_{p,S}(\la_1,\nu_0)}{r(\la_1,\nu)}
+\frac{1}{r(\la_1,\nu)}\int_{\nu_0}^{\nu}|r^{1-\frac{2}{p}}(|\eta(\de)|+|(\nabb_A\de)_A|\right.\nn\\
&&\ \ \ \left.+|\tr\chi X\!\c\de|)|_{p,S}+1\right.\bigg)\eql{3.33qw}
\eea
Observe that $\eta_a=O(\varepsilon r^{-1})$, $\de^a\leq cr^{-1}|\de|$, it follows
that
$\eta(\de)=O(\varepsilon r^{-2}|\de|)$, $|\tr\chi X\c\de|\leq c O\!\left({\ep}{r^{-2}}|\de|\right)$, therefore
\bea
|r^{1-\frac{2}{p}}\eta(\de)|_{p,S}\leq c\frac{\varepsilon}{r^2}|r^{1-\frac{2}{p}}|\de||_{p,S}
\ ,\ |r^{1-\frac{2}{p}}\tr\chi X\c\de|_{p,S}\leq c\frac{\varepsilon}{r^2}|r^{1-\frac{2}{p}}|\de||_{p,S}\ .\ \ \ \ 
\eea
Moreover, as we are free to assigne $\nabb\de$ along $C_0$, we require
\bea
|r^{1-\frac{2}{p}}(\nabb_A\de)_A|_{p,S}\leq c\frac{\varepsilon}{r}|r^{1-\frac{2}{p}}|\de||_{p,S}\ .\nn
\eea
In conclusion we expect that $\bigg(\!\frac{|\de|^2}{2}\!+\!\de\c\!X\!\bigg)$ stays bounded along $C_0$.\footnote{To pass from the $|\c|_{p,S}$ norm
estimates to the pointwise ones we need the analogous estimates for the first tangential derivatives which can be obtained exactly in the same
way, provided we have sufficient norm estimates for the derivatives of the connection coefficients relative to the $\oom$ foliation.} In fact
\bea
|T|_{p,S}(\la_1,\nu)&\leq&c\!\left(1+\frac{1}{r(\la_1,\nu)}\left(|r^{1-\frac{2}{p}}T|_{p,S}(\la_1,\nu_0)+\varepsilon\log
r(\la_1,\nu)\right)\right)\nn
\eea
As
$|\de\c X|\leq |\de||{X}|\leq |\de|^2+O(\ep^2 r^{-2})$ it follows that $|\de|^2=O(1)$ which implies $\de^a=O(r^{-1})$ and $\de\c
X=O(\varepsilon r^{-1})$.
\smallskip

 The estimate $\si=O(r^{-1})$ follows immediately from the second equation. Finally, writing for the third equation an
estimate analogous to \ref{3.33qw}, we obtain, recalling \ref{3.30qw}, denoting $|W|^2=\ga_{ab}W^aW^b$ and proceeding as in Chapter 4 of
\cite{Kl-Ni:book}, 
\bea
|W|_{p,S}(\la_1,\nu)\!\!\!&\leq&\!\!\!
c\!\left(\!\frac{|r^{1-\frac{2}{p}}|W||_{p,S}(\la_1,\nu_0)}{r(\la_1,\nu)}\right.\eql{3.33qwe}\\
&+&\left.\frac{1}{r(\la_1,\nu)}\!\int_{\nu_0}^{\nu}|r^{1-\frac{2}{p}}\left[\oom^2\lapp X^a-\left(\frac{\oom\tr\chi}{2}
X^c+\eta^c\right)\!\partial_cX^a\right]|_{p,S}\!\right)\nn\\
&\leq&\frac{c}{r(\la_1,\nu)}\left(|r^{1-\frac{2}{p}}|W||_{p,S}(\la_1,\nu_0)+\varepsilon\int_{\nu_0}^{\nu}|r^{1-\frac{2}{p}}r^{-3}|_{p,S}\right)\
.\nn
\eea
This estimate together with the analogous one for the first tangential derivative of $W$ implies that
\[|W|=|\ga_{ab}W^aW^b|^{\frac{1}{2}}=O(\varepsilon r^{-1})\ .\]
Therefore $W^a=O(r^{-2})$ and as $X^a=O(\ep r^{-2})$ and $\de^a=O(\ep r^{-1})$, it follows that $(X'^a-\de^a)=O(r^{-2})$ . 
\smallskip

 Once we have solved equations \ref{3.28wq} we have obtained, on $C_0$, the first derivatives of the harmonic coordinates written in terms of the
$\oom$-coordinates, see \ref{3.25}. This implies  that we have the metric components in the harmonic coordinates, once we have the corresponding
quantities written in the $\oom$-coordinates. To have the whole initial data required to solve locally the characteristic problem in the harmonic
coordinates we need also the first derivatives of the metric along the null direction ``orthogonal" to $C_0$. In principle one can rely, for
instance, on the Muller Zum Hagen result, see \cite{Muller}, where these derivatives are obtained using the constraint equations, but in the present
case this would be a useless repetition\footnote{In fact even more complicated, see for instance \cite{rendall:charact}.} as we have already done
it in the $\oom$-coordinates. Therefore the only thing to do is to obtain the second derivatives of the harmonic coordinates with respect to the
$``\oom"$ ones. We achieve this result proceeding as done before for the first derivatives. We start deriving equations \ref{3.27a}
\bea
{\tilde g}^{\ro\si}\partial_{\ro}\partial_{\si}(\partial_{\tau}x^{\mu})\!-\!
{\tilde g}^{\ro\si}\Ga^{\la}_{\ro\si}\partial_{\la}(\partial_{\tau}x^{\mu})\!-\!(\partial_{\tau}{\tilde
g}^{\ro\si})\Ga^{\la}_{\ro\si}\partial_{\la}x^{\mu}\!-\!{\tilde
g}^{\ro\si}(\partial_{\tau}\Ga^{\la}_{\ro\si})\partial_{\la}x^{\mu}=0\ \ \ \ \ .\eql{3.32qw}
\eea
and observing that, from equations \ref{3.25} we have
\bea
&&\frac{\partial^2 x^1}{\partial\ub^2}=-8\oom^3\om\ \ ,\ \ 
\frac{\partial^2 x^1}{\partial u\partial\ub}=\frac{\partial T}{\partial\ub}\ \ ,\ \ 
\ \frac{\partial^2 x^1}{\partial u^2}=\frac{\partial T}{\partial u}\nn\\
&&\frac{\partial^2 x^1}{\partial\om^a\partial\ub}=4\oom^3\nabb_a\log\oom\ \ ,\ \ \frac{\partial^2
x^1}{\partial\om^a\partial\om^b}=\frac{\partial\de_b}{\partial\om^a}\
\ ,\ \ 
\frac{\partial^2 x^1}{\partial\om^a\partial u}=\frac{\partial T}{\partial\om^a}\nn\\
&&\frac{\partial^2 x^2}{\partial\ub^2}=0\ \ ,\ \ \frac{\partial^2 x^2}{\partial u\partial\ub}=-(\oom\tr\chi)\si
\ \ ,\ \ \frac{\partial^2 x^2}{\partial u^2}=\frac{\partial\si}{\partial u}\nn\\
&&\frac{\partial^2 x^2}{\partial\om^a\partial\ub}=0\ \ ,\ \ \frac{\partial^2 x^2}{\partial\om^a\partial\om^b}=0
\ \ ,\ \ \frac{\partial^2 x^2}{\partial\om^a\partial u}=\frac{\partial\si}{\partial\om^a}\nn\\
&&\frac{\partial^2 x^a}{\partial\ub^2}=0\ \ ,\ \ \frac{\partial^2 x^a}{\partial u\partial\ub}=\frac{\partial W^a}{\partial\ub\ }
\ \ ,\ \ \frac{\partial^2 x^a}{\partial^2 u}=\frac{\partial W^a}{\partial u\ }\nn\\
&&\frac{\partial^2 x^a}{\partial\om^b\partial\ub}=0\ \ ,\ \ \frac{\partial^2 x^a}{\partial\om^b\partial\om^c}=0\ \ ,\ \ 
\frac{\partial^2 x^a}{\partial\om^b\partial u}=\frac{\partial W^a}{\partial \om^b\ }\nn
\eea
All these second derivatives are easily obtained with the exception of 
\[\frac{\partial^2 x^1}{\partial u^2}=\frac{\partial T}{\partial u}\ ,\ \frac{\partial^2 x^2}{\partial u^2}=\frac{\partial\si}{\partial u}
\ ,\ \frac{\partial^2 x^a}{\partial^2 u}=\frac{\partial W^a}{\partial u\ }\ .\]
To obtain and control these derivatives we have to use equation \ref{3.32qw} with $\tau=u$. It is easy to see that they provide, for these
quantities, evolution equations along $C_0$  of the same type as equations \ref{3.28wq} which can be solved in the same way. This completes the
knowledge of the second derivatives
$\frac{\partial^2 x^{\mu}}{\partial y^{\ro}\partial y^{\si}}$, where $\{y^{\tau}\}=\{u,\ub,\om^a\}$.

 Once we know these second derivatives we can express the first partial derivatives of the metric components with respect the $\{u,\ub,\om^a\}$
coordinates in terms of the first partial derivatives of the metric components $g'_{\mu\nu}$ with respect the harmonic coordinates:
\bea
\frac{\partial}{\partial y^{\la}}g_{\ro\si}
=\!\left(\frac{\partial^2x^\mu}{\partial y^{\la}\partial y^{\ro}}\frac{\partial x^{\nu}}{\partial y^{\si}}\!+\!
\frac{\partial x^{\mu}}{\partial y^{\ro}}\frac{\partial^2x^\nu}{\partial y^{\la}\partial y^{\si}}\right)\!g'_{\mu\nu}
\!+\!\frac{\partial x^{\mu}}{\partial y^{\ro}}\frac{\partial x^{\nu}}{\partial y^{\si}}\frac{\partial x^{\tau}}{\partial y^{\la}}
\left(\!\frac{\partial}{\partial x^{\tau}}g'_{\mu\nu}\!\right)\ \ \ \ \ \eql{3.34qw}
\eea
As $\{\frac{\partial x^{\mu}}{\partial y^{\ro}}\frac{\partial x^{\nu}}{\partial y^{\si}}\frac{\partial x^{\tau}}{\partial y^{\la}}\}$ is an
invertible matrix $64\times 64$ we can solve the linear system \ref{3.34qw} obtaining the initial data for the characteristic probem in the
harmonic coordinates, the derivatives along the direction ``orthonormal" to $C_0$ satisfying the constraints.

{\bf Remark:} A detailed examination along the same lines of Lemma \ref{L3.1} would allow to control the decay of these quantities along $C_0$.
Nevertheless this is not relevant here as we will use the harmonic coordinates only to prove the local existence of the vacuum Einstein
characteristic problem in a small region. 
\subsubsection{The harmonic coordinates on the incoming cone.}
We define on $\Cb_0$ the metric components, following Rendall, \cite{rendall:charact},
\bea
\tilde{\ggg}|_{\Cb_0}
\!=\!g'_{11}d{x^1}d{x^1}\!+\!g'_{12}(dx^1dx^2\!+\!dx^2dx^1)\!+\!g'_{1a}(d{x^1}d{x^a}\!+\!d{x^a}d{x^1})\!+\!\ga'_{ab}d{x^a}d{x^b}.\ \ \ \eql{3.16aa}
\eea
Proceeding as before, the analogous of equations \ref{3.23a} are
\bea
&&\Lb=\frac{\partial}{\partial x^2}\nn\\
&&L^*=\overline{\si}\frac{\partial}{\partial x^1}+\overline{X}'\eql{3.39a}\\
&&e'_A={e'_A}^a\frac{\partial}{\partial x^a}\ .\nn
\eea
and for consistency with the previous case we require on $S_0$, $\overline{\si}=2\oomb^2\ , \overline{X}'=0$. The relation between this null frame
and the $\oomb$-frame is
\bea
&&\Lb=(2\oomb)^{-1}e_3\nn\\
&&L^*=\oomb e_4+\frac{|{\overline\de}|^2}{4\oomb}e_3+{\overline\de}_Ae_A\eql{3.40a}\\
&&e'_A=e_A+\frac{{\overline\de}_A}{2\oomb}e_3\nn
\eea
The analogous of relations \ref{3.22qa} are
\bea
&&\frac{\partial}{\partial u}=(2\oom^2-{\underline X}_A\overline{\de}_A)\frac{\partial}{\partial x_2}-{\underline
X}_Ae'^a_A\frac{\partial}{\partial x^a}\nn\\ 
&&\frac{\partial}{\partial\ub}=\overline{\si}\frac{\partial}{\partial
x^1}+\frac{|\overline{\de}|^2}{2}\frac{\partial}{\partial x_2}+\left(\overline{X}'_A-\overline{\de}_A\right){e'_A}^a\frac{\partial}{\partial
x^a}\eql{3.22qb}\\
&&\frac{\partial}{\partial\om^a}=-\theta^A_a{\overline{\de}_A}\frac{\partial}{\partial x^2}
+\theta^A_a{e'_A}^c\frac{\partial}{\partial x^c}\ \nn
\eea
and immediately, 
\bea
&&\frac{\partial x^1}{\partial\ub}=\overline{\si}\ \ ,\ \ \frac{\partial x^2}{\partial\ub}=\frac{|\overline{\de}|^2}{2}\ \ ,\ \ \frac{\partial
x^a}{\partial\ub}=(\overline{X}'^a-\overline{\de}^a)\nn\\
&&\frac{\partial x^1}{\partial u}=0\ \ ,\ \ \frac{\partial x^2}{\partial u}=(2\oom^2-\underline{X}_A\overline{\de}_A)\ ,\
\frac{\partial x^a}{\partial u}=-\underline{X}^a\nn\\
&&\frac{\partial x^1}{\partial\om^b}=0\ \ ,\ \
\frac{\partial x^2}{\partial\om^b}=-\theta^A_b{\overline{\de}_A}\ \ ,\ \ \frac{\partial x^a}{\partial\om^b}=\theta^A_b{e'_A}^a\ .\eql{3.25}
\eea
Equations \ref{3.29} take the form
\bea
&&\left(\frac{\partial}{\partial u}+{\underline X}\right){\overline T}+\frac{\oom\tr\chib}{2}{\overline
T}+\left[\eta(\overline{\de})+\oom^2(\nabb_A\overline\de)_A\right]+\oom\tr\chi(\oom^2-{\underline X}\c\overline{\de})=0\nn\\
&&\left(\frac{\partial}{\partial u}+{\underline X}\right){\overline\si}+\frac{\oom\tr\chib}{2}{\overline\si}=0\eql{3.42wq}\\
&&\left(\frac{\partial}{\partial u}+{\underline X}\right){\overline W}^a+\frac{\oom\tr\chib}{2}{\overline W}^a
+\oom^2\ga^{cb}{\!\ ^{(2)}\!\Ga^a_{cb}} -\eta^a=0\nn
\eea
where
\bea
&&{\overline T}=\frac{|\overline{\de}|^2}{2}\ \ ,\ \ {\overline W}^a=(\overline{X}'_A-\overline{\de}_A){e'}_A^a\ \ \nn\\
&&{\overline\de}={\overline\de}^a\frac{\partial}{\partial\om^a}\ , \ {\underline X}={\underline X}^a\frac{\partial}{\partial\om^a}\ , \ \
{\overline X}'={{\overline X}'}^a\frac{\partial}{\partial x^a}\eql{3.43wq}
\eea
As in the case of the outgoing cone we can control the norms of the solutions of \ref{3.42wq}. The main difference is that in this case moving
forward in time along $\Cb_0$, the radius of the leaves is decreasing, therefore we do not have the analogous of the previous decay estimates.
Moreover, as discussed in the previous section, where the strategy for the global existence proof is discussed, the harmonic coordinates and
the initial data expressed through them are needed only to obtain a local existence proof of the characteristic problem, therefore the way the
harmonic null frame deviates from the $\oom$-null frame moving along the whole cones $C_0$ and $\Cb_0$ is not relevant here.

\begin{Le}\label{L3.2} Let us assume we control the norm $|\c|_{p,S}$, $p\in [2,4]$, for the connection coefficients relative to the
$\oom$-foliation and their first derivatives. On $S_0$ we require that
\[{\overline\de}|_{S_0}=0,\ {\overline\si}|_{S_0}=2\oomb^2,\ \ {\overline X}'^a|_{S_0}=0\ ,\]
Then  equations
\ref{3.42wq} have a solution along the whole $\Cb_0$ such that the following estimates hold:
\bea
{\overline\si}=O(\frac{c_0}{r})\ ,\ {\overline\de}^a=O(\frac{c_0}{r})\ ,\ W^a=O(\frac{\varepsilon c_0}{r^{2}})\ ,
({\overline X}'^a-{\overline\de}^a)=O(\frac{\varepsilon c_0}{r^{2}})\ \ \ \ 
\eea
Moreover ${\overline\de}\c\!X=O(\frac{\ep c_0}{r})$ and $(\nabb_A\overline\de)_A=0$.
\smallskip

{\bf Remark:} Here $c_0$ depends on $r_0=r(\la_1,\nu_0)$.
\end{Le}
{\bf Proof:} The proof goes basically as in the previous Lemma \ref{L3.1} and we do not report it here. We only note that in the $\oomb$-null
frame $\left(\frac{\partial}{\partial u}+{\underline X}\right)=\oomb e_3$ and the transport equations along $\Cb_0$ and their estimates are of the
same type as in Chapter 4 of \cite{Kl-Ni:book}.
\section[Appendix]{Appendix}
\subsection{The construction of the background metric ${\tilde{\ga}}_{ab}$.}
In this subsection we define in a explicit way the background metric  ${\tilde{\ga}}_{ab}$ which is used as the starting point for the ${\cal B}$
map in Lemma \ref{L2.1} and allows also to define explicitely the norm we use in the proof of the ${\cal B}$-fixed point existence.

 Let us start defining $C_0$ as an embedded hypersurface in $\{R^4,\tilde{\ggg}\}$ in the following way:\footnote{The discussion here justifies
the definition \ref{1.36b} of subsection \ref{S2.1}.}
we introduce, in $R^4$, the coordinates $\{\vb,u,\theta,\phi\}$, such that the hypersurface $C_0$ is defined by $u=\la_1$,
\[C_0=\{p\in R^4|u(p)=\la_1\}.\]
We assume that, in these coordinates, the metric $\tilde{\ggg}$ restricted to $C_0$  has the form:
\bea
\tilde{\ggg}|_{C_0}=-\frac{1}{2}(d\vb du+dud\vb)+r'^2(d\theta^2+\sin^2\!\theta d\phi^2)\ ,
\eea
we denote $\tilde{g}'_{\mu\nu}$ the components of $\tilde{\ggg}|_{C_0}$ in these coordinates and $\{{\om'}^a\}=\{\theta,\phi\}$. Therefore
\bea
&&\tilde{g}'_{uu}=\tilde{g}'_{\vb,\vb}=\tilde{g}'_{\vb a}=\tilde{g}'_{ua}=0\\
&&\tilde{g}'_{u\vb}=-\frac{1}{2}\ ,\ \tilde{g}'_{\theta\theta}={r'}^2\ ,\ \tilde{g}'_{\phi\phi}={r'}^2\sin^2\theta\ ;\ r'=\frac{1}{2}(\vb-u)\ \
.\nn
\eea
The inverse metric satisfies
\bea
&&{\tilde{g}'\ \!\!}^{uu}={\tilde{g}'\ \!\!}^{\vb,\vb}={\tilde{g}'\ \!\!}^{\vb a}={\tilde{g}'\ \!\!}^{ua}=0\nn\\
&&{\tilde{g}'\ \!\!}^{u\vb}=-2\ \ ,\ \ {\tilde{g}'\ \!\!}^{\theta\theta}={r'\ \!\!}^{-2}\ \ ,\ \
{\tilde{g}'\ \!\!}^{\phi\phi}={r'\ \!\!}^{-2}\sin^{-2}\theta\ .\nn
\eea
which implies that the eikonal equation $g^{\mu\nu}\partial_{\mu}w\partial_{\nu}w=0$ is satisfied by the functions $w(p)=u$ and $w(p)=\vb$.
Therefore the level surface $C_0$ is a null hypersurface and the tangent vector field of the null geodesics generating
$C_0$ is
\bea
L=2\frac{\partial}{\partial\vb}
\eea
Analogously on $C_0$, the tangent field for the incoming null geodesics is 
\bea
\Lb=2\frac{\partial}{\partial u}\ .\eql{5.5ww}
\eea
Given these vector fields we define a null orthonormal frame on $C_0$, $\{e'_3,e'_4,e_A\}$ as
\bea
e'_3=\Lb\ ,\  e'_4=L\ ,\ e'_A={e'_A}^a\frac{\partial}{\partial\om'^a}\ .
\eea
Let us define on $C_0$ the foliation made by the two dimensional surfaces
\bea
S_0'(\nu')=\{p\in C_0|\vb(p)=\nu'\}\ .
\eea
It follows immediately that the second null (outgoing) fundamental form of the $S_0'(\nu')$ surfaces is
\bea
\tilde\chi'_{ab}\equiv{\tilde{\ggg}}|_{C_0}(D_{\!\!\frac{\partial}{\partial\om^a}}e'_4,\frac{\partial}{\partial\om^b})=
\frac{\partial\mu_{ab}}{\partial\vb}=\frac{1}{2}\tr\tilde\chi'\mu_{ab}
=\frac{1}{r'}\mu_{ab}\eql{5.8} 
\eea
where $\mu$ is the standard metric of $S^2$:
\bea
\mu_{ab}d{\om'}^ad{\om'}^b=r^2(d\theta^2+\sin^2\!\theta d\phi^2)\ .
\eea
From \ref{5.8} it follows that $\tr\tilde\chi'$ satisfies the following equation, with $\oom_0=\frac{1}{2}$,
\bea
\frac{\partial\tr\tilde\chi'}{\partial\vb}+\frac{1}{4}{\tr\tilde\chi'\
\!}^2=\frac{\partial\tr\tilde\chi'}{\partial\vb}+\frac{{\oom}_0}{2}{\tr\tilde\chi'\ \!}^2=0\ .
\eea
Let us introduce on $C_0$ a scalar function ${\oom}={\oom}(\vb,\theta,\phi)$, and assume that, on $C_0$, $\oom$ stays
near to ${\oom}_0$, that is 
\bea
\sup_{C_0}|(1+\vb^2)^{\frac{1}{2}}({\oom}-{\oom}_0)|\leq \varepsilon\ .
\eea
We define on $C_0$ the function $\ub(p)=\ub(\vb,\theta,\phi)$ in the following way:
\bea
\ub(\vb,\theta,\phi)=\nu_0+\int_0^{\vb}\!\frac{1}{4\oom^2(\vb',\theta,\phi)}d\vb'\eql{5.12w}
\eea
and on $C_0$ we define the following leaves, different from the previous $S_0'(\nu')$,
\bea
S_0(\nu)=\{p\in C_0|\ub(p)=\nu\}\ .
\eea
Let us introduce on $C_0$ the new coordinates $\{u,\ub,\om^a\}$, with $\{\om^a\}=\{\om'^a\}=\{\theta,\phi\}$ and write in these
coordinates the restriction $\tilde{\ggg}|_{C_0}$ of the metric $\tilde{\ggg}$ of $R^4$. It is clear that in this case we have some arbitrariness
as the components of the metric transform, under a change $\psi$ of coordinates, through relations which depend on the partial derivatives
$\partial_\mu\psi$, of the transformation.

 Let us consider the following change of coordinates in a neighborhood ${\cal V}\subset R^4$
containing $C_0$, \footnote{We can choose many other changes of coordinates bringing to the same conclusions.}
\bea
&&\ub=\ub(\vb,\theta,\phi)+\psi^0(u,\vb,\theta,\phi)\nn\\
&&u=u \eql{5.14}\\ 
&&{\om}^a={\om'}^a\ \ ,\nn
\eea
which reduces, on $C_0$, to changing only $\vb$ into $\ub$, if $\psi^0|_{C_0}=0$. Nevertheless as we do not require that the first derivatives of
$\psi$ be zero on $C_0$ the expression of the various components of ${\tilde{\ggg}}|_{C_0}$ in the $\{u,\ub,\om^a\}$ coordinates, $g_{\mu\nu}$, depend
on $\psi$.

 We ask then that the change of coordinates in ${\cal U}$ be such that on $C_0$ the vector fields $L^{\mu}=-g^{\mu\nu}\partial_{\nu}u$ and
$\Lb^{\mu}=-g^{\mu\nu}\partial_{\nu}\ub$  be null vector fields. Moreover we require that $L$ be proportional to $\frac{\partial}{\partial\ub}$. These
conditions imply that
\[{\tilde g}_{\ub a}={\tilde g}_{\ub\ \!\!\ub}={\tilde g}^{uu}={\tilde g}^{au}=0\ .\]
Moreover, requiring ${\tilde g}^{\ub\ \!\!\ub}=0$, it follows
that ${\tilde g}_{uu}={\tilde g}^{ab}{\tilde g}_{au}{\tilde g}_{bu}$.\footnote{In fact the following relation holds $g^{\ub
b}=-\ga^{ba}g_{au}g^{\ub u}$ which with the relation $g^{u\ub}g_{uu}=-g^{a\ub}g_{au}$ following from $g^{\ub\ub}=0$, implies the relation.} The change
of coordinates
$\ref{5.14}$ satisfy this last condition if
\bea
\frac{\partial\vb}{\partial u}=-\frac{1}{4}{\tilde g}^{ab}\frac{\partial\vb}{\partial\om^a}\frac{\partial\vb}{\partial\om^b}\ .
\eea
Therefore with this change of coordinates we have:\footnote{The second equation in \ref{5.16w} is obtained deriving equation \ref{5.12w}, where
$v=v(\nu,\theta,\phi)$,\ \ \ \ \ \ \ \ \ \ \ \ \ \ \ \ \ \ \ \ \ \    
$\nu=\ub(\vb(\nu,\theta,\phi),\theta,\phi)=\nu_0+\int_0^{\vb(\nu,\theta,\phi)}\!\frac{1}{4\oom^2(\vb',\theta,\phi)}d\vb'$.}
\bea
&&{\tilde g}_{u\ub}=-\frac{1}{2}\frac{\partial\vb}{\partial\ub}=-2{\oom}^2\nn\\
&&{\tilde g}_{ua}=-\frac{1}{2}\frac{\partial\vb}{\partial\om^a}=
-\frac{{\oom}^2}{2}\int_0^{\vb}\frac{1}{2{\oom}^2}\partial_a\log{\oom}\ \!d\vb'\equiv -{\tilde X}_a\nn\\
&&{\tilde g}_{uu}=-\frac{\partial\vb}{\partial u}={\tilde g}^{ab}{\tilde g}_{au}{\tilde g}_{bu}=|{\tilde X}|^2\eql{5.16w}
\eea
and the metric has, on $C_0$, the following expression:
\bea
\tilde{\ggg}|_{C_0}(\c,\c)=|{\tilde X}|^2du^2-2{\oom}^2(dud\ub+d\ub du)-{\tilde X}_a(dud\om^a+d\om^a du)+{\tilde\ga}_{ab}d\om^ad\om^b\ \ \ \
\eql{5.17sw}
\eea
where ${\tilde\ga}_{ab}={\tilde g}_{ab}$ and ${\tilde g}^{ab}={\tilde\ga}^{ab}$. 
With these definitions we have on $C_0$ \footnote{Observe that while
$L$ is the tangent vector field of the null geodesics generating $C_0$, $\Lb={(2\oom^2)}^{-1}(\frac{\partial}{\partial u}+{\tilde X})$ is a null
vector field defined on $C_0$ different from the vector field $\Lb$ defined in \ref{5.5ww}. This is due to the fact that now $\Lb$ is orthogonal to
the vector fields tangent to the leaves $S_0(\nu)$ instead of $S'_0(\nu')$. Of course $\Lb$ can be considered as the vector field of the null incoming
geodesics starting on $C_0$.}
\bea
L=\frac{1}{2\oom^2}\frac{\partial}{\partial\ub}\ ,\ \Lb=\frac{1}{2\oom^2}\left(\frac{\partial}{\partial u}+{\tilde X}\right)
\eea
and the null orthonormal frame $\{e_4,e_3,e_A\}$, with 
\bea
e_4=2\oom L\ ,\ e_3=2\oom\Lb\ ,\ e_A=e_A^a\frac{\partial}{\partial\om^a}
\eea
and ${\tilde\ga}_{ab}e_A^ae_B^b=\de_{AB}$. Observe also that once we have performed the coordinate change, the vector field
$\frac{\partial}{\partial\om^a}$ has to be interpreted as a tangent vector to the surface $S_0(\nu)$, which implies that the partial derivative is
made keeping
$u$ and $\ub$ constants. The result is that $\tilde{\ggg}|_{C_0}(\frac{\partial}{\partial\om^a},\Nb)=0$, but $\Lie_{\Nb}\frac{\partial}{\partial\om^a}=
[\Nb,\frac{\partial}{\partial\om^a}]\neq 0$, with $\Nb=(\frac{\partial}{\partial u}+X)$. 

 The global effect is that the second null fundamental form of $S_0(\nu)$ relative to $e_4$, $\tilde\chi_{ab}$, is connected in a simple
way to $\tilde\chi'_{ab}$,\footnote{The one relative to $e_3$ has a more complicated expression.}
\bea
{\tilde\chi}_{ab}=\frac{1}{2{\tilde\oom}}\frac{\partial\tilde{\ga}_{ab}}{\partial\ub}
=\frac{1}{2{\oom}}4{\oom}^2\frac{\partial\tilde{\ga}_{ab}}{\partial\vb}=2{\oom}\frac{\partial\tilde{\ga}_{ab}}{\partial\vb}
=2{\oom}{\tilde\chi}'_{ab}
\eea
From the evolution equation satisfied by ${\tilde\chi}'_{ab}$ it follows immediately the equation satisified by $\tr{\tilde\chi}$:
\bea
\frac{\partial\tr{\tilde\chi}}{\partial\ub}+\frac{{\oom}}{2}(\tr{\tilde\chi})^2+2{\oom}{\om}\tr{\tilde\chi}=0
\eea
where $\om$ is defined as 
\[{\om}=-\frac{1}{2\oom}\frac{\partial\log\oom}{\partial\ub}\ \ .\]

\subsection{Proof of Lemma \ref{L2.1}}
The result is obtained in two steps: first we prove the existence of a solution of the system \ref{1.43b} we rewrite in the following way
\bea
&&\frac{\partial\ga}{\partial\nu}-{\oom\tr\chi}\ga-2\oom\chih=0\eql{2.20z}\\
&&\frac{\partial}{\partial\nu}(\oom^{-1}\tr\chi)+\frac{\oom^2}{2}(\oom^{-1}\tr\chi)^2+\oom^{-1}|\chih|^2_{\ga}=0\nn
\eea
which satisfies the following inequalities
\bea
\|\tilde{r}^{-2}\ga-\tilde{r}^{-2}\tilde{\ga}\|\leq c{\varepsilon}\ ,\ 
\|\tr\chi-\tr\tilde{\chi}\|\leq c\frac{\varepsilon\log\tilde{r}}{\tilde{r}^2}\ ,\eql{1.50c}
\eea
with $\!\tr\tilde{\chi}$ the trace of the second fundamental form relative to the metric $\tilde{\ga}$ and the norm$\|\c\|$ is defined, for a
covariant tensor $f$, as
\bea
\|f\|\equiv\sup_{\nu\in C_0}\big(\sup_{\{a_1...a_k\}}|f_{a_1...a_k}|\big)\ .
\eea
Second we show that, given the solution $(\ga,\tr\chi)$, equations \ref{2.20z} provide also an estimate for the
Sobolev  $|\c|_{p,S}$ norms of $\ga$ and $\tr\chi$, relative to the metric $\tilde\ga$,
\footnote{$|f|=|f|_{\tilde\ga}=f_{ab}f_{cd}\tilde{\ga}^{ac}\tilde{\ga}^{bd}$} 
\[|f|_{p,S}(\nu)\equiv\bigg(\int_{S_0(\nu)}|f|^p_{\tilde\ga}d\mu_{\tilde\ga}\bigg)^{\frac{1}{p}}.\]

\NI{\bf Proof of the first step:}
To solve \ref{2.20z} we define a map ${\cal B}$ bringing $(\ga',\tr\chi')$ into $(\ga,\tr\chi)$:
\smallskip

\NI{\bf The map ${\cal B}$:} Given $(\ga',\tr\chi')$, $(\ga,\tr\chi)\equiv{\cal B}[(\ga',\tr\chi')]$ is a solution of the equations
\bea
&&\frac{\partial\ga}{\partial\nu}-{\oom\tr\chi'}\ga-2\oom\chih=0\eql{2.21z}\\
&&\frac{\partial}{\partial\nu}(\oom^{-1}\tr\chi)+\frac{\oom^2}{2}(\oom^{-1}\tr\chi)^2+\oom^{-1}|\chih|^2_{\ga'}=0\ .\nn
\eea
with initial conditions \footnote{On $S_0(\nu_0)$, ${\tilde r}(\nu_0)=r(\nu_0)$. Recall that on $C_0$ ${\tilde r}(\nu)$ is defined, see Lemma
\ref{L2.1}, through the equation $4\pi|{\tilde r}(\nu)|^2=|S_0(\nu)|_{\tilde\ga}$.}
\bea
\ga(\nu_0)(\c,\c)={\tilde\ga}(\nu_0)(\c,\c)={\tilde r}^2(\nu_0)(d\theta^2+\sin^2\theta d\phi^2)
\ \ ,\ \ \tr\chi(\nu_0)=\tr{\tilde\chi}(\nu_0)\ .\ \ \ \ \ 
\eea
The solution of \ref{2.20z} is obtained proving the existence of a fixed point for the map ${\cal B}$. To do it we
show first that ${\cal B}$ sends the points $(\ga',\tr\chi')$ of a closed set $\cal U$ in other points of the same set, where, with $\mu$ the
standard metric of $S^2$,
\bea
{\cal U}=\left\{(\ga,\tr\chi)\bigg| \|{\tilde r}^{-2}\ga-{\mu}\|\leq {\de_2{\varepsilon}}\ ,\ \big\|\tr\chi-\frac{2}{\tilde
r}\big\|\leq\de_1{\varepsilon}\frac{{\log{\tilde r}}}{{\tilde r}^2}\ ,\right\}
\eea
and fixed $\de_1,\de_2>0$.

 Second we show that a contraction holds, namely that, with $\si<1$,
\bea
&&\|{\tilde r}^{-2}(\ga_n-\ga_{n-1})\|\leq \si\|{\tilde r}^{-2}({\ga_{n-2}}-{\ga_{n-3}})\|\nn\\
&&\|\tr\chi_{n-1}-\tr\chi_{n-2}\|\leq \si\|\tr\chi_{n-3}-\tr\chi_{n-4}\|\ .\eql{2.24z} 
\eea
Let us start proving that $\tr\chi\in{\cal U}$ assuming $\ga'\in{\cal U}$. We look at the second equation of \ref{2.21z}
\bea
\frac{\partial}{\partial\nu}(\oom^{-1}\tr\chi)+\frac{\oom^2}{2}(\oom^{-1}\tr\chi)^2+\oom^{-1}|\chih|^2_{\ga'}=0\ .\nn
\eea
This equation is still a non linear one. To look for a solution of it we define
$u=(\oom^{-1}\tr\chi)^{-1}$ and rewrite the previous equation as,
\bea
\frac{\partial u}{\partial\nu}=\frac{\oom^2}{2}+\oom^{-1}|\chih|^2_{\ga'}u^2\ .
\eea
This equation can be solved again by a fixed point method. Define the map $\cal T$ in the following way
\bea
u'\rightarrow u\equiv{\cal T}(u')
\eea
where $u$ is the solution of the equation
\bea
\frac{\partial u}{\partial\nu}=\frac{\oom^2}{2}+\oom^{-1}|\chih|^2_{\ga'}{u'}^2\ ,\eql{2.26z}
\eea
with initial conditions at $\nu_0$. The ball in the function space where $u$ remains after the application of $\cal T$, is defined, with a given
$\de_0>0$, as 
\bea
{\cal O}_{\de_0}\!=\!\left\{\!f\!\in C^{\la}([\nu_0,\infty))\bigg| \sup_{\nu\in C_0}|{\log{\tilde r}}^{-1}(f(\nu)-\oom^2{\tilde r}(\nu))|\leq
\de_0{\varepsilon},f(\nu_0)\!-\!\oom_0^2{\tilde r}(\nu_0)=c_0\varepsilon\!\right\}\ \ \eql{2.27t}
\eea
 In fact let $u'\in {\cal O}_{\de_0}$ then, solving
\ref{2.26z}, we obtain
\bea
|u(\nu)-\oom^2{\tilde r}(\nu)|\!&\leq&\! (c_0+c_1+{\hat c}){\varepsilon}\log{\tilde r}(\nu)
+\left|\int_{\nu_0}^{\nu}\frac{O({\varepsilon}^2)}{{\tilde r}^{5+\ep}}{\tilde r}^2\right|\nn\\
\!&\leq&\! (c_0+c_1+{\hat c}){\varepsilon}\log{\tilde r}(\nu)+C{\varepsilon}^2
\eea
where $c_1$ and $\hat c$ satisfy
\bea
\sup_{\nu\in C_0}|{\log{\tilde r}}^{-1}({\tilde r}(\nu)-\frac{\nu-\la_1}{2})|\leq c_1\varepsilon\ ,\ 
\sup_{\nu\in C_0}|{\tilde r}(\frac{1}{4}-\oom^2)|\leq {\hat c}\varepsilon\eea and choosing ${\varepsilon}$ sufficiently small and
$\de_0>c_0+c_1+{\hat c}$ the right hand side is less than
$\de_0{\varepsilon}\log{\tilde r}(\nu)$, as needed. The contraction mapping is proved in the same way. Integrating the equation for the difference
$u_k-u_{k-1}$, we obtain
\bea
&&\|u_k-u_{k-1}\|\equiv\sup_{\nu\in[\nu_0,\infty)}|u_k-u_{k-1}|(\nu)\leq \int_{\nu_0}^{\nu}\frac{O({\varepsilon}^2)}{{\tilde
r}^{5+\ep}}|u_{k-1}+u_{k-2}||u_{k-1}-u_{k-2}|\nn\\
&&\leq \left(c{\varepsilon}^2 \int_{\nu_0}^{\nu}\frac{1}{{\tilde r}^{4+\ep}}\right)\|u_{k-1}-u_{k-2}\|\leq \si\|u_{k-1}-u_{k-2}\|\ ,
\eea
with $\si<1$. All this amounts to conclude that, given $\ga'\in{\cal U}$ the solution of the second equation of \ref{2.21z} satisfies
\bea
\tr\chi\leq \frac{\oom^{-1}}{{\tilde r}(\nu)}\left(1+\de_0{\varepsilon}\frac{\log{\tilde r}(\nu)}{{\tilde r}(\nu)}\right)\leq
\frac{2}{{\tilde r}(\nu)}\left(1+(c_2+\de_0){\varepsilon}\frac{\log{\tilde r}(\nu)}{2{\tilde r}(\nu)}\right)
\eea
where $c_2$ is defined through the inequality
\bea
|\oom^{-1}-2|\leq \frac{c_2\varepsilon}{\tilde r}\ .
\eea
Therefore, $\tr\chi\in {\cal U}$, choosing $2\de_1\!>c_2+\de_0$. The proof that $\ga\in{\cal U}$ once
$(\ga',\tr\chi')\in{\cal U}$ is immediate. In fact 
\bea
\frac{\partial}{\partial\nu}({\tilde r}^{-2}{\ga})&=&{\tilde r}^{-2}\frac{\partial\ga}{\partial\nu}
-2{\tilde r}^{-3}\frac{\partial{\tilde r}}{\partial\nu}\ga={\tilde r}^{-2}\frac{\partial\ga}{\partial\nu}
-{\tilde r}^{-2}\overline{\oom\tr{\tilde\chi}}\ga\\
&=&({\oom\tr\chi'}-\overline{\oom\tr{\tilde\chi}})({\tilde r}^{-2}\ga)+2\oom\chih{\tilde r}^{-2}\ .\nn
\eea
As $\tr{\chi'}\in{\cal U}$, $({\oom\tr\chi'}-\overline{\oom\tr{\tilde\chi}})\leq {2\de_1{\varepsilon}}{\tilde r}^{-2}\log{\tilde r}$ an
application of Gronwall's Lemma gives
\bea
\sup_{ab}|({\tilde r}^{-2}\ga_{ab})(\nu)-({\tilde r}^{-2}\ga_{ab})(\nu_0)|\leq c_0{\varepsilon}
\eea
which, assuming $\de_2>c_0$, implies that $\ga\in{\cal U}$. To prove the contraction for the $\ga_n$'s we write the
equations satisfied by $\ga_n-\ga_{n-1}$:
\bea
\frac{\partial}{\partial\nu}(\ga_n-\ga_{n-1})-{\oom\tr\chi_{n-1}}(\ga_n-\ga_{n-1})-(\oom\tr\chi_{n-1}-\oom\tr\chi_{n-2})\ga_{n-1}=0\nn
\eea
which we rewrite, as an equation for ${\tilde r}^{-2}(\ga_n-\ga_{n-1})$ in the following way:
\bea
&&\frac{\partial}{\partial\nu}[{\tilde r}^{-2}(\ga_n-\ga_{n-1})]-({\oom\tr\chi_{n-1}}-\overline{\oom\tr{\tilde\chi}})[{\tilde
r}^{-2}(\ga_n-\ga_{n-1})]\nn\\
&&-(\oom\tr\chi_{n-1}-\oom\tr\chi_{n-2})({\tilde r}^{-2}\ga_{n-1})=0
\eea
As $(\ga_n-\ga_{n-1})(\nu_0)=0$, proceeding as before we obtain
\bea
&&|{\tilde r}^{-2}(\ga_n-\ga_{n-1})|(\nu)
\leq c\int_{\nu_0}^{\nu}\frac{1}{{\tilde r}^2}|{\tilde r}^2(\oom\tr\chi_{n-1}-\oom\tr\chi_{n-2})|\\
&&\leq c\sup_{\nu}|{\tilde r}^2(\oom\tr\chi_{n-1}-\oom\tr\chi_{n-2})|\equiv
c\|{\tilde r}^2(\oom\tr\chi_{n-1}-\oom\tr\chi_{n-2})\|\ \ \ \ \ \nn
\eea 
and from it, recalling that $\oom$ is bounded, with a different constant $c$,
\bea
\|{\tilde r}^{-2}(\ga_n-\ga_{n-1})\|\leq c\|\oom^{-1}{\tilde r}^{2}(\tr\chi_{n-1}-\tr\chi_{n-2})\|\ .\eql{2.37t}
\eea

 The difference $\oom^{-1}(\tr\chi_{n-1}-\tr\chi_{n-2})$ satisfies the following equation:
\bea
&&\frac{\partial}{\partial\nu}[\oom^{-1}(\tr\chi_{n-1}-\tr\chi_{n-2})]
+\frac{\oom}{2}(\tr\chi_{n-1}+\tr\chi_{n-2})[\oom^{-1}(\tr\chi_{n-1}-\tr\chi_{n-2})]\nn\\
&&=-\left(\oom^{-1}|\chih|^2_{\ga_{n-2}}-\oom^{-1}|\chih|^2_{\ga_{n-3}}\right)\ .
\eea
As $\tr\chi_n$ and $\tr\chi_{n-1}$ belong to $\cal U$ we can rewrite the previous equation, denoting
$V_n\equiv[\oom^{-1}(\tr\chi_{n-1}-\tr\chi_{n-2})]$, as
\bea
&&\frac{\partial}{\partial\nu}V_n+\overline{\oom\tr\tilde\chi}V_n+
+[\frac{\oom}{2}(\tr\chi_{n-1}+\tr\chi_{n-2})-\overline{\oom\tr\tilde\chi}]V_n\nn\\
&&=-\left(\oom^{-1}|\chih|^2_{\ga_{n-2}}-\oom^{-1}|\chih|^2_{\ga_{n-3}}\right)\ .
\eea
From it immediately
\bea
&&\frac{\partial}{\partial\nu}({\tilde r}^2V_n)+O({\varepsilon}\frac{\log{\tilde r}}{{\tilde r}^2})({\tilde r}^2V_n)
+c{\tilde r}^2(|\chih|^2_{\ga_{n-2}}-|\chih|^2_{\ga_{n-3}})=0\ .
\eea
Applying Gronwall's Lemma we obtain
\bea
&&|({\tilde r}^2V_n)|(\nu)\leq \frac{c}{2}\int_{\nu_0}^{\nu}{\tilde r}^2\chih^{ac}\chih^{bd}
\left[({\ga_{n-2}}_{ac}+{\ga_{n-3}}_{ac})({\ga_{n-2}}_{bd}-{\ga_{n-3}}_{bd})\right.\nn\\
&&\left.+({\ga_{n-2}}_{ac}-{\ga_{n-3}}_{ac})({\ga_{n-2}}_{bd}+{\ga_{n-3}}_{bd}))\right]\nn\\
&&\leq \frac{c}{2}\int_{\nu_0}^{\nu}{\tilde r}^4(\sup_{ab}|\chih^{ab}|)^2|\ga|\sup_{ab}|
({\tilde r}^{-2}{\ga_{n-2}}_{ab}-{\tilde r}^{-2}{\ga_{n-3}}_{ab})|\nn\\
&&\leq\left(\frac{c}{2}\int_{\nu_0}^{\nu}{\tilde r}^4(\sup_{ab}|\chih^{ab}|)^2|\ga|\right)
\|{\tilde r}^{-2}({\ga_{n-2}}-{\ga_{n-3}})\|\\
&&\leq \left(\frac{c}{2}\int_{\nu_0}^{\nu}{\tilde r}^6\frac{{\varepsilon}^2}{{\tilde r}^{9+2\de}}\right)
\|{\tilde r}^{-2}({\ga_{n-2}}-{\ga_{n-3}})\|\leq c{\varepsilon}^2\|{\tilde r}^{-2}({\ga_{n-2}}-{\ga_{n-3}})\|\ .\nn
\eea
 Substituting in \ref{2.37t} we obtain 
\bea
&&\|{\tilde r}^{-2}(\ga_n-\ga_{n-1})\|\leq c\|\oom^{-1}{\tilde r}^{2}(\tr\chi_{n-1}-\tr\chi_{n-2})\|\nn\\
&&\leq c{\varepsilon}^2\|{\tilde r}^{-2}({\ga_{n-2}}-{\ga_{n-3}})\|\ .\eql{2.37tz}
\eea
Therefore we have proved the following inequalities with $\si<1$, chosing $\varepsilon$ sufficiently small,
\bea
&&\|{\tilde r}^{-2}(\ga_n-\ga_{n-1})\|\leq \si\|{\tilde r}^{-2}({\ga_{n-2}}-{\ga_{n-3}})\|\nn\\
&&\|\tr\chi_{n-1}-\tr\chi_{n-2}\|\leq \si\|\tr\chi_{n-3}-\tr\chi_{n-4}\|\ .
\eea
They imply existence of a fixed point for the map ${\cal B}$ and from it a solution of \ref{2.20z}.
\medskip

 Let us prove the second step of our result. The estimate for $\ga$ satisfying \ref{2.20z} is of the following type:
\bea
|\tilde{r}^{(-2-\frac{2}{p})}\ga|_{p,S}(\la_1,\nu)\!&\leq\!&
c_0\left(|\tilde{r}^{(-2-\frac{2}{p})}\ga|_{p,S}(\la_1,\nu_0)+\int_{\nu_0}^{\nu}|\tilde{r}^{(-2-\frac{2}{p})}\chih|_{p,S}d\nu'\right)\nn\\
\!&\leq\!&c_0|\tilde{r}^{(-2-\frac{2}{p})}\ga|_{p,S}(\la_1,\nu_0)+c_1\varepsilon\ .
\eea
This is obtained rewriting the first equation of \ref{2.20z} as
\bea
\frac{\partial\ga_{ab}}{\partial\nu}-{\oom\tr\tilde{\chi}}\ga_{ab}
+\left[\oom(\tr\tilde{\chi}-\tr\chi)\ga_{ab}-2\oom\chih_{ab}\right.\bigg]=0\ \ \ \ \
\eea
and observing that the term $\oom(\tr\tilde{\chi}-\tr\chi)$ behaves, due to the results
of step one, as $O(\varepsilon\log{\tilde r}\ \!\tilde{r}^{-2})$.
To get an estimate for $\tr\chi$ we use the second equation in \ref{2.20z} which we rewrite as
\bea
\frac{\partial\tr\chi}{\partial\nu}+\frac{\oom\tr\tilde{\chi}}{2}\tr\chi
+\big(2\oom\om+\frac{\oom}{2}(\tr{\chi}-\tr\tilde{\chi})\big)\tr\chi+|\chih|^2_{\ga}=0\ \ .
\eea
Again the term $\left(2\oom\om+\frac{\oom}{2}(\tr{\chi}-\tr\tilde{\chi})\right)$ behaves, due to the results of
step one, as $O(\varepsilon\log{\tilde r}\ \!\tilde{r}^{-2})$. From it, applying standard techniques and Gronwall's Lemma we obtain:
\bea
|\tilde{r}^{(1-\frac{2}{p})}\tr\chi|_{p,S}(\la_1,\nu)\!&\leq\!&
c_0\left(|\tilde{r}^{(1-\frac{2}{p})}\tr\chi|_{p,S}(\la_1,\nu_0)+\int_{\nu_0}^{\nu}|\tilde{r}^{(1-\frac{2}{p})}|\chih|_{\ga}^2|_{p,S}d\nu'\right)\nn\\
\!&\leq\!&c_0|\tilde{r}^{(1-\frac{2}{p})}\tr\chi|_{p,S}(\la_1,\nu_0)+c_1\varepsilon^2\ .\eql{1.54c}
\eea
with $c_0=1+c\varepsilon$. From the second inequality in \ref{1.50c}, defining $r(\nu)\equiv |S_0(\nu)|_{\ga}$, it follows immediately that
\bea
\frac{dr}{d\tilde{r}}=\frac{dr}{d\nu}(\frac{d\tilde{r}}{d\nu})^{-1}=1+\frac{1}{\overline{\oom\tr\tilde{\chi}}}\overline{\oom(\tr\chi-\tr\tilde{\chi})} 
\eea
and from it
\bea
\left|\frac{dr}{d\tilde{r}}\right|\leq 1+c\varepsilon\frac{\log{\tilde r}}{\tilde{r}}
\eea
implying that there exist constants $c_1$, $c_2$ bounded by $1+c\varepsilon$, such that
\bea
c_1\tilde{r}\leq r\leq c_2\tilde{r}\ . \eql{1.57c} 
\eea
From \ref{1.50c} and \ref{1.57c} inequalities \ref{1.48c} follow. Moreover we also obtain  that on $C_0$ the following estimate holds:
\bea
|\tr\chi-\overline{\tr\chi}|=O(\varepsilon)\frac{\log r}{r^{2}}\ .
\eea
Finally to prove the last estimate of \ref{1.48c}
\[|r^{3-\frac{2}{p}}\nabb\tr\chi|_{p,S}=O(\varepsilon)\ ,\]
we write the evolution equation along $C_0$ for $\nabb\tr\chi$, see \cite{Kl-Ni:book}, Chapter 4, equation (4.3.4), and obtain the result from the
assumptions of Lemma \ref{L2.1} and an application of Gronwall's lemma.\footnote{The $\de>0$ in the decay assumptions for $\nabb\log\oom$ and
$\nabb\dddd_4\log\oom$ is crucial for this result.}

\bibliographystyle{math}
\bibliography{math}
\end{document}